\documentclass[fleqn,useAMS,usenatbib]{mnras}

\usepackage{newtxtext,newtxmath}
\usepackage[T1]{fontenc}
\usepackage{graphicx}
\usepackage{caption}
\usepackage{tabularx}

\title[CGM anisotropy]{Predictions for anisotropic X-ray signatures in the circumgalactic medium:\\
imprints of supermassive black hole driven outflows}
\author[Truong et al.]{
Nhut Truong$^{1}$\thanks{E-mail: truong@mpia-hd.mpg.de}, Annalisa Pillepich$^{1}$, Dylan Nelson$^{2}$, Norbert Werner$^{3}$, and Lars Hernquist$^{4}$ 
\\
$^{1}$Max-Planck-Institut f{\"u}r Astronomie, K{\"o}nigstuhl 17, 69117 Heidelberg, Germany\\
$^{2}$Universit\"{a}t Heidelberg, Zentrum f\"{u}r Astronomie, Institut f\"{u}r theoretische Astrophysik, Albert-Ueberle-Str. 2, 69120 Heidelberg, Germany\\
$^3$Department of Theoretical Physics and Astrophysics, Faculty of Science, Masaryk University, Kotl\'a\v{r}sk\'a 2, Brno, 611 37, Czech Republic \\
$^4$Institute for Theory and Computation, Harvard-Smithsonian Center for Astrophysics, 60 Garden Street, Cambridge, MA 02138, USA
}

\volume{{\rm submitted}}
\date{}
\pubyear{2021}

\begin{document}
\label{firstpage}
\pagerange{\pageref{firstpage}--\pageref{lastpage}}
\maketitle

\begin{abstract}
The circumgalactic medium (CGM) encodes signatures of the galaxy-formation process, including the interaction of galactic outflows driven by stellar and supermassive black hole (SMBH) feedback with the gaseous halo. Moving beyond spherically symmetric radial profiles, we study the \textit{angular} dependence of CGM properties around $z=0$ massive galaxies in the IllustrisTNG simulations. We characterize the angular signal of density, temperature, and metallicity of the CGM as a function of galaxy stellar mass, halo mass, distance, and SMBH mass, via stacking. TNG predicts that the CGM is anisotropic in its thermodynamical properties and chemical content over a large mass range, $M_*\sim10^{10-11.5}M_\odot$. Along the minor axis directions, gas density is diluted, whereas temperature and metallicity are enhanced. These feedback-induced anisotropies in the CGM have a magnitude of $0.1-0.3$ dex, extend out to the halo virial radius, and peak at Milky Way-like masses, $M_*\sim10^{10.8}M_\odot$. In TNG, this mass scale corresponds to the onset of efficient SMBH feedback and the production of strong outflows. By comparing the anisotropic signals predicted by TNG versus other simulations -- Illustris and EAGLE -- we find that each simulation produces distinct signatures and mass dependencies, implying that this phenomenon is sensitive to the underlying physical models. Finally, we explore X-ray emission as an observable of this CGM anistropy, finding that future X-ray observations, including the eROSITA all-sky survey, will be able to detect and characterize this signal, particularly in terms of an angular modulation of the X-ray hardness.
\end{abstract}

\begin{keywords}
galaxies: evolution -- galaxies: formation -- galaxies: haloes -- galaxies: circumgalactic medium--galaxies: supermassive black holes --- X-ray: galaxies --- methods: numerical
\end{keywords}

\section{Introduction}

The circumgalactic medium (CGM) is the gaseous atmosphere that extends far beyond the stellar body of a galaxy, up to hundreds of kiloparsecs in galactocentric distance, and plays an essential role in the process of galaxy evolution (see \citealt{tumlinson.etal.2017} for a review). The CGM is a unique laboratory where the cosmic baryon cycle of gas inflows and gas outflows plays out. While gas accretion provides the fuel for future star formation and the overall growth of galaxies, gaseous outflows act in opposition, removing baryons from the dense star-forming interstellar medium. Galactic-scale outflows, also referred to as galactic winds, are sourced primarily by stellar feedback in low-mass galaxies \citep{zhang.2018}, active galactic nuclei (AGN) feedback in more massive systems \citep{morganti.2017}, or a combination of the two at intermediate masses. 

AGN feedback, powered by gas accretion onto supermassive black holes (SMBHs), is an important ingredient in current theoretical models of galaxy formation and evolution (\citealt{springel.2005,sdh.2005,booth.schaye.2009,sijacki.etal.2015,weinberger.etal.2017,dave.etal.2019}). 
It enables cosmological simulations to reproduce various observed galaxy properties, both of the stellar body as well as their gaseous atmospheres. It is thought to be the physical mechanism that quenches star formation in massive galaxies \citep{silk.rees.1998,dsh.2005,man.belli.2019}. 
Without AGN feedback, simulations produce overly-massive galaxies inconsistent with observations \citep{khalatyan.etal.2008,lebrun.etal.2014,pillepich.etal.2018}. In addition, AGN feedback is necessary for simulations to reproduce the observed thermodynamical properties of halo gas, from galaxies to groups and cluster mass scales (\citealt{McCarthy.etal.2010,lebrun.etal.2014,henden.etal.2018,nelson.etal.2018b,truong.etal.2018,truong.etal.2020}). 

Observations have inferred that most, if not all, massive galaxies contain a SMBH at their centers \citep{kormendy.ho.2013,van.den.bosch.2016}. Nonetheless, both observationally and theoretically, the exact nature of how AGN feedback interacts with the galaxy, much less the surrounding CGM, remains unclear. 

Observationally, it is suggested that AGN feedback appears in two modes: ``quasar'' and ``radio'' \citep{croton.etal.2006,cattaneo.etal.2009,fabian.2012,McNamara.nulsen.2012,harrison.2017,morganti.2017}. The quasar (or radiative) mode is mostly found in galaxies hosting luminous AGN, where the SMBH has a high accretion rate $\gtrsim1-10\%$ of  Eddington. In this case, feedback energy is mainly released in the form of radiation from the accretion disk. In contrast, a radio mode may instead operate for SMBHs with low accretion rates. In this state, observations suggest that feedback is released in the form of mechanical energy by means of bubbles, cavities, or relativistic jets, with large amounts of energy ($\sim10^{60}$ erg/s) deposited into the surrounding environments via e.g. turbulence and/or shocks (\citealt{nulsen.etal.2009,randall.etal.2011, zhuravleva.etal.2014,hlavacek-larrondo.etal.2015}).

From the perspective of galaxy evolution simulations, successful models for SMBH feedback and galaxy quenching are non-trivial. First, finite numerical resolution implies that this multi-scale feedback physics must be captured in large part by sub-resolution models (e.g., \citealt{schaye.etal.2015}). Different implementations can differ in a number of ways: the number of feedback channels or modes, the form of feedback energy (e.g. thermal versus kinetic or mechanical), the time variability (e.g. continuous or bursty), and the directionality (e.g. isotropic versus bipolar). The diversity of SMBH feedback models among current cosmological simulations, as well as the diversity of their outcomes, makes it clear that the underlying physics and any subsequent regulation of galaxy growth are still uncertain \citep{naab.ostriker.2017}.

The CGM, also called the intracluster medium or ICM for higher galaxy or halo masses, provides an ideal laboratory to study the impact and consequences of SMBH feedback. Some fraction of the feedback energy and momentum injected by a galaxy is expected to be absorbed by this gaseous halo, and galactic winds (or outflows) can modify the global thermodynamical properties of the CGM \citep{werner.etal.2019,eckert.etal.2021}. In addition, outflows also enrich the gaseous halo by transporting heavy elements synthesised within the star-forming regions. Early X-ray observations provided strong evidence for SMBH-powered ultra-fast outflows (e.g., \citealt{reeves.etal.2009, tombesi.etal.2011, tombesi.etal.2012}). They also showed that the distribution of heavy elements in the cores of galaxy clusters can be spatially variable, often aligned with radio jets and cavity axes (\citealt{simionescu.etal.2008,simionescu.etal.2009,kirkpatrick.etal.2009, osullivan.etal.2011,kirkpatrick.etal.2011}).   

In this paper, we use the IllustrisTNG cosmological hydrodynamical simulations (\url{https://www.tng-project.org}/) to quantify the spatially-anisotropic distribution of CGM properties. We devote special attention to massive galaxies, where SMBH feedback is the predominant mechanism that drives galactic outflows \citep{nelson.etal.2019b,pillepich.etal.2021}. In practice, we study if and how the properties of the halo gas depend on its angular position with respect to the central galaxy; i.e. its stellar minor axis. We characterize the CGM structure across the galaxy population by stacking large samples of simulated galaxies, and examine how anisotropies in the CGM can differentiate among different SMBH feedback models, by comparing IllustrisTNG predictions with two other publicly available cosmological simulations: Illustris and EAGLE. Finally, we identify observables with which the predicted CGM anisotropy can be tested. 

Our focus on the IllustrisTNG simulation model is motivated by the fact that this model has been shown to contain many baryonic signatures in the gaseous halos of galaxies, while at the same time broadly reproducing many observed features of these atmospheres. The latter include, for example, the abundances of highly ionized oxygen in OVI, OVII, and OVIII \citep{nelson.etal.2018b}, the X-ray properties of hot halos \citep{truong.etal.2020,truong.etal.2021}, the cool-core fractions of clusters at intermediate redshifts \citep{barnes.2018}, the abundance of cool MgII-traced gas around luminous red galaxies \citep{nelson.2020} and MgII emission from the CGM \citep{nelson.2021}, and the high redshift Lyman-$\alpha$ halos in emission \citep{byrohl.2020}. 
Crucially, according to the IllustrisTNG models, galactic outflows \citep{nelson.etal.2019b} and the CGM metallicity of low-mass galaxies \citep{peroux.etal.2020} are anisotropic, with larger mass outflow rates and higher gas metallicities along the minor axis of galaxies. Furthermore, IllustrisTNG predicts the presence of galactic center-like bubbles in Milky Way/M31-mass galaxies (\citealt{pillepich.etal.2021}) and that SMBH feedback carves lower-density regions in the CGM around massive galaxies, particularly along their minor axis \citep{nelson.etal.2019b, martin-navarro.etal.2021}. 

The paper is arranged as follows. We first introduce the simulations and our analysis methods in Section~\ref{sec:method}. In Section~\ref{sec:fiducial_11_0} we present our main result on anistropic CGM properties: density, temperature, and metallicity. We then study how this signal varies as a function of galaxy properties (Section~\ref{sec:mass_dependence}), its connection to SMBH activity (Section~\ref{sec:smbh_connection}), and related signatures in other gas properties: pressure and entropy (Section~\ref{sec:pressure_entropy}). We compare the three different cosmological simulations in Section~\ref{sec:comparison}, and conclude by discussing X-ray emission predictions in Section~\ref{sec:Xray_implication}. Our main results are summarized in Section~\ref{sec:conclusion}. 


\section{Methods}
\label{sec:method}

\subsection{The TNG100 simulation of IllustrisTNG}
\label{sec:tng100}

We primarily use simulated data from TNG100, a large-volume cosmological magneto-hydrodynamical (MHD) simulation with an extent of $\sim110.7$ cMpc per side. TNG100 is one of the three primary simulations of the IllustrisTNG project (hereafter TNG, \citealt{nelson.etal.2018,volker.etal.2018,marinacci.etal.2018,naiman.etal.2018,pillepich.etal.2018b}). The simulations are run with {\sc arepo} code (\citealt{springel.2010}), in which dark matter, baryons (gas and stars), and black holes are evolved self-consistently according to the equations of self-gravity and MHD within a $\Lambda$CDM universe with cosmological parameters consistent with Planck constraints (\citealt{planck.2016}). All three TNG simulations are publicly available \citep{nelson19a}.

The volume covered by TNG100 is compatible with current X-ray observations of nearby galaxies (within 100 Mpc, see \citealt{truong.etal.2020} for a comparison between TNG100 and current X-ray observations) and also well within the grasp of future all-sky survey of eROSITA X-ray telescope (\citealt{oppenheimer.etal.2020}). Furthermore, the relatively good resolution of TNG100 (baryon mass resolution $m_{\rm baryon}=1.4\times10^6 M_\odot$) allows us to comfortably study CGM gas properties in galaxies down to stellar mass $M_*\sim10^{10}M_\odot$. 

The TNG model for galaxy formation physics (\citealt{weinberger.etal.2017,pillepich.etal.2018}) implements wide-ranging astrophysical processes that are particularly relevant to studies of CGM properties: radiative microphysics including primordial and metal-line radiative cooling in the presence of ionizationing UV background radiation field; star formation and stellar evolution, including chemical enrichment and supernovae feedback \citep{pillepich.etal.2018}; the formation and growth of SMBHs as well as their associated feedback. We briefly present the key features of the stellar and SMBH feedback in the following paragraphs.

The TNG model of stellar feedback is based on the original Illustris model (\citealt{vogelsberger.etal.2013}) with a few modifications (see \citealt{pillepich.etal.2018} for a detailed description). Stellar feedback is realized via wind particles, which are launched isotropically from the star-forming gas cells. The feedback energy carried by the wind is proportional to the local, instantaneous star-forming rate and contains a kinetic as well as thermal component.

In TNG, SMBHs are seeded with a mass of $M_{\rm BH, seed} \sim 10^6 \,\rm{M}_\odot$ in halos that exceed a total mass of $7\times10^{10} \rm{M}_\odot$. SMBHs then grow via gas accretion governed by a Eddington-limited Bondi model (see the governing Equations 1-3 in \citealt{weinberger.etal.2018}), and/or via merging with other SMBHs, following the mergers of their host galaxies and haloes. The TNG model adopts a dual-mode SMBH feedback mechanism, injecting thermal and kinetic energy at high and low accretion rates, respectively \citep{weinberger.etal.2017}. In both modes the amount of feedback enery is proportional to the accretion rate (as described by Equations 7-9 in \citealt{weinberger.etal.2017}). SMBHs transition to the low-state kinetic mode when their Eddington ratio falls below a threshold of $\chi = {\rm min}[0.002(M_{\rm BH} / 10^{8}M_\odot)^2,0.1]$. Due to the black hole galaxy stellar mass relation, this occurs at a characteristic mass scale $M_\star \sim 10^{10.5} \rm{M}_\odot$ corresponding to hosted SMBHs with $M_{\rm BH} \sim 10^8 \rm{M}_\odot$ \citep{nelson.etal.2018,weinberger.etal.2018}. This transition results in the quenching of galaxies, halting their star-formation while also shaping the thermodynamical properties of their gaseous halos \citep{nelson.etal.2018b,zinger.etal.2020,truong.etal.2020,enelson.2021}. 

\subsection{Galaxy selection and gas properties}
\label{sec:properties}

In this work we consider only central galaxies at $z=0$ and with halo mass $M_{\rm 200c}\gtrsim10^{11.5}M_\odot$, where $M_{200c}$ is the total mass within a spherical region with mean density equal to 200 times the critical density of the Universe ($\rho_{\rm crit}$) at the considered redshift: $M_{\rm 200c}=4/3\pi R_{\rm 200c}^3 200\rho_{\rm crit}(z)$ where $R_{200c}$ is the radius of the sphere. At $z=0$, this halo mass selection corresponds to stellar mass $M_*\gtrsim10^{10}M_\odot$, where we define $M_*$ as the mass of stars within a spherical aperture of physical radius $30$ kpc. We also define the following relevant galaxy properties:

\begin{itemize}
    \item {\it Star formation status.} The star formation status of a galaxy is defined according to its relative distance, based on the instantaneous star formation rate (SFR) versus stellar mass plane, from the star-forming main sequence. Following \cite{pillepich.etal.2019} we adopt:
    \begin{itemize}
        \item Star-forming: $\Delta\log_{10}{\rm SFR}>-0.5$.
        \item Quenched: $\Delta\log_{10}{\rm SFR}\leq-1.0$.
    \end{itemize}
    
    \item {\it Galaxy shape.} We use stellar morphology to define galaxy shape via two parameters: the minor-to-major axis ratio (s) and the middle-to-major axis ratio (q). Following \cite{pillepich.etal.2019} (see their Figure~8) we define disk galaxies as those that satisfy two criteria: a minor-to-major ratio $s<0.35$ and a middle-to-major ratio $q>0.65$; otherwise, galaxies are classified as non-disks.\\
    
    \item {\it SMBH kinetic feedback energy}. For each galaxy, $E_{\rm kin}$ is defined as the accumulated feedback energy ever released by its central SMBH while in the kinetic mode.
\end{itemize}

To derive the gas properties we select all non star-forming gas cells that belong to a given halo, as identified by the FoF algorithm. As a result, we include not only gas cells that are gravitationally bound to the central galaxy, but also those that belong to the halo's satellites as well as gas in winds and outflows that are not necessarily gravitationally bound to galaxies. This best mimics observations where it is generally non-trivial to separate e.g. X-ray emission arising from satellites or from high-velocity outflows. In addition to density, temperature, and metallicity, which are direct simulation outputs and that determine the X-ray luminosity of the volume-filling gas, we also explicitly quantify:

\begin{itemize}
    \item Gas pressure: $P_{\rm e}=n_{\rm e}k_{\rm B}T$, where $n_{\rm e}$ is the number density of electrons, and $T$ is the temperature of the gas cell.\\
    
    \item Gas entropy: $K=k_{\rm B}T/n_{\rm e}^{2/3}$.\\
    
    \item X-ray quantities: For most of the theoretical quantification provided in this paper, we compute the intrinsic X-ray emission on a cell-by-cell basis, starting from the gas density, temperature, and metal abundances predicted by the simulations and using the emission model APEC from XSPEC (\citealt{smith.etal.2001}). Namely, for the results of Figures~\ref{fig:Xray}, \ref{fig:Xray_hardness}, \ref{fig:a3}, and \ref{fig:a2}, we compute intrinsic X-ray emission and do not mock observational effects such as galactic absorption or instrumental response. However, to make actual predictions for X-ray observations (i.e. in Section~\ref{sec:detectability} and Figure~\ref{fig:Xray_brightness}), we also create mock photon counts as observed with the eROSITA telescope (\citealt{merloni.etal.2012}). For this task we employ the APEC model plus galactic absorption (i.e. ['wabs(apec)']) with a column density of ${\rm n_H=10^{20}cm^{-2}}$. The mock spectrum, i.e. photon counts as a function of energy, is generated via the \texttt{fakeit} procedure implemented in XSPEC taking into account the eROSITA instrument response files\footnote{\url{https://wiki.mpe.mpg.de/eRosita/erocalib_calibration}}.

\end{itemize}

\subsection{Edge-on projections}
\label{sec:projection}

Throughout the paper, we analyze galaxies and their CGM in edge-on projections. We first rotate each galaxy edge-on, then we create a 2D map of gas properties for all gas within a cube of size $2\times R_{\rm 200c}$ centered on the point with minimum gravitational potential. The edge-on rotation is based on the stellar minor axis of a galaxy, chosen as the total stellar angular momentum of stars within twice the stellar-half mass radius. Our fiducial map resolution is $30\times30$ pixels, which always has a fixed size in units of the virial radius, and each pixel has the same fraction size of $R_{\rm 200c}$. For reference, in our lowest (highest) mass bin, $M_*\sim10^{10.0}M_\odot$ and $R_{\rm 200c}\sim100$ kpc ($M_*\sim10^{11.4}M_\odot$ and $R_{\rm 200c}\sim 570$ kpc), the size of each pixel is $\sim 6$ ($\sim 40$) kpc. For the projection, gas column densities and X-ray fluxes are summed over gas along the line of sight, while other averaged quantities, e.g. gas temperature and metallicity, are taken as the mass-weighted mean of the gas cells along the line of sight. We also examine the X-ray emission-weighted averages of the gas temperature and metallicity and present the results in Appendix~\ref{sec:app_b}.

\subsection{Stacked maps}
\label{sec:stacking}

For stacked maps, the individual galaxy maps are combined pixel-by-pixel via median stacking. We discarded either mean or mode stacking as they emphasize the existence of sub-structure (i.e. satellite galaxies) that exist in only one halo of the stack, confusing our study of the average properties of the diffuse CGM gas.

\subsection{Characterization of the CGM anisotropy}
\label{sec:ratio}

Throughout the paper, the angular location of CGM gas is defined with respect to the stellar minor axis of galaxies (Section~\ref{sec:projection}), so that an azimuthal angle of 0 degrees denotes alignment with a galaxy's major axis, whereas an azimuthal angle of 90 degrees signifies alignment with the minor axis. We make no distinction among the four quadrants of a galaxy's edge-on projection, which are then ``stacked'' together, and a galaxy does not need to be late-type or disk-shaped for a minor or major axis to be defined. 

To characterize the level of anisotropy of the CGM, we use the ratio between properties measured along the minor versus major axes at a given galactocentric distance. Starting from the edge-on projected maps of galaxies, either individual or stacked (Sections~\ref{sec:projection} and \ref{sec:stacking}), we measure the minor (major) axis value for gas at azimuthal angles, in each quadrant, of $>81$ ($<9$) degrees and within a thin shell of $0.04R_{\rm 200c}$ thickness. We label the ratio of these two values the ``minor-to-major ratio'' for CGM properties such as density, temperature, and metallicity. A minor-to-major ratio close to 0 (in the log) denotes isotropy; a minor-to-major ratio larger (smaller) than 0 indicates that a give property is enhanced (suppressed) along the minor axis. 
We measure the minor-to-major metric on a galaxy-by-galaxy basis. Namely, we start from individual galaxy maps and record the minor-to-major ratio of the gas properties for each, at various galactocentric distances. Results for galaxy samples are given in terms of the median ratios across the considered galaxies, and galaxy-to-galaxy variation is represented by e.g. the 32-th and 68-th percentiles of the ratio distribution. In all plots, shaded areas represent the locus of such percentiles, i.e. the $\sigma/2$ galaxy-to-galaxy scatter in the anisotropy. Note that in selected instances, where noted, we will comment on the median minor-to-major ratios that are obtained instead from stacked images of galaxy subsamples. 

\subsection{Illustris and EAGLE Simulations}
\label{sec:othersims}

We compare TNG100 to two other publicly-available cosmological simulations: the original Illustris simulation and EAGLE, which  both have comparable simulated volumes and mass resolutions. 

\begin{itemize}
    \item \textbf{Illustris.} We use the Illustris simulation (\citealt{vogelsberger.etal.2014b,vogelsberger.etal.2014a, genel.etal.2014, sijacki.etal.2015}), which is the predecessor of IllustrisTNG. Illustris is a series of large-scale hydrodynamical simulations, which were also run with AREPO code (\citealt{springel.2010}). We use the highest resolution Illustris-1 run, which simulates a volume of $(106.5 {\rm Mpc})^3$ with baryon mass resolution $m_{\rm baryon}=1.26\times10^6 M_\odot$. The Illustris model of galaxy formation includes radiative cooling (both primordial and metal-line cooling); star formation, stellar evolution, metal enrichment, and stellar feedback from supernovae; SMBHs seeding, growth, as well as their feedback in radio, quasar, and radiative modes. A detailed description of the model is provided in \cite{vogelsberger.etal.2013}.\\
    
    \item \textbf{EAGLE.} We also use the EAGLE simulation (\citealt{schaye.etal.2015,crain.etal.2015}). The EAGLE project is a suite of cosmological hydrodynamical simulations of various simulated volumes and resolutions. For our study, we make use of the flagship run, referred as 'Ref-L0100N1504', which simulates a cosmic box with a comoving side-length of 100 Mpc and a baryon mass resolution of $m_{\rm baryon}=1.81\times10^6 M_\odot$ (\citealt{Mcalpine.etal.2016}). The EAGLE simulations are performed with a modified version of the GADGET-3 smoothed particle hydrodynamics (SPH) code \citep{springel.2005}. They implement a wide range of physical processes including radiative cooling and photoheating, star formation, stellar mass-loss and thermal stellar feedback; procedures for SMBH seeding, growth, as well as SMBH thermal feedback, which we discuss in more detail later. We refer to \cite{schaye.etal.2015} for the EAGLE model description.
\end{itemize}

\begin{figure*}
  \centering
    
    \includegraphics[width=0.32\textwidth]{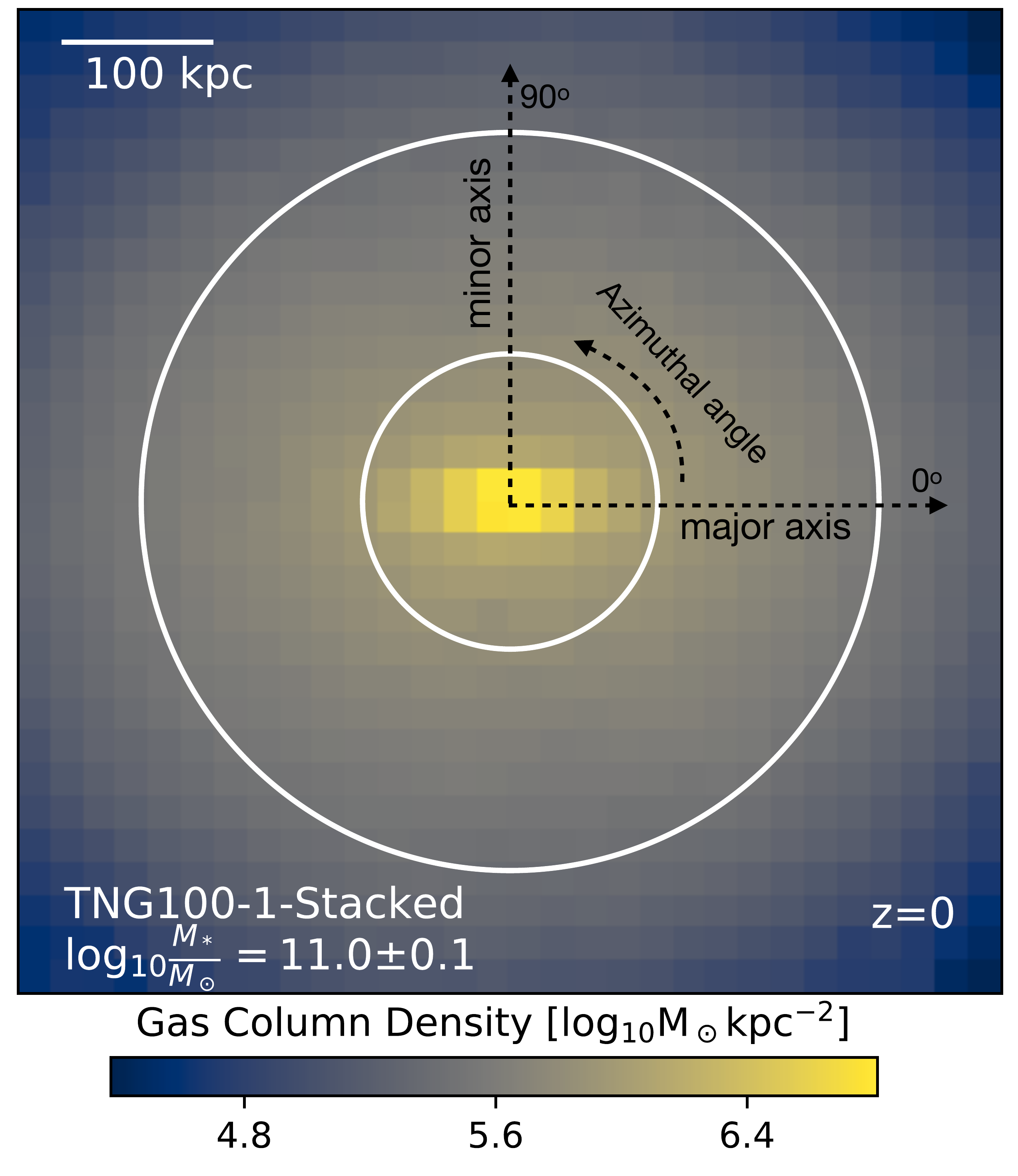}
    \includegraphics[width=0.32\textwidth]{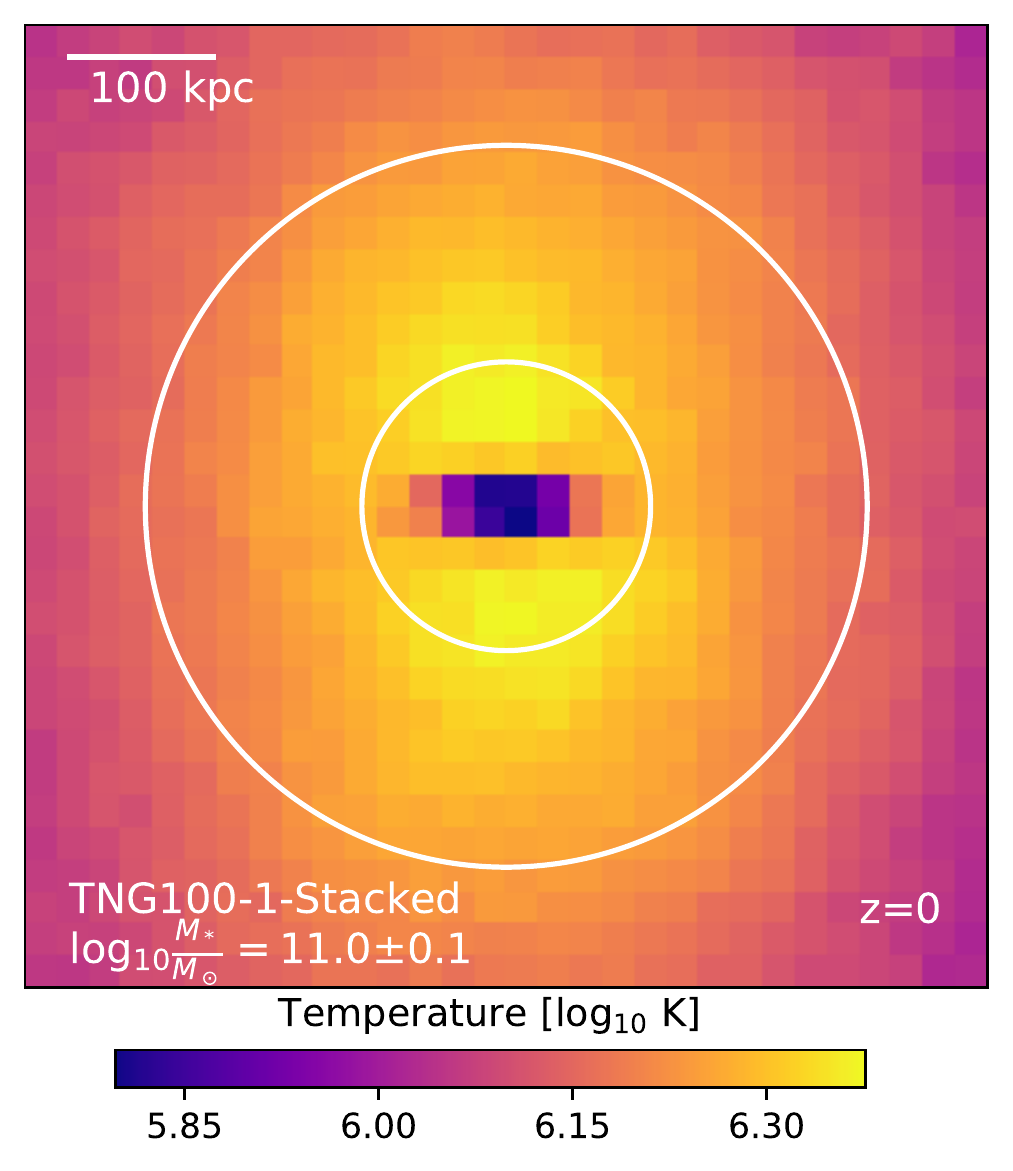}
    \includegraphics[width=0.32\textwidth]{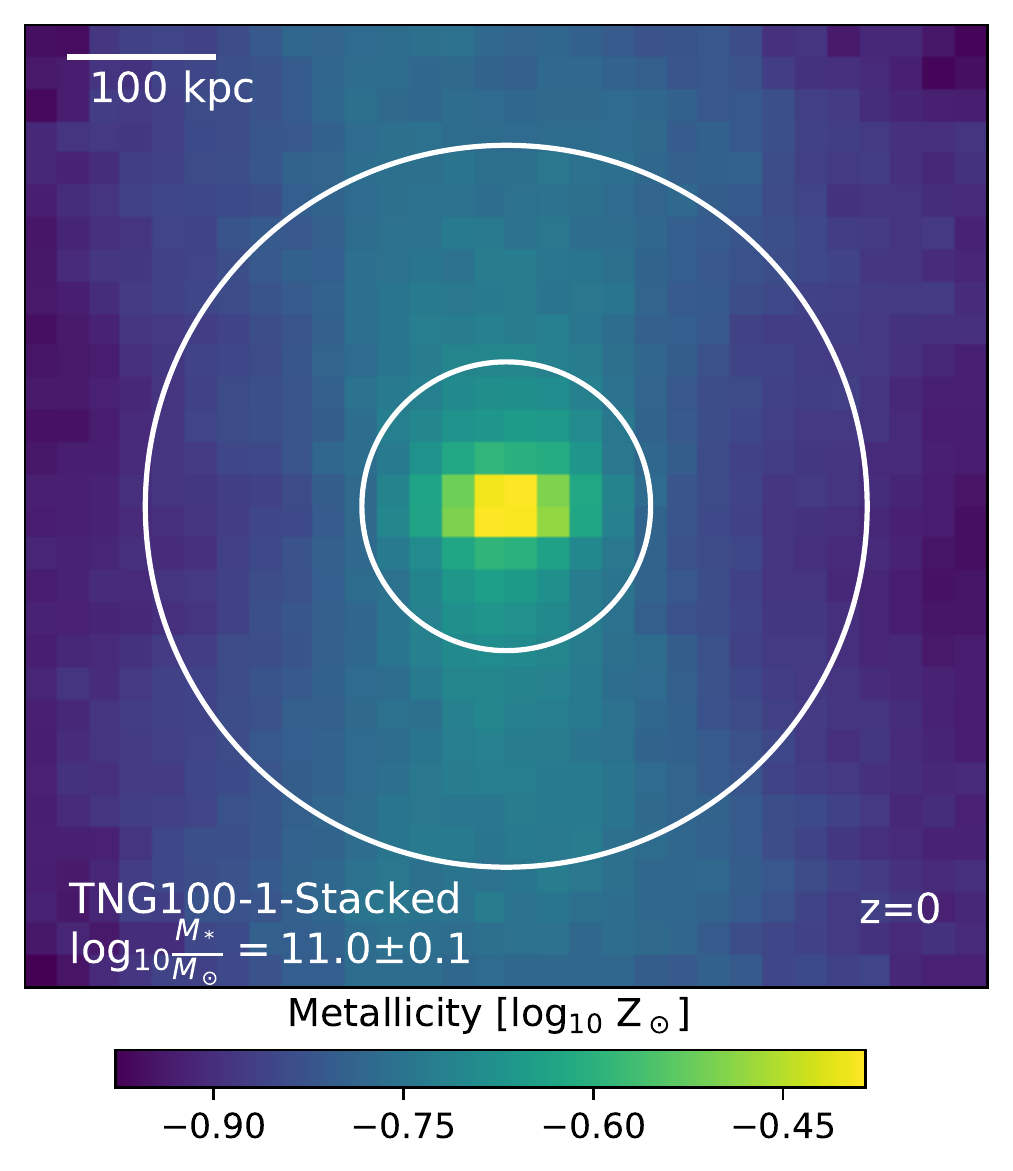}
    \includegraphics[width=0.32\textwidth]{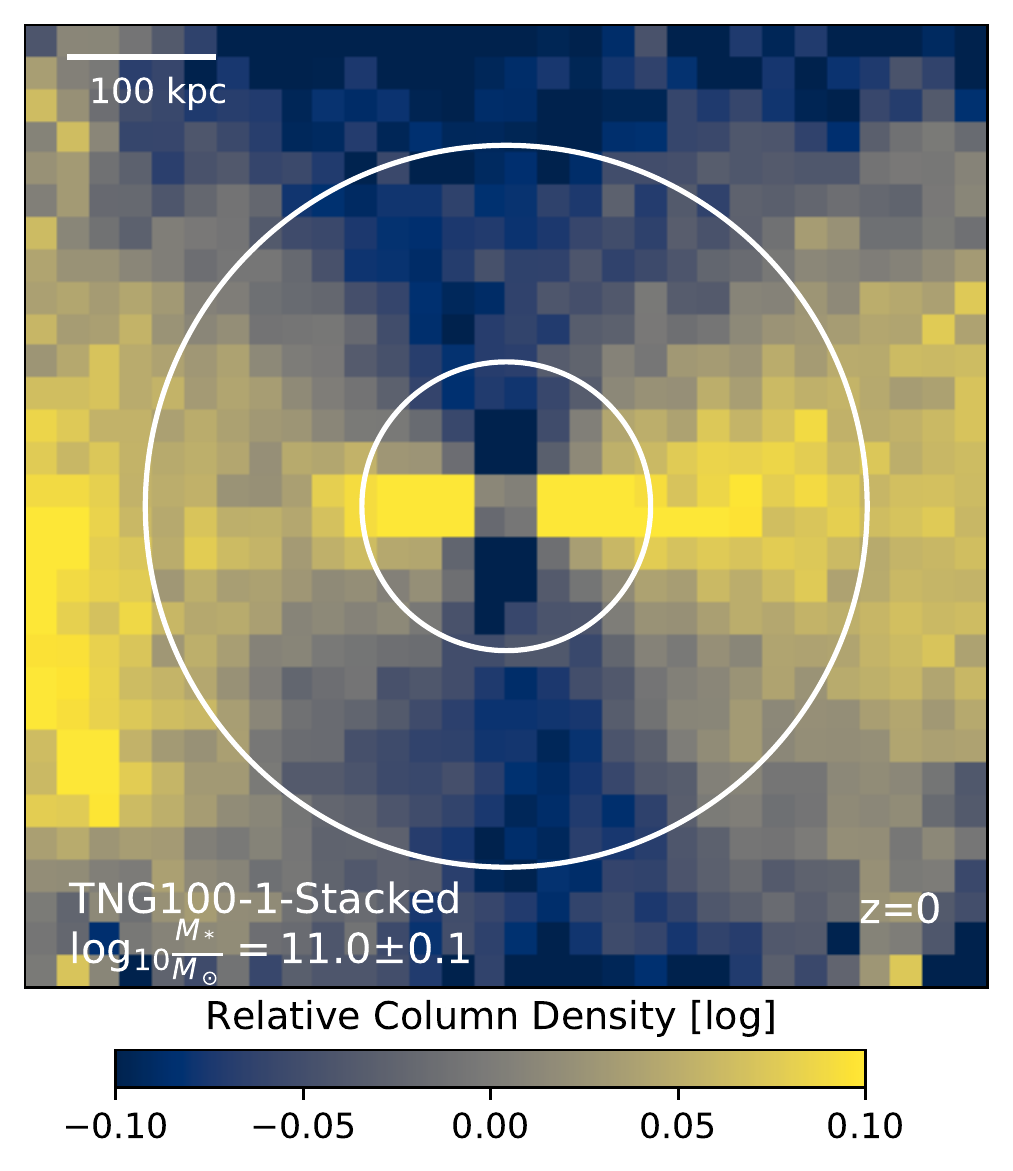}
    \includegraphics[width=0.32\textwidth]{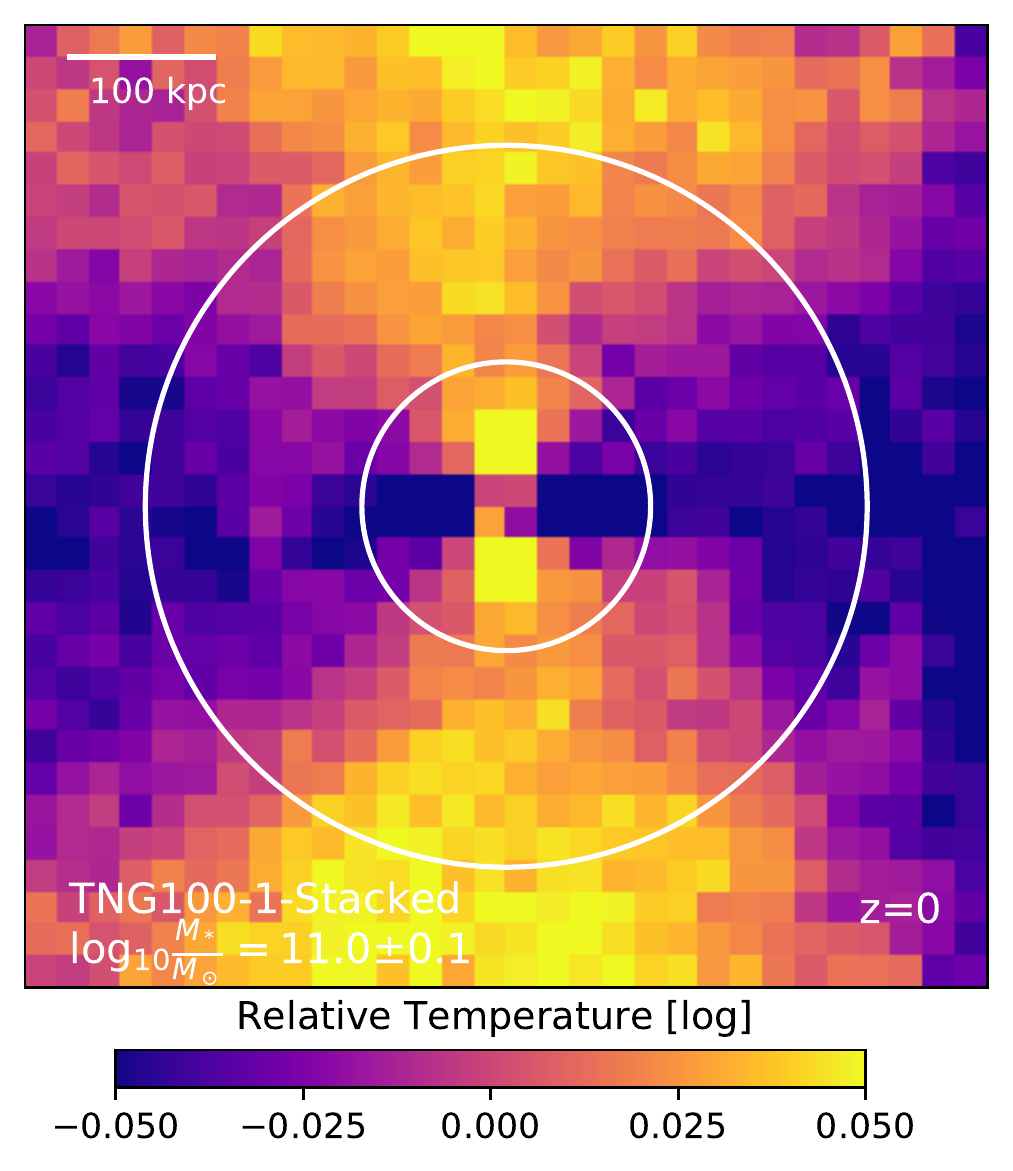}
    \includegraphics[width=0.32\textwidth]{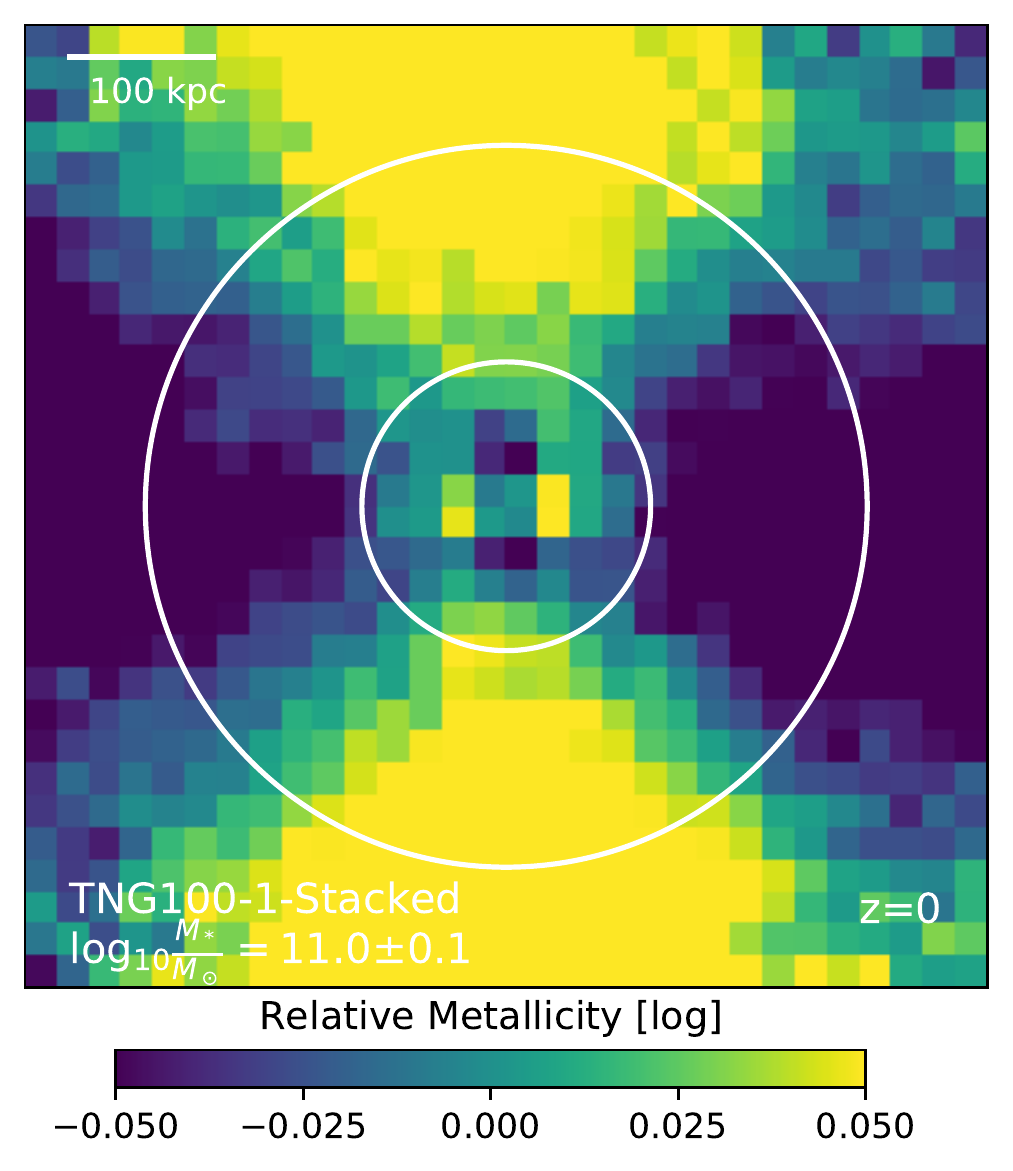}
    \caption{Anisotropy of the CGM properties according to the TNG model at $z=0$. The results are obtained by stacking 277 central galaxies with $M_*=10^{11.0\pm0.1}M_\odot$ from TNG100 (see Sections~\ref{sec:projection} and \ref{sec:stacking} for details). \textbf{\textit{Top row:}} Stacked maps of gas properties: from left to right, gas column density, gas temperature, and gas metallicity. The stacks are made for edge-on galaxies, rotated based on their stellar component, and the maps span a cubic region of $2R_{\rm 200c}$ per side. The two concentric circles show galactocentric distances of $0.25R_{\rm 200c}$ and $0.75R_{\rm 200c}$. \textbf{\textit{Bottom row:}} Relative stacked maps, whereby the gas properties are normalized to their azimuthally-averaged value as a function of distance. Brighter colors indicate an enhancement of the gas properties relative to their azimuthal average. Massive galaxies have angle-dependent CGM properties: the gas is less dense, hotter, and more enriched along the minor-axis directions.}
    \label{fig:signals_11_0}
\end{figure*}
\section{Angular dependence of CGM Properties}

\subsection{CGM anisotropy according to IllustrisTNG}
\label{sec:fiducial_11_0}

\begin{figure*}
  \centering
  \includegraphics[width=0.33\textwidth]{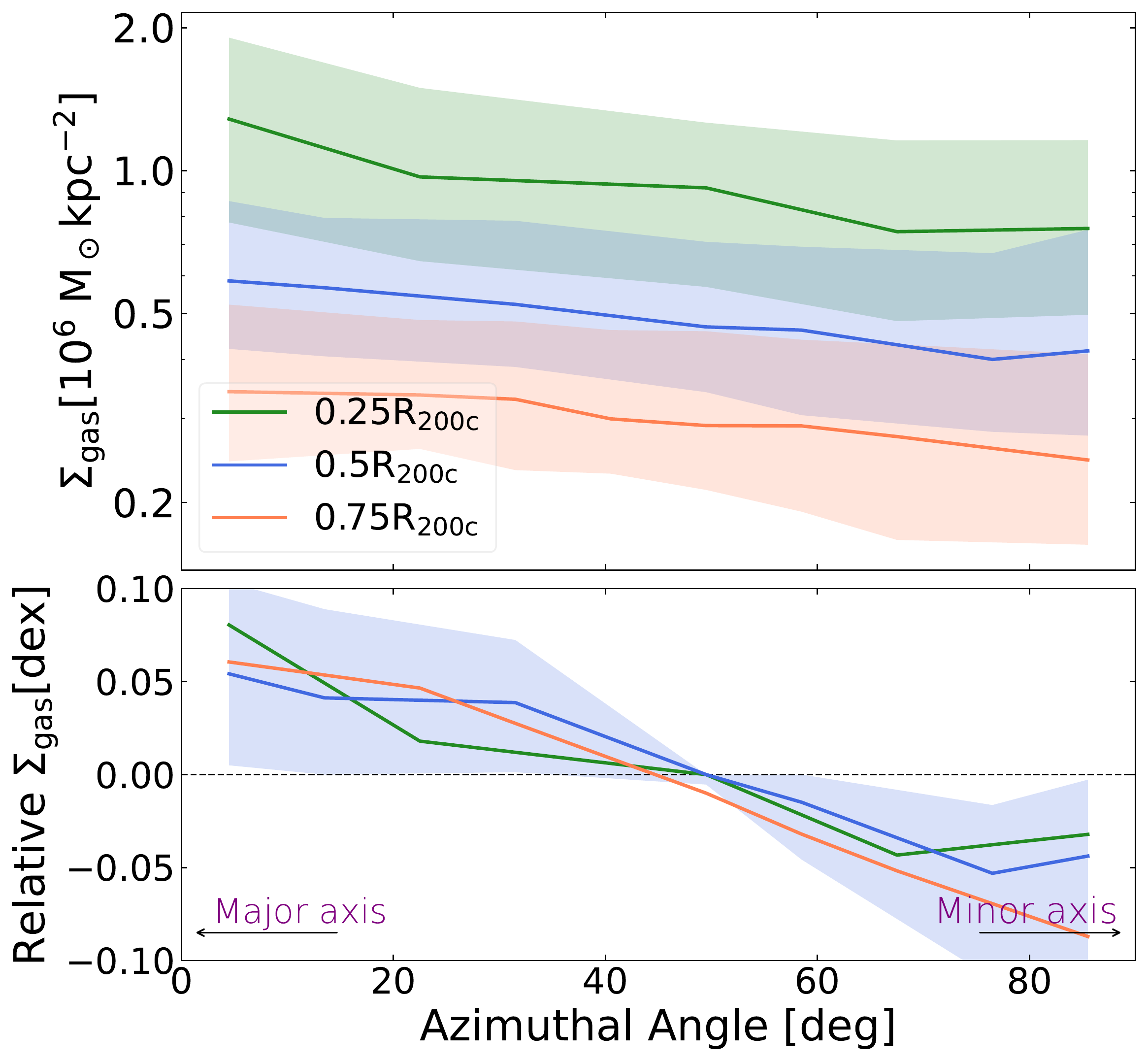}
  \includegraphics[width=0.33\textwidth]{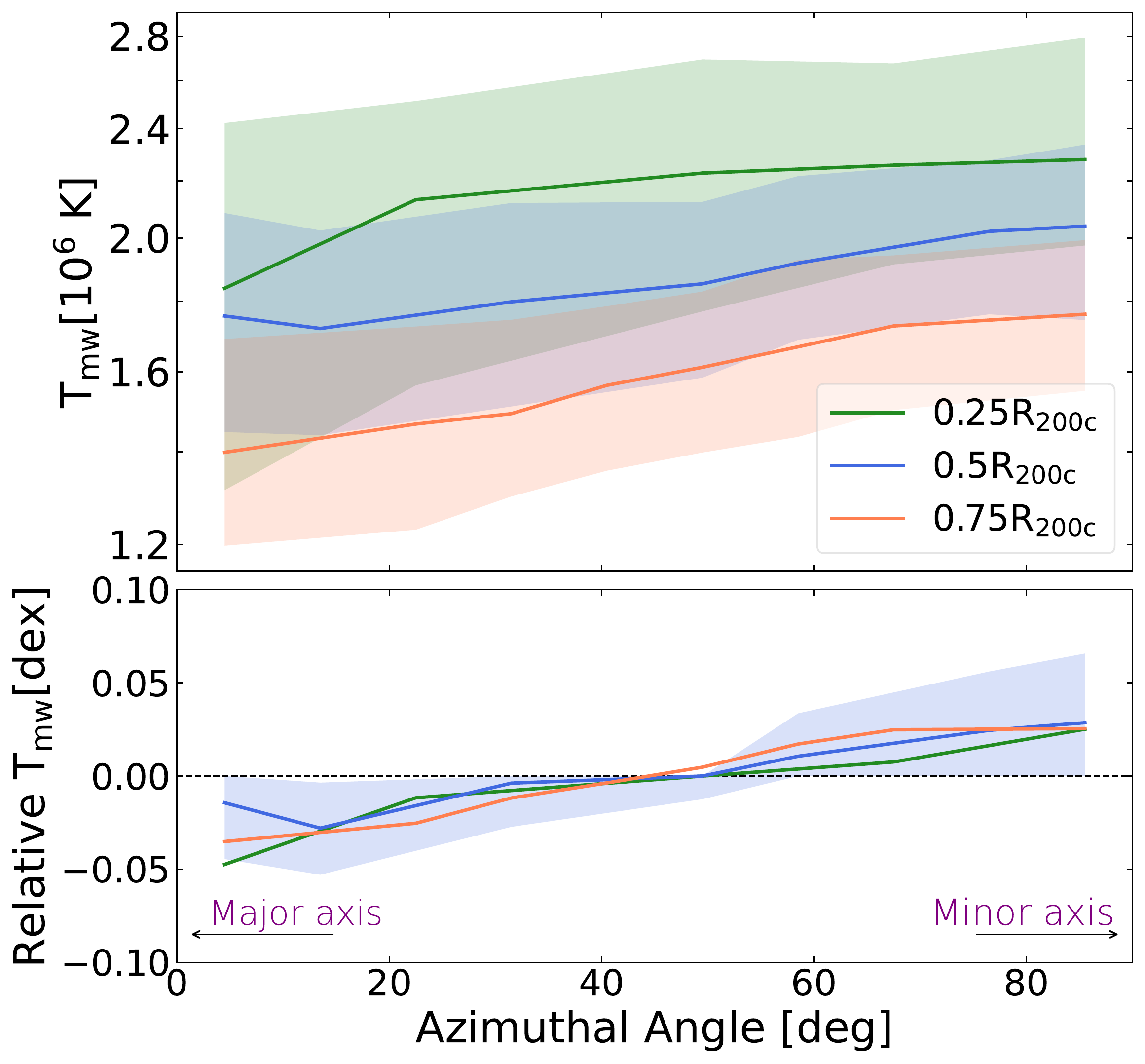}
  \includegraphics[width=0.33\textwidth]{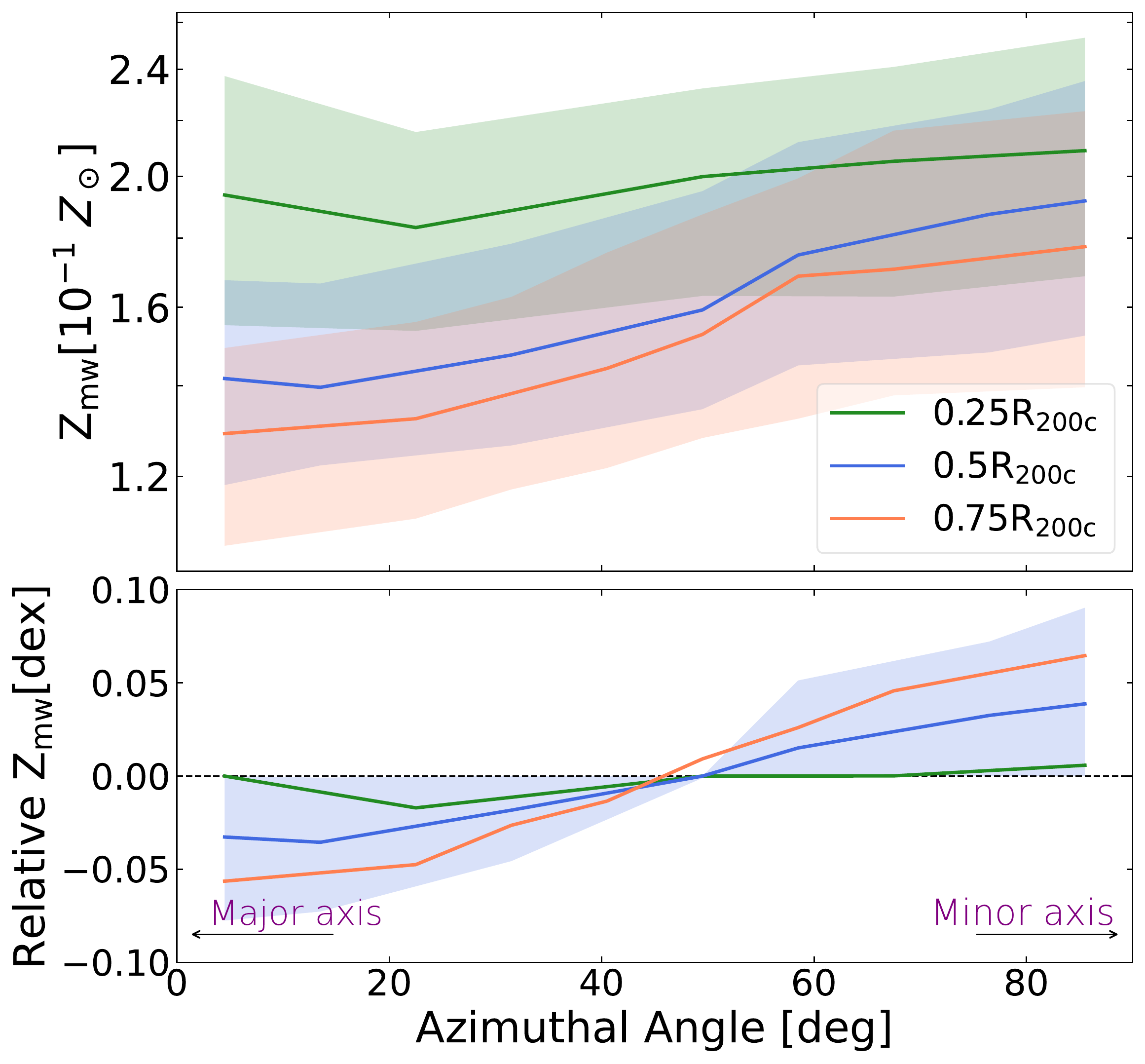}
  \caption{Quantification of the CGM anisotropy predicted by TNG100 and visualized in Figure~\ref{fig:signals_11_0}. We show gas properties as a function of azimuthal angle at varying galactocentric distances (top) and the corresponding relative properties with respect to their azimuthal median (bottom), for TNG100 galaxies at $z=0$ with $M_*=10^{11.0\pm0.1}M_\odot$. The solid lines are the population median, which are derived from individual galaxy maps, while the shaded areas specify the 32-th and 68-th percentiles; i.e. $\sim\sigma/2$ galaxy-to-galaxy intrinsic variation. TNG predicts an angular dependence for CGM properties in massive galaxies, with the gas being more diluted, hotter, and more chemically enriched along the minor axes.}
  \label{fig:signals_11_0b}
\end{figure*}

Figure~\ref{fig:signals_11_0} visualizes how the CGM properties of TNG galaxies at $z=0$ vary with angular location. Results are given for a subsample of TNG100 central massive galaxies with stellar mass $M_*=10^{11\pm0.1}M_\odot$. The {\it top row} presents stacked maps of gas thermodynamical properties; i.e. gas column density ($\Sigma_{\rm gas}$), mass-weighted temperature ($T_{\rm mw}$), and mass-weighted metallicity ($Z_{\rm mw}$). The maps are shown edge-on, with the vertical direction aligned along the minor axis of galaxies (i.e. with respect to their stellar angular momentum -- see Section~\ref{sec:stacking}), whereas the horizontal direction lies on the same plane of the galactic ``disks''. To best visualize the spatially anisotropic distribution of the CGM properties, we then normalize each quantity with respect to its azimuthal average (bottom row), where brighter colors indicate an enhancement of the gas properties with respect to their azimuthal average. 

To quantify these trends, Figure~\ref{fig:signals_11_0b} shows how the gas properties within a projected, thin shell at a certain galactocentric distance vary as a function of azimuthal angle. The azimuthal dependence is computed from individual galaxy maps, and the trends represent the population median behavior, with shaded areas denoting the $\sim\sigma/2$ intrinsic galaxy-to-galaxy scatter.

Figures~\ref{fig:signals_11_0} and \ref{fig:signals_11_0b} show that the TNG model predicts that the CGM of massive galaxies is anisotropic. This anisotropy arises with different dependencies, and to different degrees, for different gas properties. The CGM gas density is lower along the minor axis of galaxies, whereas gas temperature and metallicity are higher. Quantitatively, the amplitude of the anisotropy is largest for gas density and metallicity, reaching on average $\sim 0.1$ dex ($\sim 25\%$), although these two trend with azimuthal angle are in opposite directions. Gas temperature varies with angle less strongly, by $\sim0.05$ dex ($\sim12\%$).

The spatial anisotropies extend from the central regions of haloes ($r\sim 0.25R_{\rm 200c}$) all the way to the canonical halo boundaries ($r\sim R_{\rm 200c}$). According to TNG, the relative anisotropy of gas and temperature does not depend on galactocentric distance, at this mass, whereas the angular dependence of the gas metallicity is stronger at larger distances. We note that the galaxy-to-galaxy scatter is comparable or even larger than the amplitude of the signals, hinting that the degree of anisotropy may depend on further galaxy properties beyond mass. For the sake of clarity, we opt to report the intrinsic scatter in term of the interval between the 32th and 68th percentile elements.  

In TNG, the relative under-density of CGM gas along the minor axis of $\sim10^{11} M_\odot$ galaxies -- with these regions being simultaneously also hotter and more metal-enriched -- is a consequence of galactic outflows. Although the energy injection from both stellar and SMBH feedback is isotropic in the TNG model, galactic-scale outflows tend to emerge along paths of least resistance creating bipolar structures oriented along the minor axis of galaxies \citep{nelson.etal.2019b}. When advancing to larger galactocentric altitudes, such outflows push gas out of the central regions of haloes \citep{terrazas.etal.2020, zinger.etal.2020, davies.etal.2020} while at the same time thermalizing and heating up the gas via shocks \citep{weinberger.etal.2017, pillepich.etal.2021}. Metals that have been previously produced in the central star-forming regions of galaxies are carried by outflows, thereby enriching the gas above and below the galactic disks \citep{pillepich.etal.2021}. The radial dependence of the CGM metallicity anisotropy can be explained by the changing contrast of metal-poor gas inflows, preferentially along the galactic planes, and metal-rich outflows \citep{peroux.etal.2020}.

\subsection{Mass trends of the CGM anisotropy}
\label{sec:mass_dependence}
\begin{figure*}
  \includegraphics[width=0.33\textwidth]{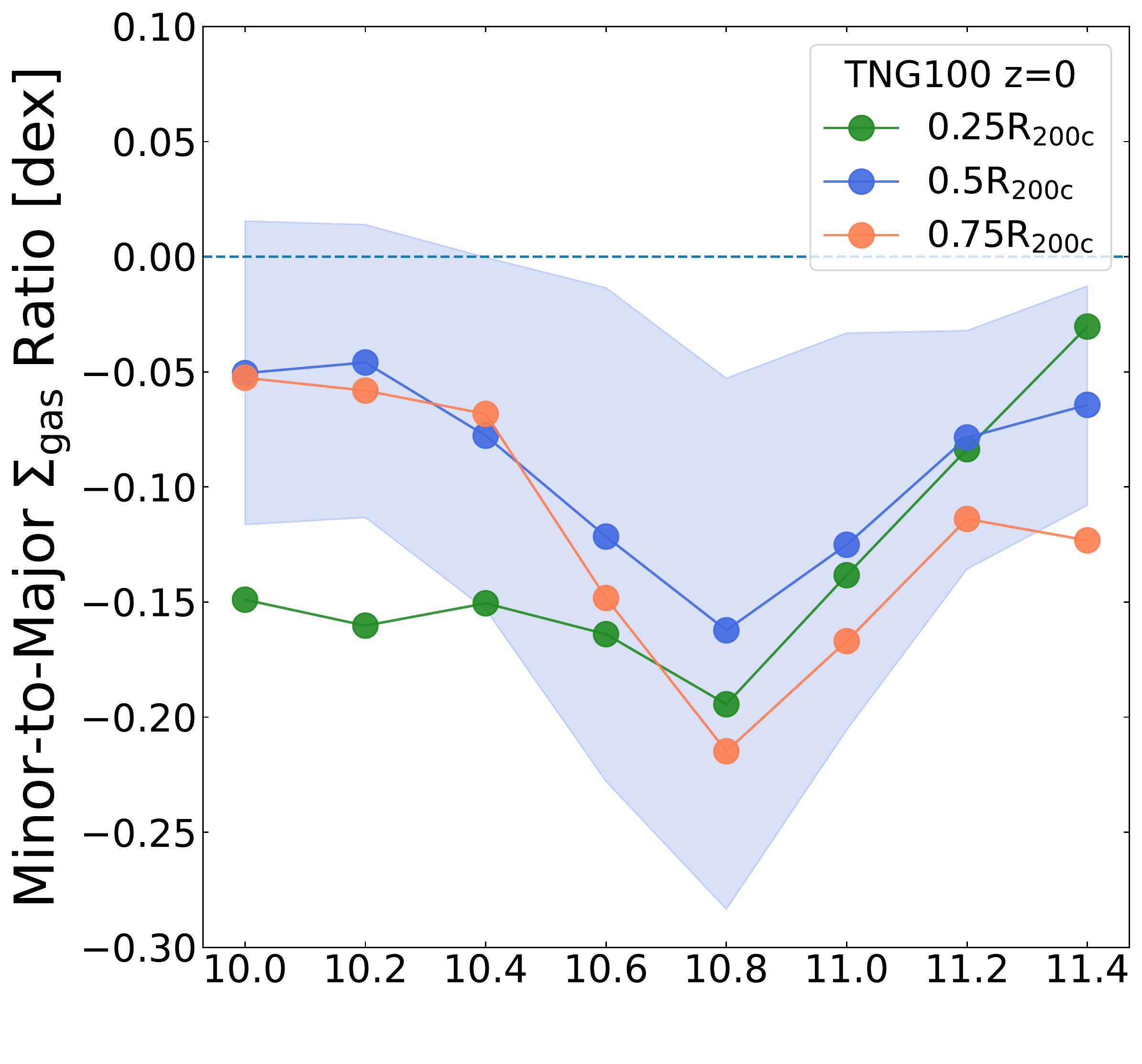}
  \includegraphics[width=0.33\textwidth]{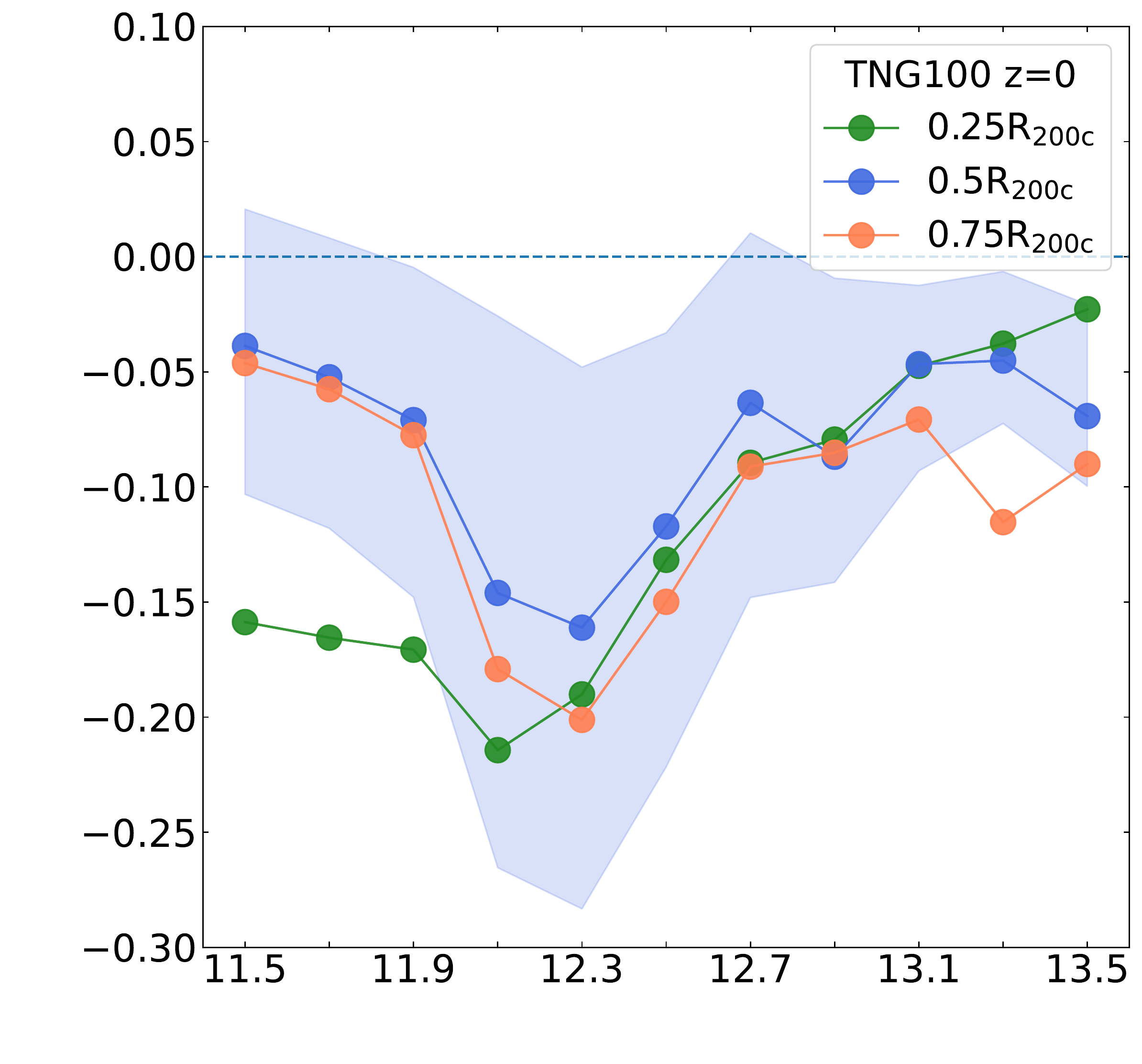}  
  \includegraphics[width=0.33\textwidth]{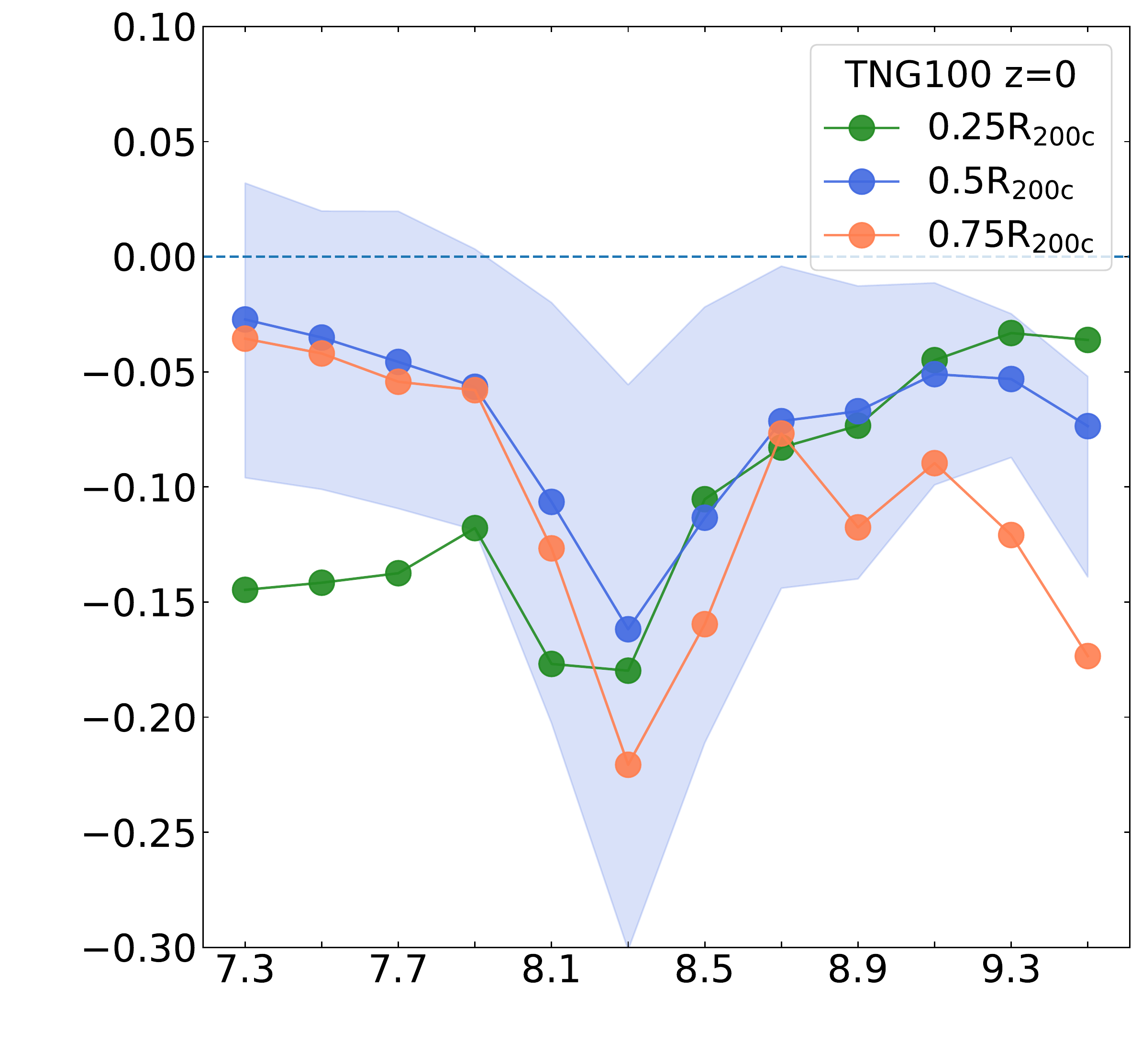}
  \includegraphics[width=0.33\textwidth]{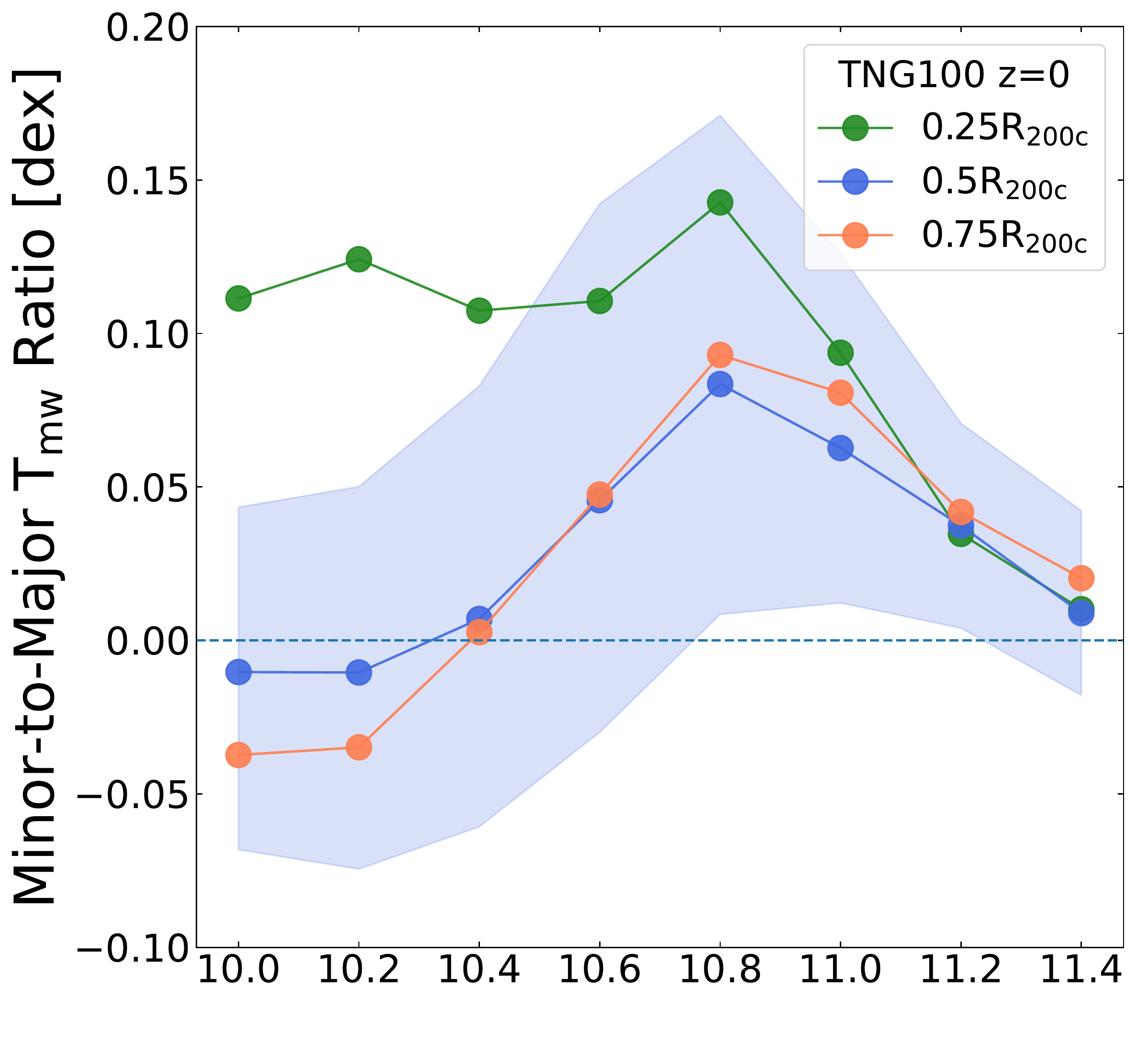}
  \includegraphics[width=0.33\textwidth]{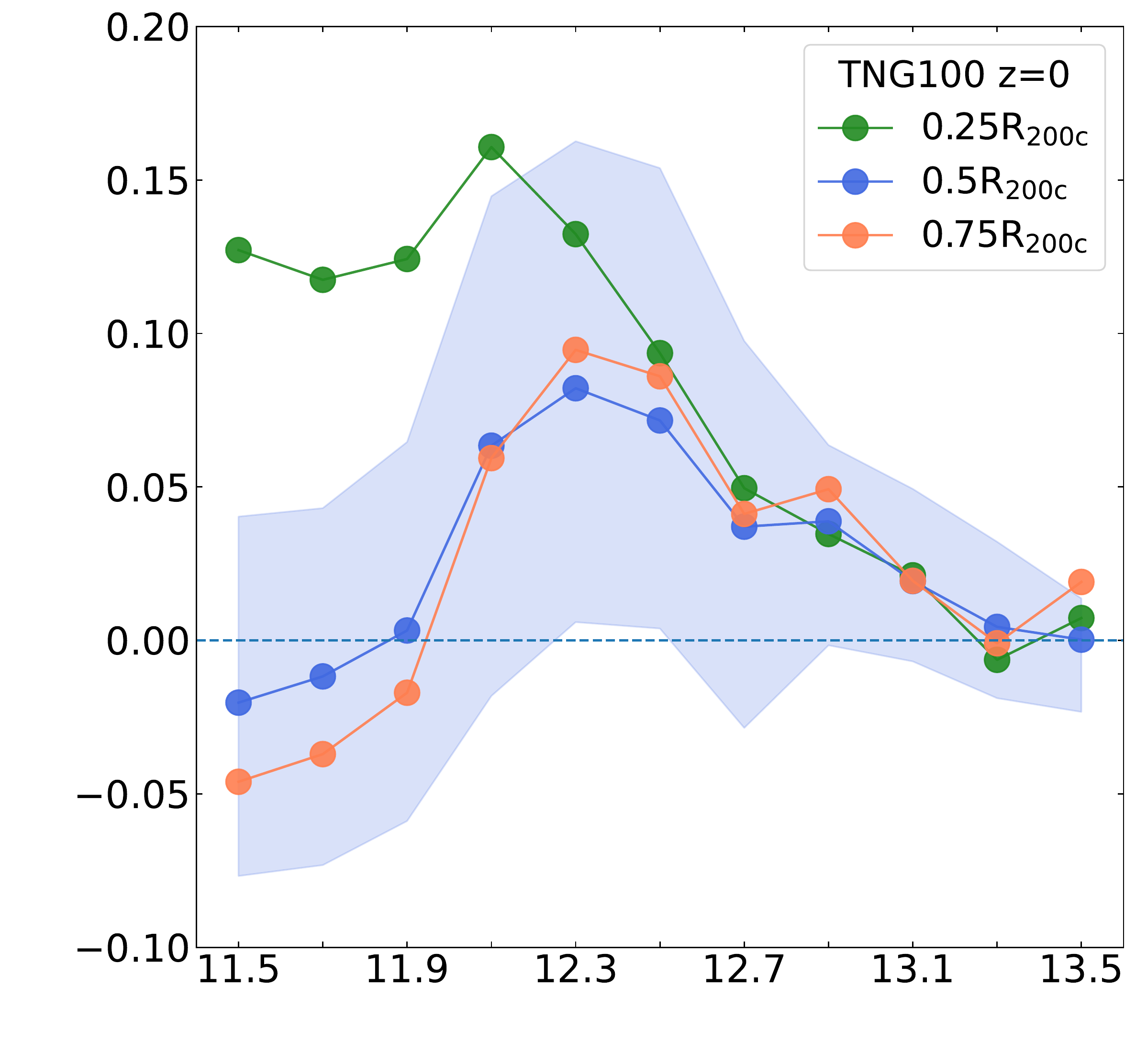}  
  \includegraphics[width=0.33\textwidth]{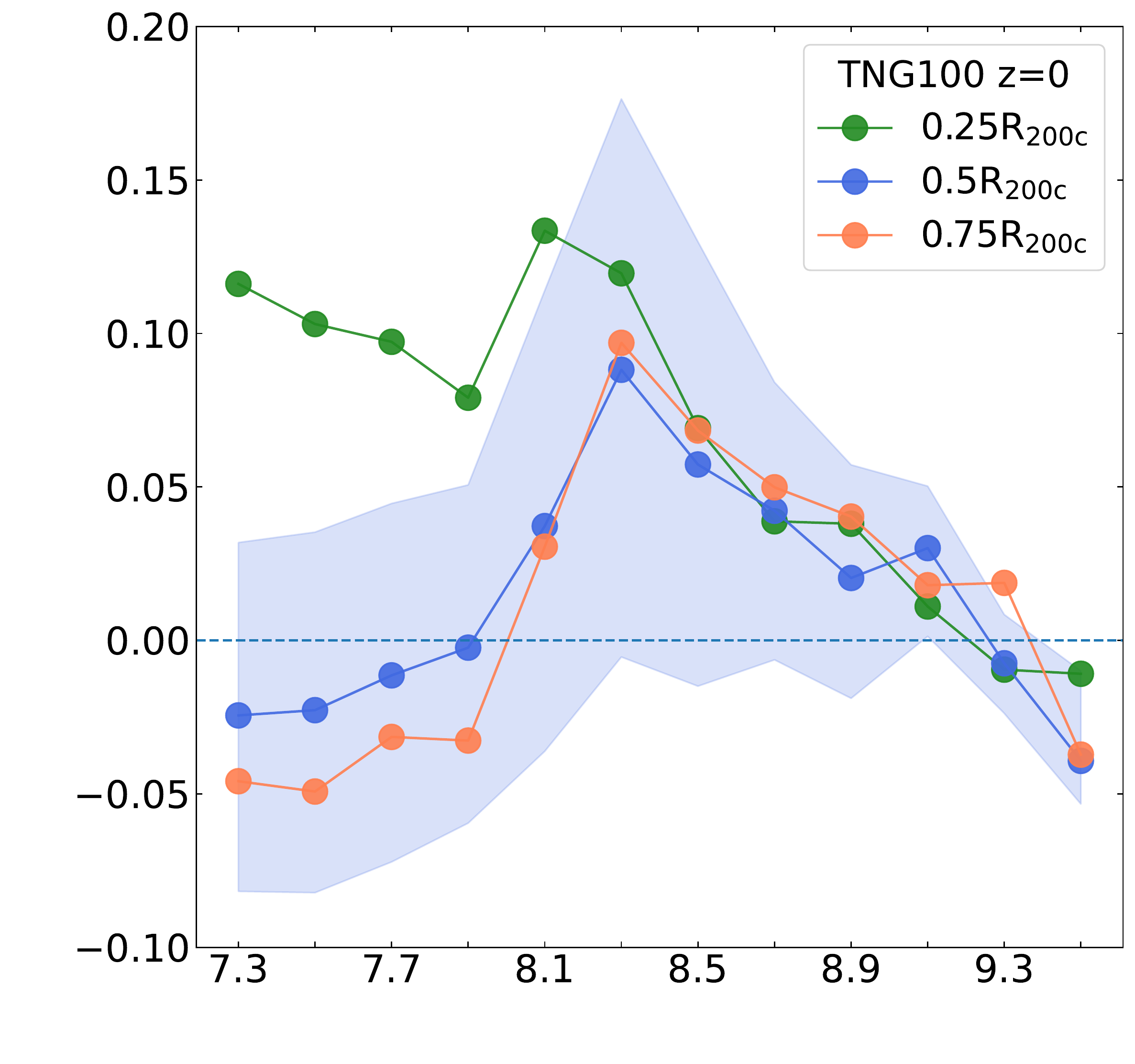}
  \includegraphics[width=0.33\textwidth]{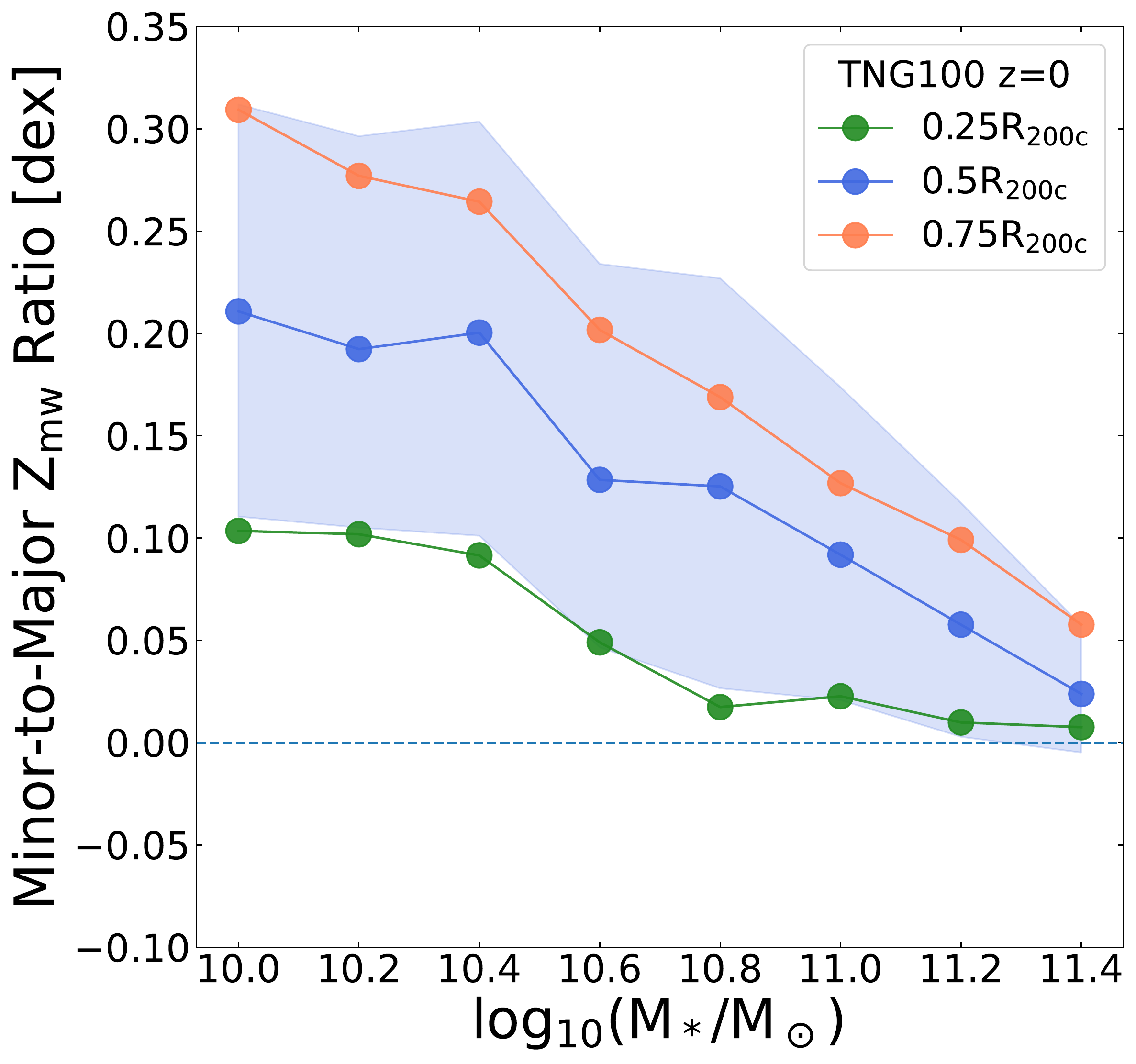}
  \includegraphics[width=0.33\textwidth]{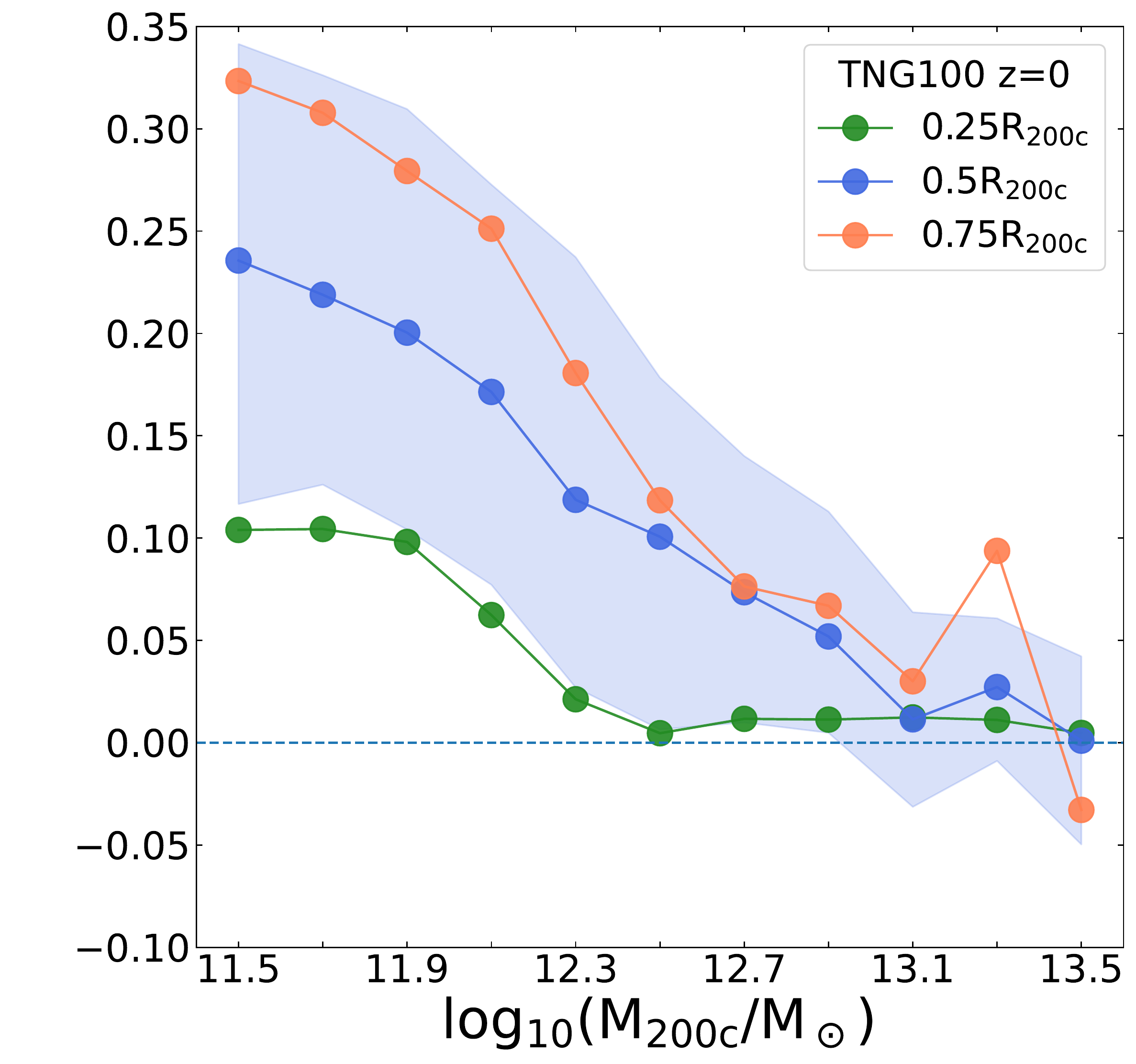}  
  \includegraphics[width=0.33\textwidth]{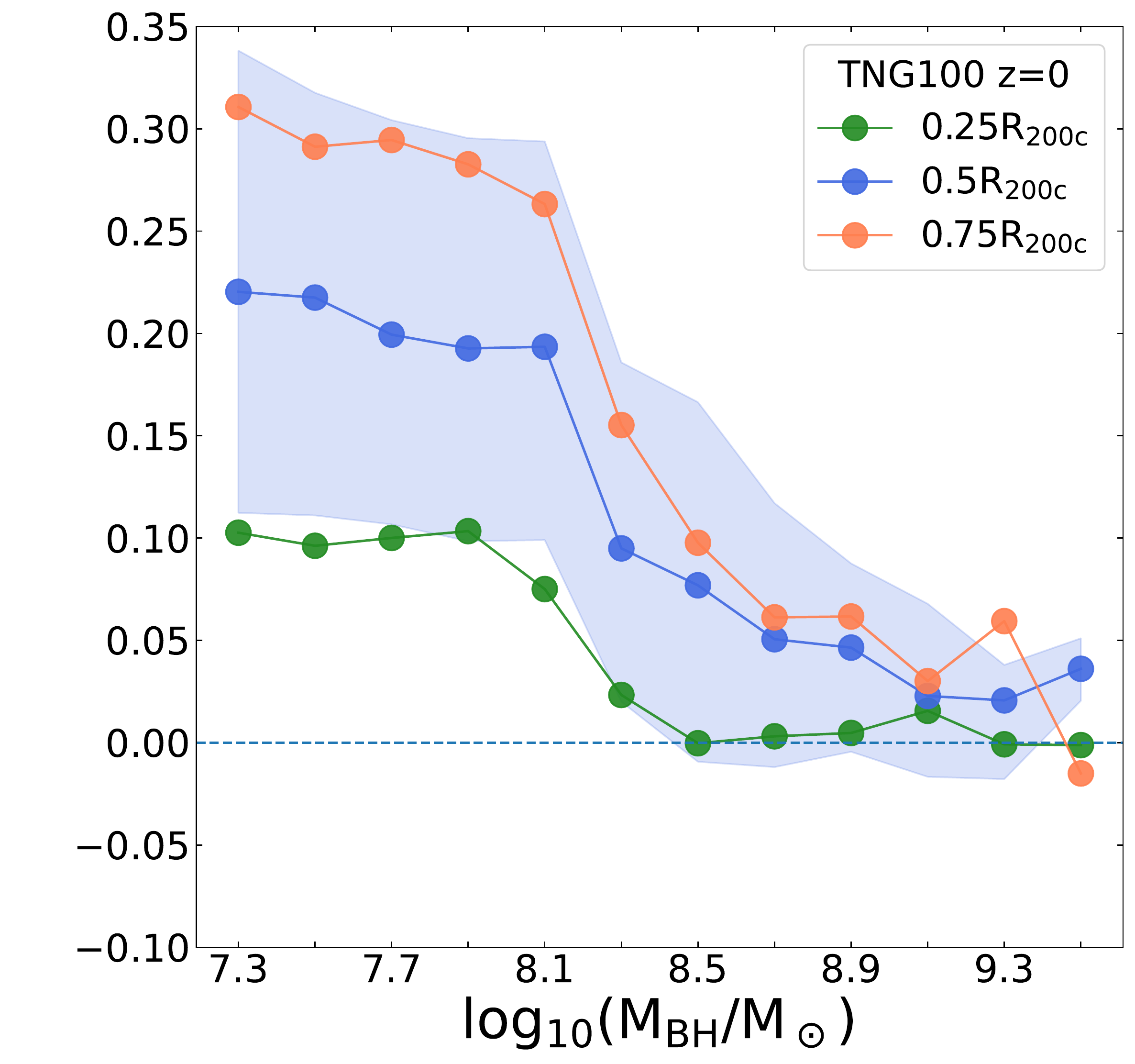}

  \caption{Mass dependence of the CGM anisotropy for TNG100 galaxies at $z=0$, quantified with the minor-to-major ratio (see Section~\ref{sec:ratio}) of galaxies in a given mass bin, measured at various galactocentric distances. From top to bottom, we show results for gas column density, mass-weighted gas temperature, and mass-weighted gas metallicity. Different columns give the dependence on different masses: galaxy stellar mass ($M_*$, {\it left}), total halo mass ($M_{\rm 200c}$, {\it middle}), and SMBH mass ($M_{\rm BH}$, {\it right}). The symbols and solid lines represent the galaxy-population median, whereas the grey shaded area is the 32th-to-68th percentile interval (i.e. $\sim\sigma/2$) for the measurement at $r=0.5R_{\rm 200c}$, i.e. representing the galaxy-to-galaxy variation. According to TNG, the CGM anisotropy of gas density and gas temperature is maximal at the transitional mass scale of $M_*\simeq10^{10.5-11}M_\odot$, i.e. $M_{\rm 200c}\simeq10^{12.1-12.5}M_\odot$ and $M_{\rm BH}\simeq10^{8.0-8.5}M_\odot$.}
 \label{fig:mass_dependence}
\end{figure*}
We proceed to quantify how the CGM anisotropy varies with galaxy mass, for central galaxies with stellar mass larger than $M_*\sim10^{10}M_\odot$ (see Section~\ref{sec:properties}), using the minor-to-major ratio, which is derived from individual galaxy maps as described in Section~\ref{sec:ratio}. 

Figure~\ref{fig:mass_dependence} shows the level of anisotropy of gas density ({\it first row}), temperature ({\it second row}), and metallicity ({\it third row}) as a function of various masses: galaxy stellar mass ($M_*$, {\it first column}), total halo mass ($M_{\rm 200c}$, {\it second column}), and SMBH mass ($M_{\rm BH}$, {\it third column}), at three galactocentric distances.  

According to the TNG model, the anisotropy of gas density and temperature is non-monotonic with mass: the angular modulation increases with mass at the low-mass end, peaks at a maximum, and then decreases with mass at the high-mass end, irrespective of galactocentric distance. The anisotropy is maximized at the same galaxy stellar mass for both density and temperature, $M_*\sim10^{10.5-11}M_\odot$. However, as shown in the previous Section, the gas density and temperature display anisotropic signals in opposite directions, and this pattern is consistent across the considered mass range. For instance, at the peak of CGM anisotropy, TNG galaxies exhibit under-dense halo gas along the minor axis (by $\sim0.15-0.20$ dex), whereas the gas temperature is higher (by $\sim0.05-0.15$ dex). The level of density and temperature angular dependence is roughly distance independent at the high-mass end (consistent with the results of Figure~\ref{fig:signals_11_0} for $M_*\sim10^{11}M_\odot$ galaxies), whereas at the low-mass end gas density and temperature are more anisotropic at regions closer to the galaxies center (see e.g. the green curves at $r\sim0.25 R_{\rm 200c}$).
\begin{figure*}
  \includegraphics[width=0.8\textwidth]{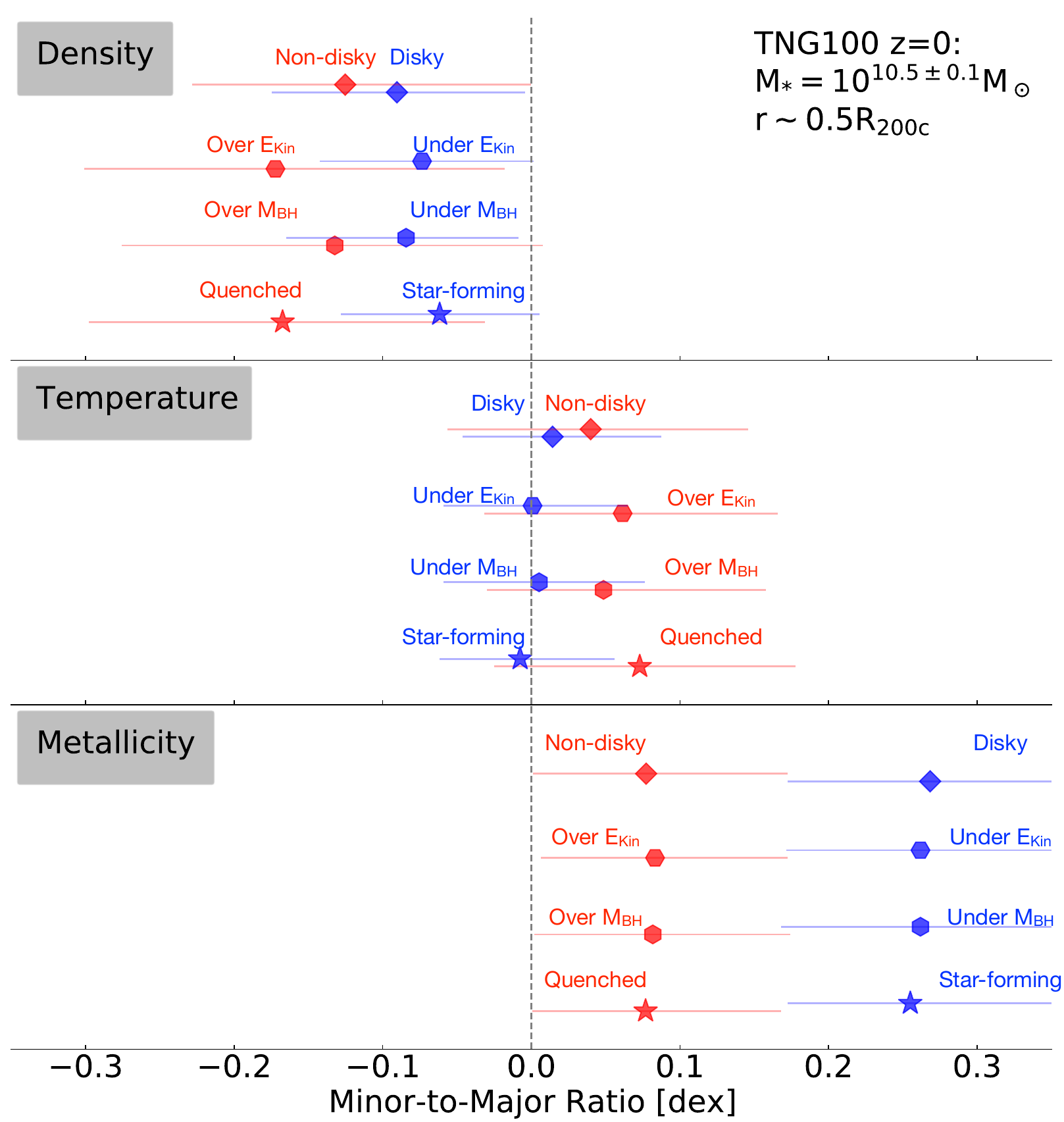}
  \caption{Dependence of the CGM anisotropy on various galaxy properties according to the TNG model. From {\it top} to {\it bottom}, we give results for $\Sigma_{\rm gas}$, $T_{\rm mw}$, and $Z_{\rm mw}$, respectively. The considered galaxy properties are: star formation status, SMBH mass, accumulated energy in SMBH kinetic feedback (${\rm E_{kin}}$), and galaxy stellar morphology. The star formation status of a galaxy is defined based on its distance from the star-forming main sequence, whereas for $M_{\rm BH}$ and $E_{\rm kin}$, the `under' and `over' quantities are defined relative to the median values in the considered mass range (see Sections~\ref{sec:properties} and \ref{sec:smbh_connection} for details). Here we show the median results from 742 $M_*\sim10^{10.5}M_\odot$ TNG100 galaxies (filled, large symbols), at the galactocentric distance $r\sim0.5 R_{\rm 200c}$. The error bars represent $\sim\sigma/2$ galaxy-to-galaxy variation. According to the TNG model, massive galaxies that are quenched, host over-massive SMBHs, and/or whose SMBHs have injected relatively more kinetic feedback energy, exhibit stronger angular modulations of their CGM density and temperature than e.g. their star-forming counterparts. These trends are reversed for the metallicity anisotropy.}
  \label{fig:BH_dependence}
\end{figure*}
On the other hand, the anisotropy of the CGM metallicity is strongest at the low-mass end ($M_*\sim10^{10}M_\odot$), where gas metallicities vary by $0.1-0.3$ dex between directions along the minor vs. major axis, before monotonically decreasing for more massive systems. The radial dependence of the metallicity anisotropy, as found in the previous section at the mass bin of $M_*\sim10^{11}M_\odot$, is preserved across the considered mass range, with stronger angular variations at larger galactocentric distances, and more so around low-mass galaxies. 

For gas density and temperature, the galaxy stellar mass at which the CGM anisotropy is maximal ($M_*\sim10^{10.5-11}M_\odot$, left column of Figure~\ref{fig:mass_dependence}) corresponds to total halo masses of $M_{\rm 200c}\sim10^{12.1-12.5}M_\odot$ and SMBH masses of $M_{\rm BH}\sim10^{8.0-8.5}M_\odot$ (mid and right columns Figure~\ref{fig:mass_dependence}, respectively). This suggests that the CGM anisotropy may be connected to SMBH feedback activity; i.e. to SMBH-driven outflows, as this SMBH-mass range corresponds to the scale at which, in the TNG model, SMBHs typically switch from the thermal to the kinetic feedback mode \citep{weinberger.etal.2017} and to stellar mass scales ($M_*\sim10^{10.5-11}M_\odot$) above which SMBH kinetic feedback becomes the dominant channel in galaxies at low redshifts \citep{weinberger.etal.2018}. We expand on the connection between CGM anisotropy and SMBH feedback activity in the next Section. It is worth mentioning that the distribution of both CGM density and temperature, except for the measurements at small radii ($r\sim0.25R_{\rm 200c}$), is more isotropic at the both low-mass and high-mass ends. We speculate that the CGM anisotropy is most prominent at the transitional mass range around the Milky Way's mass because here two physical conditions are in place: i) the occurrence of sufficiently-strong, feedback-driven outflows; ii) the presence of a gaseous disk in the inner regions of galaxies that can (re)direct outflows in bi-polar directions. At higher masses, outflows become more isotropic in the absence of such a gaseous disk; towards the lower-mass end ($\lesssim M_{*}\sim10^{10}M_\odot$), where outflows are driven predominantly by stellar feedback and by SMBHs in thermal mode, they appear to be ineffective at large galactocentric distances (see also Section 6.2 in \citealt{pillepich.etal.2021} for a relevant discussion). 

On the other hand, we speculate that the different trends for the metallicity anisotropy could be explained by the decrease of metal content of massive galaxies. \cite{torrey.etal.2019} showed that the gas-phase metallicity-stellar mass relation in the TNG simulations flattens around $M_*\sim10^{10.5}M_\odot$ and declines towards the high-mass end due to the suppression of star formation. As a consequence, the gas carried out by outflows in more massive galaxies may be less metal rich than that removed from the central regions of less-massive and more star forming galaxies (and progenitors). Furthermore, the gaseous halo in massive galaxies is already enriched at early times and it is likely more well-mixed compared to low-mass galaxies. As a consequence, for massive galaxies the CGM metallicity along the minor axis is more similar to that of the azimuthally-averaged halo gas, even at a mass range where outflows are strongest.

\subsection{Connection to SMBH activity}
\label{sec:smbh_connection}

What is the physical cause for the CGM anisotropy around TNG galaxies, specifically massive ones? In Figure~\ref{fig:BH_dependence}, we show the dependence of the angular modulation at $\sim0.5 R_{\rm 200c}$ on various galaxy properties. We consider the CGM density ({\it top row}), temperature ({\it middle row}), and metallicity ({\it bottom row}), for $z=0$ TNG galaxies with $M_*=10^{10.5\pm0.1}M_\odot$. We choose this stellar mass bin because at this mass scale, both in observations and in the TNG simulations \citep[see e.g.][]{Donnari20}, the number of star-forming and quenched galaxies is comparable, i.e. about 370 galaxies in each category in TNG100. For each galaxy property, we subdivide the sample into two subgroups based on:

\begin{itemize}
    \item Star formation status: quenched vs. star-forming (Section~\ref{sec:tng100}).
    \item SMBH mass: over-massive vs. under-massive $M_{\rm BH}$ relative to the median $M_{\rm BH}$ value.
    \item SMBH accumulated kinetic feedback (${\rm E_{kin}}$): over-energetic vs. under-energetic ${\rm E_{kin}}$ relative to the median ${\rm E_{kin}}$ value.
    \item Galaxy shape: disky vs. non-disky, as defined in Section~\ref{sec:tng100}.
\end{itemize}

For each CGM property, we evaluate the anisotropy from the gas maps on a galaxy-by-galaxy basis; i.e. without stacking. We display the median minor-to-major ratio of the sample (large symbols in Figure~\ref{fig:BH_dependence}) while the error bars represent the galaxy-to-galaxy variation across the 32th and 68th percentiles of the considered galaxy sample.

First, we consider the CGM temperature anisotropy and its dependence on the galaxy star formation status. It has been extensively shown that in TNG the star formation in massive galaxies is quenched due to ejective effect of the SMBH kinetic feedback (\citealt{weinberger.etal.2017, weinberger.etal.2018,nelson.etal.2018, terrazas.etal.2020, davies.etal.2020}). Therefore, quiescence can be considered as a proxy for the (past) action of SMBH feedback. As visible in the middle row of Figure~\ref{fig:BH_dependence}, at the fixed galaxy stellar mass of $\sim10^{10.5}M_\odot$, quenched galaxies exhibit systematically elevated minor-to-major $T_{\rm mw}$ ratios ($\sim0.1$dex) compared to the star-forming counterparts: for the latter, in fact, the gas temperature is, on average, isotropic -- minor-to-major temperature ratio $\simeq 0$. In terms of the impact of SMBHs, galaxies with over-massive SMBHs (``over $M_{\rm BH}$'') exhibit a similar level of temperature anisotropy, whereas those with under-massive SMBHs (``under $M_{\rm BH}$'') show on average no temperature angular modulation.
\begin{figure}
  \includegraphics[width=0.42\textwidth]{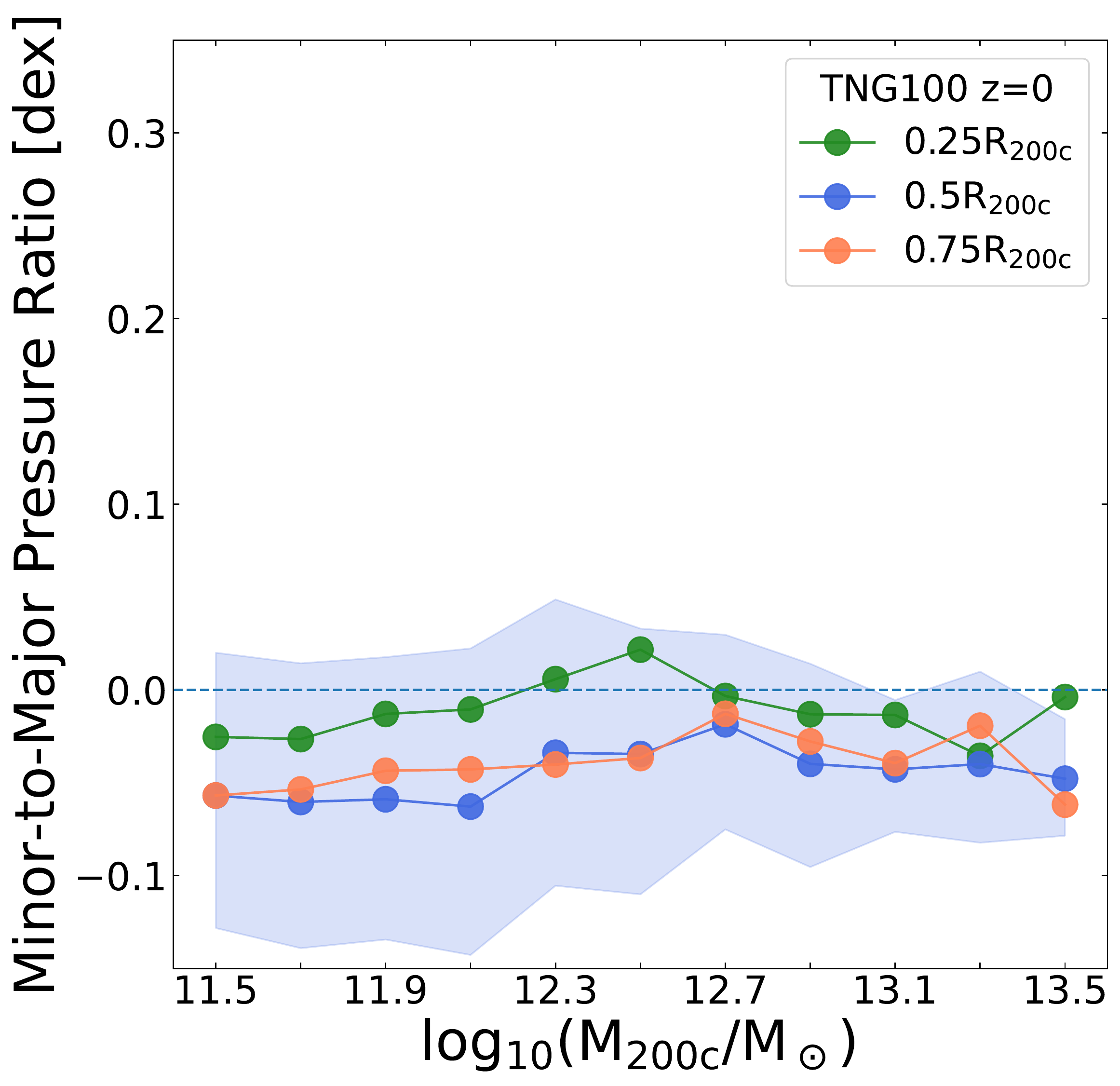}
  \includegraphics[width=0.42\textwidth]{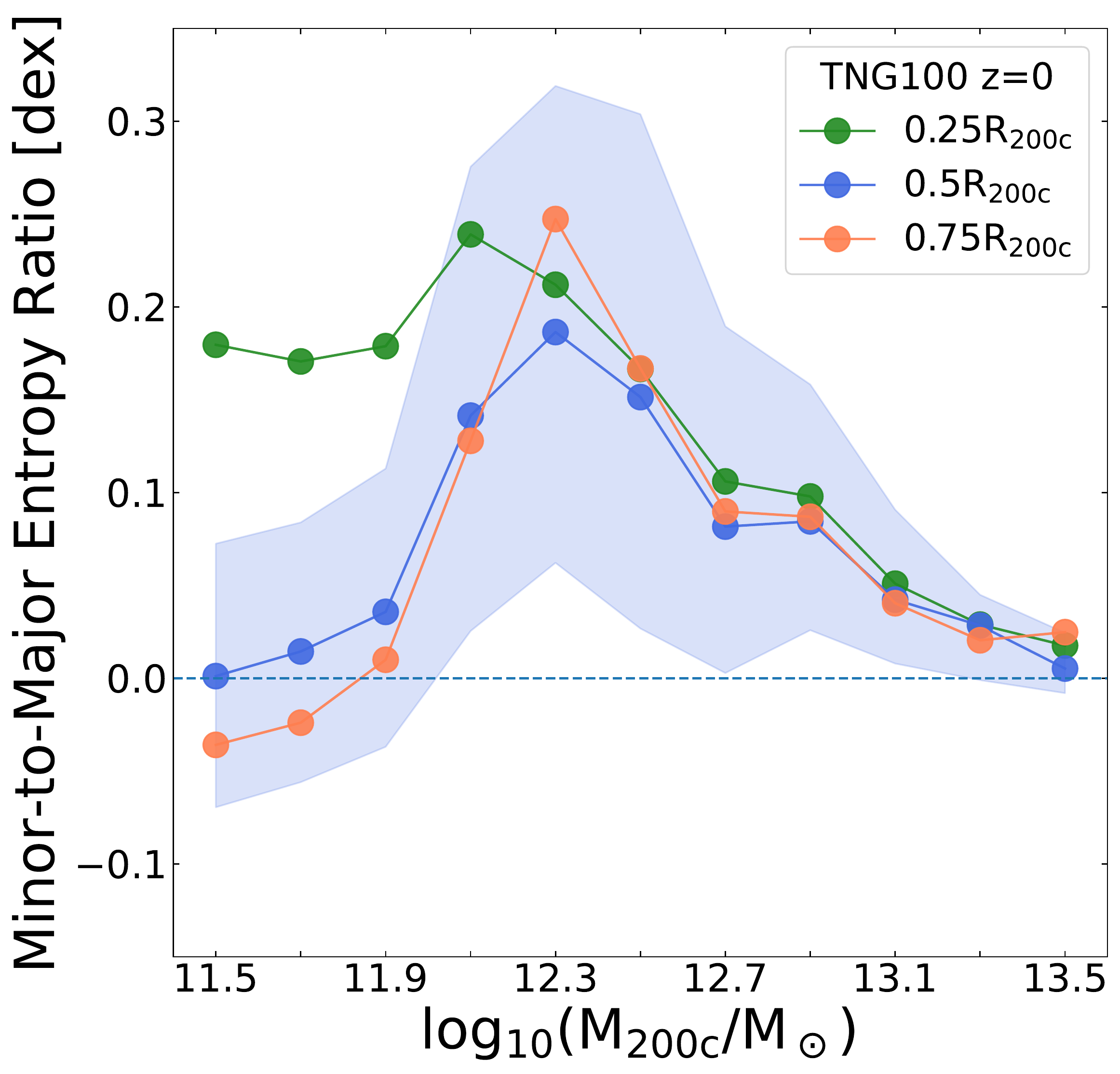}
  \caption{CGM anisotropy in mass-weighted gas pressure ({\it top}) and gas entropy ({\it bottom}) for TNG100 galaxies at $z=0$. The symbols and solid lines represent the median minor-to-major ratio of galaxies in a given mass bin, whereas the grey shaded area shows the 32th to 68th percentiles (i.e. $\sim\sigma/2$) for the $r=0.5R_{\rm 200c}$ curve. The CGM entropy captures in full the angular modulations of the underlying gas properties (i.e. density and temperature), whereas pressure is essentially isotropic.}
  \label{fig:P_K}
\end{figure}
Upon closer inspection of the subsamples, we notice that for galaxies with over $M_{\rm BH}$, at this mass range, the vast majority of their SMBHs ($\gtrsim90\%$) have already switched to the kinetic mode, whereas, in the under $M_{\rm BH}$ subsample, most of the SMBHs ($\gtrsim60\%$) are still in the thermal mode. By inspecting the variations across galaxies with different levels of SMBH feedback energy ever released in SMBH-driven winds, we find that galaxies with above-average $E_{\rm kin}$ exhibit more pronounced temperature anisotropies. Interestingly, considering the dependences on stellar morphology, we find that the non-disky subsample appears to have only marginally stronger anisotropy in $T_{\rm mw}$. Namely, the distinction between the two morphologically-defined subsamples is less significant than that associated to SMBH-related diagnostics.

The level of anisotropy of the CGM density depends on the considered galaxy properties similarly as for gas temperature. Namely, the angular modulation of the gas density is stronger in quenched galaxies, galaxies with over massive SMBHs or larger-than average amounts of energy injected via SMBH-driven winds. The anisotropy appears less sensitive to the stellar morphology, though it is slightly stronger in non-disky galaxies. 
 
On the other hand, the CGM metallicity anisotropy exhibits opposite trends with galaxy and SMBH properties in comparison to gas density and temperature. The angular dependence of the gas metallicity is stronger in star-forming, disky galaxies with relatively undermassive SMBHs and low ${\rm E_{kin}}$ values. This is consistent with the the stronger CGM metallicity anisotropy in low-mass star-forming galaxies (Figure~\ref{fig:mass_dependence}), where galactic outflows driven by stellar feedback e.g. supernovae explosions are: a) dominant over those from SMBH activity \citep{nelson.etal.2019b}; and b) able to enrich the CGM with metals originating from the inner regions of galaxies. Stellar feedback has been interpreted as the physical driver of the angular dependence of gas metallicity for $M_*\lesssim10^{10.5}M_\odot$ in TNG50 \citep{peroux.etal.2020}.

Overall our picture is that, in the TNG model, SMBH-driven outflows in the kinetic, low-accretion mode produce the observed CGM anisotropies in gas density and temperature at the high-mass end ($M_*\gtrsim10^{10.5-11}M_\odot$). This does not exclude the possibility that the angular modulation of CGM density and temperature within the central regions of haloes ($r\sim0.25 R_{\rm 200c}$), as well as the metallicity anisotropy at all radii, of lower-mass systems may be caused by other feedback mechanisms: i.e. stellar feedback and SMBH feedback in the thermal mode (\citealt{weinberger.etal.2018}).

\subsection{CGM anisotropy in pressure and entropy}
\label{sec:pressure_entropy}

Figure~\ref{fig:P_K} explores whether the anisotropic nature of the CGM density, temperature, and metallicity is imprinted in two other important thermodynamical properties: gas pressure and gas entropy, as defined in Section~\ref{sec:properties}. We show the median minor-to-major ratios of galaxies in bins of halo mass, for mass-weighted gas pressure ($P_{\rm mw}$, {\it top}) and mass-weighted gas entropy ($K_{\rm mw}$, {\it bottom}), at the same three galactocentric distances as previously.

The gas pressure exhibits only minor differences between minor vs. major axes values ($\lesssim0.05$ dex) across the whole inspected mass range. This result is a direct consequence of the gas pressure being effectively the product of density and temperature: the two quantities depend on azimuthal angle in opposite directions, as shown in Section~\ref{sec:mass_dependence}, and these trends largely cancel out.
\begin{figure}
  \centering
  \includegraphics[width=0.42\textwidth]{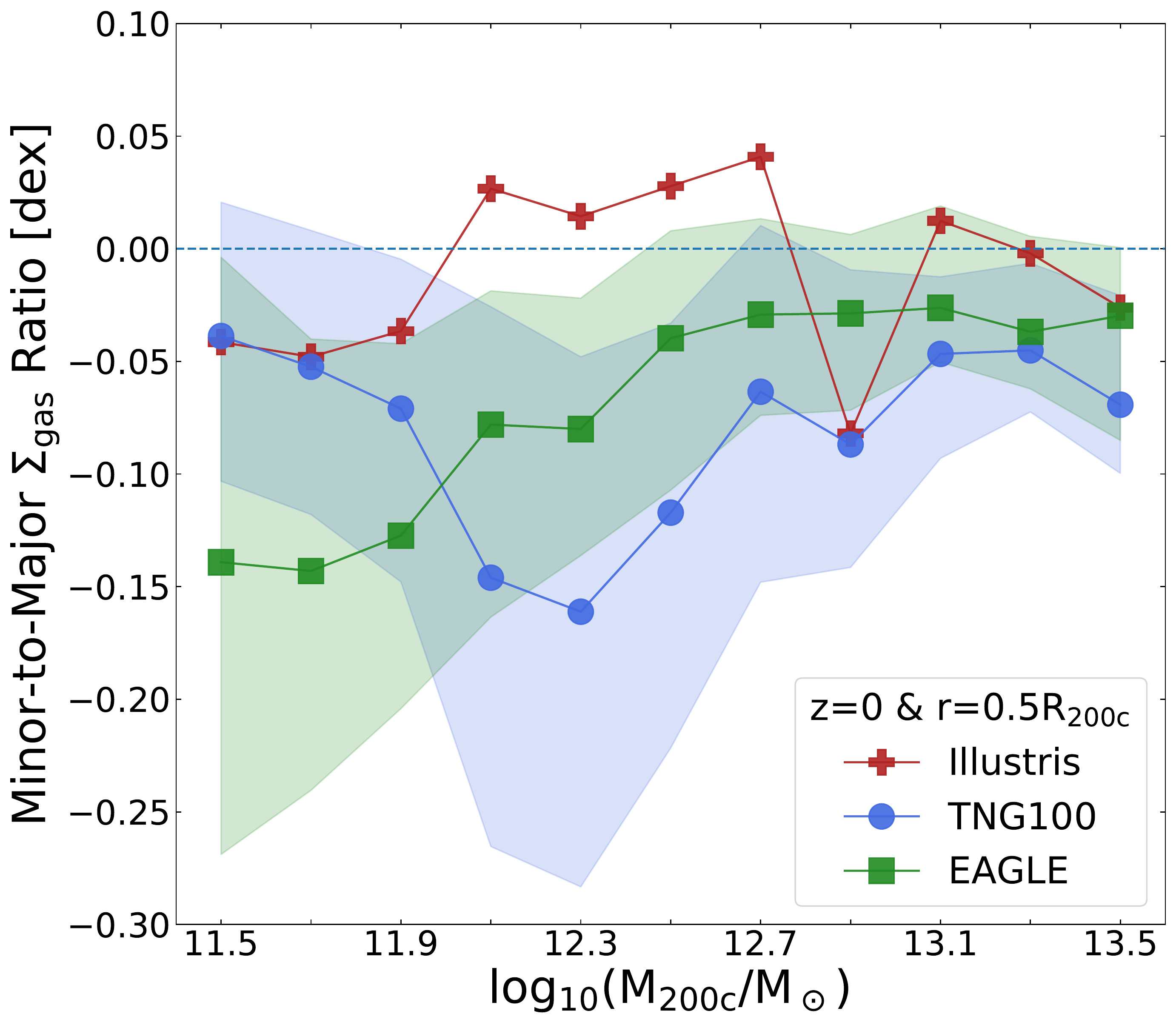}
  \includegraphics[width=0.42\textwidth]{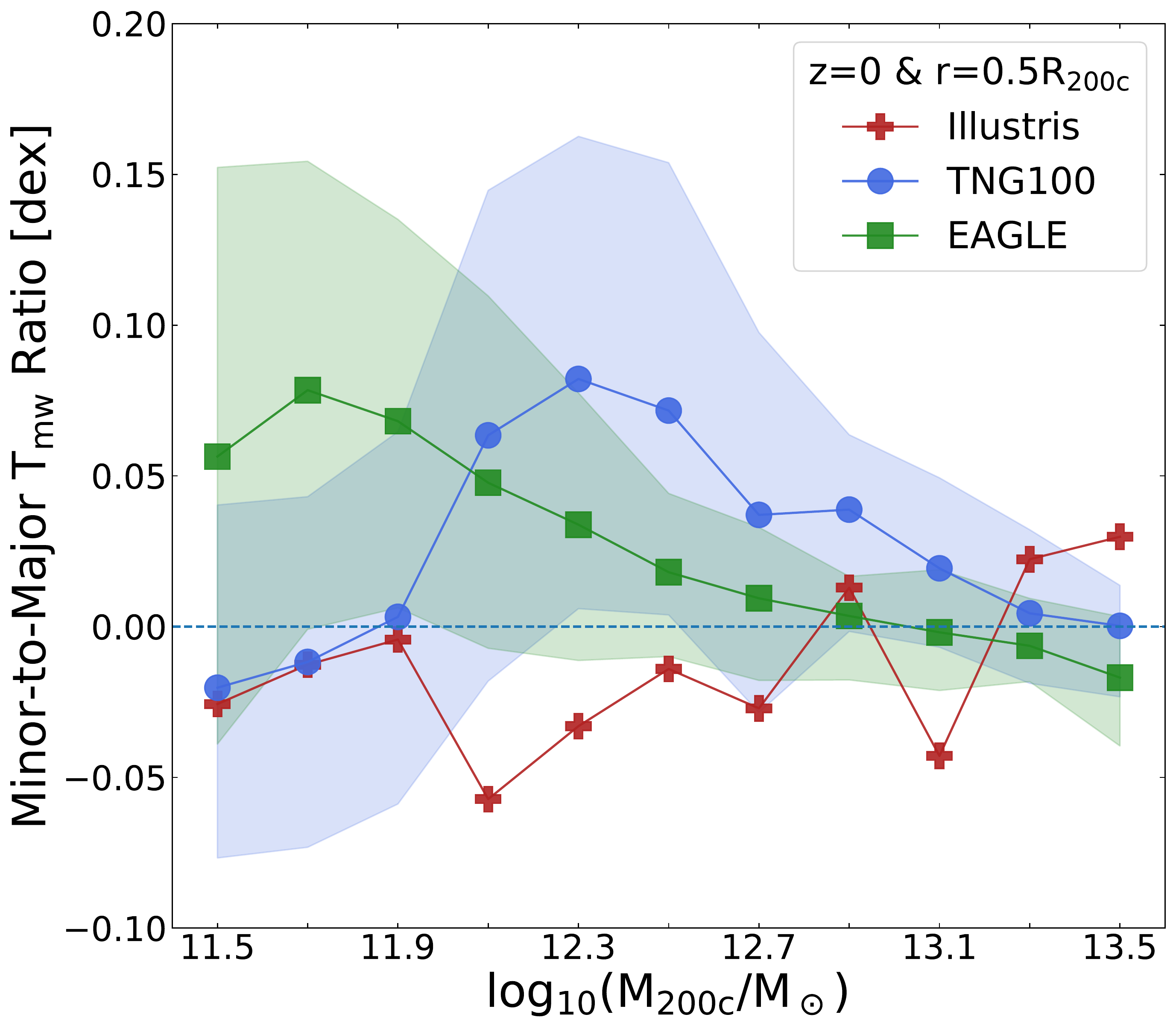}
  \includegraphics[width=0.42\textwidth]{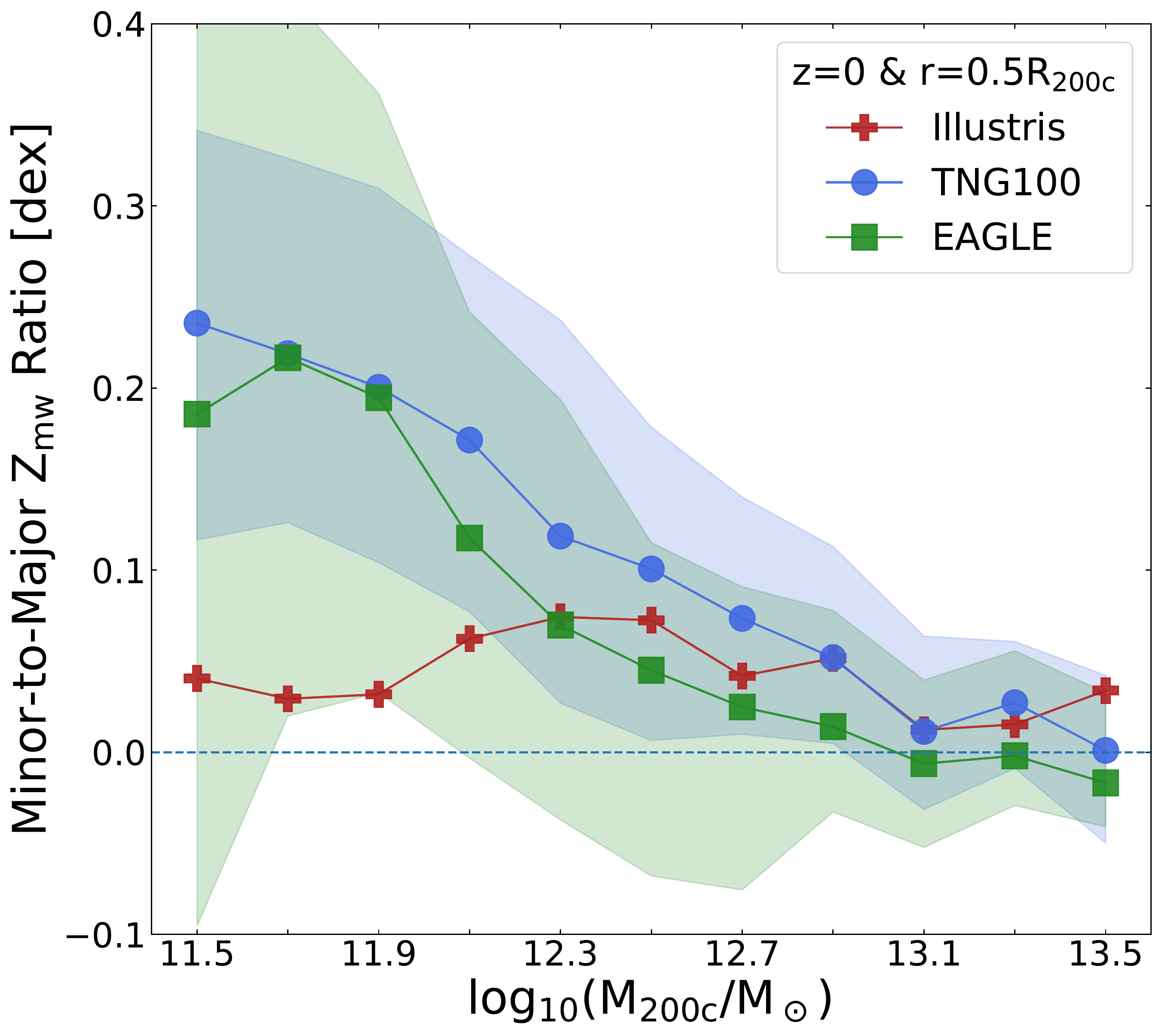}
  \caption{The CGM anisotropy according to the Illustris, TNG, and EAGLE simulations, in comparison. From top to bottom, we show results for gas column density, mass-weighted gas temperature, and mass-weighted gas metallicity, as a function of total halo mass. The minor-to-major ratios are measured for simulated galaxies at $z=0$ and at a galactocentric distance of $r=0.5 R_{\rm 200c}$. For clarity, we show the scatter only for the TNG and EAGLE simulations. Illustris predicts weaker CGM anisotropy signals than EAGLE and TNG, and the latter two models agree qualitatively but not quantitatively, with generally lower densities, higher temperatures, and higher metallicities in the CGM along the minor axes of galaxies in both models.}
\label{fig:comparison}
\end{figure}
On the other hand, the entropy of the CGM exhibits a similar anisotropy, and associated mass dependence, as the gas temperature. Namely, entropy is enhanced along the minor axis of galaxies, and this anisotropy is maximized for galaxies in $M_{\rm 200c}\sim10^{11.9-12.5}M_\odot$ haloes; i.e. the mass scale where galaxies transition from star-forming to quiescent. In fact, the angular modulation of entropy is the largest of all thermodynamical properties studied thus far, as it encodes information on both gas density ($\propto n_{\rm e}^{-2/3}$) and temperature ($\propto T$).
\begin{table*}
  \caption{\label{tb1}
  Summary of stellar and SMBH feedback models in three simulations: Illustris, TNG, and EAGLE. For the sake of brevity, we list only main features of the models (locality, stochasticity, form of feedback energy and orientation) and refer to the main text and references therein for expanded descriptions.}
 \begin{center}
 
 \begin{tabularx}{0.99\textwidth}{X|X|X}
  \huge Illustris & \huge TNG & \huge EAGLE \\
 \hline\hline \vspace{0.1em}
\bf Stellar feedback \vspace{0.3em} & & \\
- Non-local: temporary hydro decoupling &
- Non-local: temporary hydro decoupling &
- Local: direct ISM heating \\
- Timing: continuous probabilistic, $\propto$ SFR &
- Timing: continuous probabilistic, $\propto$ SFR &
- Timing: stochastic probabilistic, $\Delta T=10^{7.5}K$ \\
- Feedback energy: kinetic only (cold wind) &
- Feedback energy: kinetic + thermal (warm) &
- Feedback energy: thermal only \\ 
- Orientation: bi-polar &
- Orientation: isotropic&
- Orientation: n/a. \vspace{0.3em} \\
\hline \vspace{0.1em}
\bf SMBH feedback \vspace{0.3em} & & \\
- Dual-mode: quasar $\&$ radio &
- Dual-mode: high-state $\&$ low-state &
- Single-mode: thermal \\
- Timing (high $\dot{M}_{\rm BH}$): continuous &
- Timing (high $\dot{M}_{\rm BH}$): continuous &
- Timing (high $\dot{M}_{\rm BH}$): stochastic, $\Delta T=10^{8.5}K$ \\
- Timing (low $\dot{M}_{\rm BH}$): pulsated &
- Timing (low $\dot{M}_{\rm BH}$): pulsated &
- Timing (low $\dot{M}_{\rm BH}$): n/a \\
- Feedback energy: thermal / thermal &
- Feedback energy: thermal / kinetic &
- Feedback energy: thermal \\
- Orientation: isotropic &
- Orientation: isotropic &
- Orientation: n/a \vspace{0.3em}\\
\hline
 \end{tabularx}

 \end{center}
\end{table*}
%

As shown by \cite{zinger.etal.2020}, gas entropy is a sensitive diagnostic of feedback injection and a good proxy for quiescence, with quiescent galaxies exhibiting much larger entropy, in both their ISM and CGM, than their star-forming counterparts. In TNG this is due to the ejection and heating of gas via SMBH-driven winds. Our analysis shows that the entropy's increase due to SMBH feedback occurs preferentially along the minor axis of galaxies.


\section{Comparison of Cosmological Simulations}
\label{sec:comparison}

We have shown thus far that, in the TNG simulations, feedback from stars and SMBHs imprints its effect on the CGM in a rather unique way: by making gas density, temperature, metallicity, and entropy dependent on the angle with respect to galaxy orientation. We have focused on the TNG simulations because previous analyses had pointed towards an anisotropy of e.g. the galactic-wind mass outflow rates \citep{nelson.etal.2019b}, CGM metallicities \citep{peroux.etal.2020}, and of satellite galaxy quiescence \citep{martin-navarro.etal.2021}.

A similar outflow directionality, for example, has been found in the EAGLE simulation, with larger outflow rates occurring orthogonal to galactic disks \citep{Mitchell20}. We therefore investigate how the TNG predictions compare with results from other cosmological simulations, specifically, from the original Illustris and EAGLE simulations (see Section~\ref{sec:othersims}). All three have non-negligibly different models for both stellar and SMBH feedback, and Table \ref{tb1} gives a comparative overview.

In Figure~\ref{fig:comparison} we contrast the outcome from Illustris (red), TNG (i.e. TNG100, blue), and EAGLE (green) in the anisotropy of gas column density ({\it top}), mass-weighted temperature ({\it middle}), and mass-weighted metallicity ({\it bottom}) across a wide range of halo mass ($M_{\rm 200c}\sim10^{11.5}-10^{13.5}M_\odot$). For this comparison we select simulated galaxies at $z=0$ and measure their median minor-to-major ratios at a radial distance $r\sim0.5 R_{\rm 200c}$. We quantify and comment on the comparison at other galactocentric distances via Figure~\ref{fig:a1}. 

Illustris and EAGLE predict quite different mass dependences of the CGM density anisotropy compared to TNG, especially for galaxies with $M_{\rm 200c}\lesssim10^{12.7}M_\odot$. On the one hand, the Illustris model predicts no substantial angular dependence across the whole mass range (with amplitude $\lesssim0.05$ dex). On the other hand, the CGM density anisotropy according to EAGLE -- whereby the CGM is underdense along galaxies' minor axes, as in TNG -- is most prominent at the lowest mass bin $M_{\rm 200c}\sim10^{11.5}M_\odot$ and decreases toward higher masses. 

Similarly, the quantitative predictions for the temperature angular modulation differ across the three models. Illustris predicts minor-to-major $T_{\rm mw}$ ratios very close to unity; i.e. almost isotropic. The EAGLE model, like TNG, generally predicts hotter CGM orthogonal to galactic disks, but the anisotropy is strongest at lower masses than in TNG and the signal decreases with increasing mass.       

In agreement with the findings of \citealt{peroux.etal.2020}, EAGLE and TNG predict rather consistent CGM metallicity anisotropies, in both cases with monotonically-decreasing angular modulation with increasing mass. On the other hand, at the galactocentric distance of $r\sim0.5 R_{\rm 200c}$, Illustris galaxies exhibit no significant anisotropy in the gas metallicity across the whole mass range ($\lesssim0.05$ dex). 

Overall, the Illustris model predicts no significant anisotropy in the CGM properties at galactocentric distances of $r\sim0.5 R_{\rm 200c}$. However, interestingly, at smaller distances ($r\sim0.25R_{\rm 200c}$), Illustris predicts strong angular modulations of both gas density and temperature ($\sim0.4$ dex, see Figure~\ref{fig:a1}), particularly so for low-mass galaxies but in opposite senses compared to TNG or EAGLE. In Illustris, the CGM around low-mass galaxies is denser, colder, and metal richer along the minor axis than along the disks. We interpret these differences to be due to a combination of two aspects of the Illustris stellar feedback model which differ from TNG: i) the injection of cold winds, and ii) the bi-polar orientation of the winds at launch. The latter aspect had been discussed in the TNG methods paper \citep[][see their Figure 5]{pillepich.etal.2018}, but not fully quantified in terms of its possible effects on the CGM. It should be noted, however, that in general the Illustris feedback model is strongly disfavored based on comparisons of its outcome to galaxy demographics such as the galaxy color bimodality at the high-mass end, the galaxy stellar mass function at $z=0$, and the halo gas fraction within group-mass haloes \citep{vogelsberger.etal.2014b, genel.etal.2014}.

TNG and EAGLE qualitatively agree in terms of CGM geometry, with generally lower densities, higher temperatures, and higher metallicities along the minor axes of galaxies in both models. However, the two simulations return different mass dependences of the anisotropic signals. Whereas in TNG the CGM density and temperature anisotropies are maximal at intermediate, Milky Way-like masses of $M_*\sim10^{10.5-11}M_\odot$ i.e. $M_{\rm 200c}\sim10^{12.1-12.5}M_\odot$, EAGLE predicts monotonic mass trends for all the studied CGM properties. As already pointed out, these are the mass scales where observed galaxies transition from being mostly star-forming to being mostly quiescent and where the effects of SMBH feedback may start to be effective, in reality as well as in the models. Therefore, the differences in the CGM structure for galaxies with $M_*\gtrsim10^{10.5}M_\odot$ ($M_{\rm 200c}\gtrsim10^{12}M_\odot$) in EAGLE and TNG are undoubtedly due to the different implementations of SMBH feedback. 

Unlike in TNG, EAGLE injects SMBH feedback purely in the form of thermal energy, with non-continuous (i.e. bursty) events. SMBHs store up sufficient energy until they can change the temperature of nearby gas by a specified amount. It appears that as galaxies begin to quench in EAGLE this mechanism does not change in behavior or operation in any significant way \citep{bower17}, resulting in the monotonic mass trends described above. Importantly, the mass outflow rates of galactic winds in EAGLE massive galaxies are lower than those in TNG, at least at intermediate redshifts \citep{Mitchell20}, pointing to the possibility that the SMBH thermal feedback in EAGLE is not as ejective as the SMBH kinetic feedback in TNG. Together, these effects explain how, at the transitional mass scale of $M_{\rm 200c}\sim10^{12}M_\odot$, EAGLE can produce weaker angular modulations of the CGM density and temperature compared to TNG. Finally, we also note that at the low-mass end, stellar feedback appears to more strongly perturb CGM properties in EAGLE, in comparison to TNG. 

Overall, we conclude that the explored CGM anisotropies may be a powerful diagnostic to discriminate among different feedback models in cosmological simulations, or at least among their effects: at the low-mass end, the CGM anistropic signals are sensitive to stellar feedback, whereas at the high-mass end, they can help distinguish among different SMBH feedback models.  


\section{Tests of CGM anisotropy via X-ray Observations}
\label{sec:Xray_implication}
\begin{figure*}
  \includegraphics[width=1.0\textwidth]{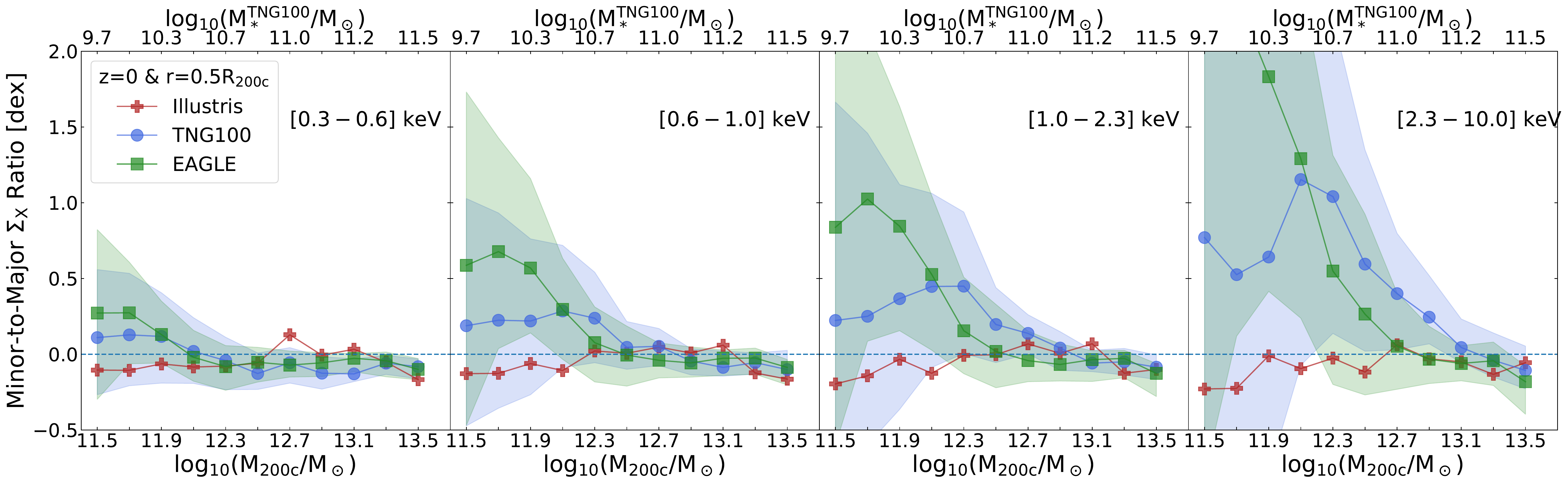}
  \includegraphics[width=0.9\textwidth]{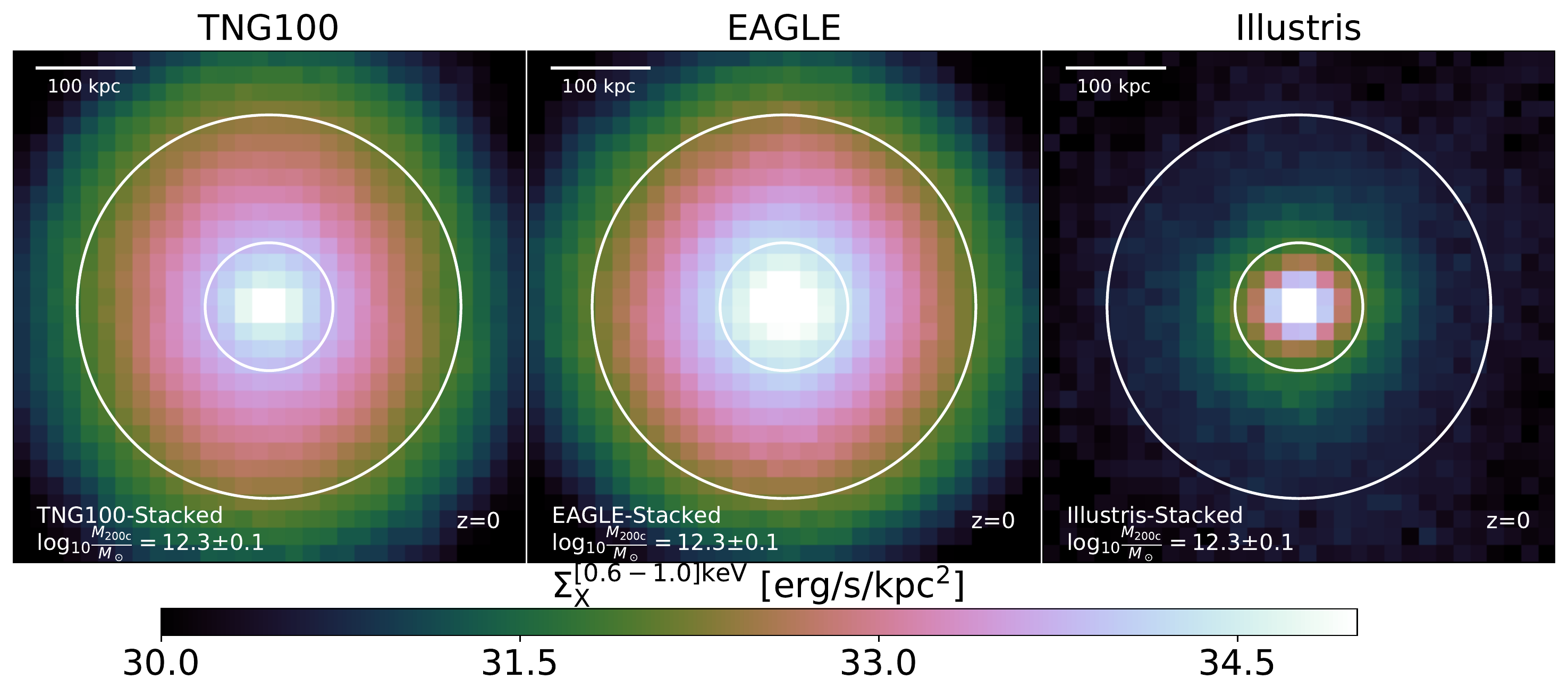}
  \caption{CGM anisotropy in X-ray surface brightness, for Illustris, TNG, and EAGLE simulated galaxies at $z=0$. {\it Top:} From {\it left} to {\it right} we show the minor-to-major ratio of the X-ray surface brightness (${\rm \Sigma_X}$) in four different energy bands: $[0.3-0.6]$ keV, $[0.6-1.0]$ keV, $[1.0-2.3]$ keV, and $[2.3-10.0]$ keV, respectively. Annotations are as in Figure~\ref{fig:comparison}. The top x-axes represent median values of stellar mass for TNG galaxies (note the scale is not linear). {\it Bottom:} Stacked edge-on maps of the X-ray surface brightness in $[0.6-1.0]$ keV for galaxies with $M_{\rm 200c}=10^{12.3\pm0.1}M_\odot$. From left to right we show: TNG100, EAGLE and Illustris. The two circles specify galactocentric distances at $0.25R_{\rm 200c}$ and $0.75R_{\rm 200c}$. In the soft bands, e.g. $[0.6-1.0]$ keV (second panel from the left in the {\it top row}), the CGM of massive galaxies ($M_{\rm 200c}\simeq10^{12.1-12.5}M_\odot$) is X-ray brighter in directions orthogonal to the stellar ``disks'', by about a factor of two. For both TNG and EAGLE, harder bands reveal stronger angular modulations of X-ray luminosity.}
  \label{fig:Xray}
\end{figure*}
 X-ray observations can probe the thermodynamical state of the gaseous atmospheres. In principle, from X-ray spectroscopic data it would be possible to derive the amount of hot gas, the gas temperature and metallicity. In that case, we could make comparison between simulations and observations on X-ray emission-weighted-like quantities, as provided for TNG100 at $z=0$ in Figure~\ref{fig:b1}. Unfortunately, such spectroscopic data is currently not available for galaxies below the group mass scale ($M_{\rm 200c}\lesssim10^{13}M_\odot$). Therefore, in the following, we explore potential tests to (in)validate the predictions for the CGM anisotropy presented in this paper by focusing on X-ray emission data. 

\subsection{CGM X-ray luminosity anisotropy}

In Figure~\ref{fig:Xray}, we quantify the predicted angular modulation of the intrinsic X-ray luminosity from the CGM gas in Illustris, TNG, and EAGLE. We show in the {\it top row} the minor-to-major ratios in X-ray surface brightness ($\Sigma_{\rm X}$) as a function of halo mass at $z=0$, at the galactocentric distance $r\sim0.5 R_{\rm 200c}$, while in the {\it bottom row} we show stacked X-ray surface brightness maps in $[0.6-1.0]$ keV band at the mass bin $M_{\rm 20c}=10^{12.3\pm0.1}M_\odot$ separately for TNG, EAGLE, and Illustris.

The resulting predictions are not trivial, as the X-ray luminosity of the halo gas depends to different degrees on gas density, temperature and metallicity, but, as shown in the previous Sections, the angular modulations of such gas physical properties may unfold in various and even opposing manners. We hence explore the anisotropy of $\Sigma_{\rm X}$ across different energy bands: from left to right, in the $[0.3-0.6]$ keV, $[0.6-1.0]$ keV, $[1.0-2.3]$ keV, and $[2.3-10.0]$ keV bands. These are energy ranges which the eROSITA telescope is sensitive to \citep{predehl.etal.2020}, each of which is sensitive to the contributions from different gas phases. 

For all the considered energy bands, the Illustris model predicts consistently no significant difference in $\Sigma_{\rm X}$ measured along the minor vs. the major axes of galaxies, across the considered mass range. 

On the other hand, for TNG and EAGLE, the X-ray emission from the CGM has non-negligible levels of angular modulation, particularly for low-mass systems ($M_{\rm 200c}\lesssim10^{12.0-12.5}M_\odot$). More intriguingly, the  X-ray anisotropy, as well as its associated scatter, become progressively stronger toward harder energy bands, for both TNG and EAGLE but more so for the latter. In fact, at the transitional mass regime of $M_*\sim10^{10.5-11}M_\odot$ or $M_{\rm 200c}\sim10^{12.0-12.5}M_\odot$ and even in the soft bands (e.g. $[0.6-1.0]$ keV), TNG predicts the X-ray surface brightness of the CGM located along the minor axis of galaxies to be about two times \emph{higher} than that of gas located along the major axis (minor-to-major ratios of $\sim0.25$ dex). The corresponding anisotropy is instead weaker, if not vanishing, according to EAGLE ($\sim0.20$ dex). Furthermore, it is interesting to notice that the anisotropy of the X-ray emission does not necessarily reflect the anisotropy of the gas density; i.e. the former can be enhanced along the minor axis where the latter is lower. 

The enhancement along the minor axis of the soft X-ray emission, e.g. in the $[0.3-0.6]$ keV band, in low-mass TNG and EAGLE haloes, as well as the lack of anisotropy in Illustris, predominantly reflects the trends and angular modulations of the CGM metallicity. In $M_{\rm 200c}\sim10^{11.5}M_\odot$ systems, the gas temperature is about $10^{5}$ K, and most of the soft X-ray emission comes from metal lines. On the other hand, the X-ray emission in harder bands is mainly driven by gas with high temperature, so that the anisotropy of the hard-band $\Sigma_{\rm X}$ instead reflects the temperature modulation: see Figure~\ref{fig:comparison}, middle panel, vs. Figure~\ref{fig:Xray}, rightmost panel. The X-ray comparisons among Illustris, TNG100, and EAGLE at different galactocentric distances are given in Appendix~\ref{sec:app_a}, Figure~\ref{fig:a2}, and are qualitatively similar to the results of Figure~\ref{fig:Xray}.

\subsection{X-ray hardness}
\begin{figure*}
    \includegraphics[width=0.45\textwidth]{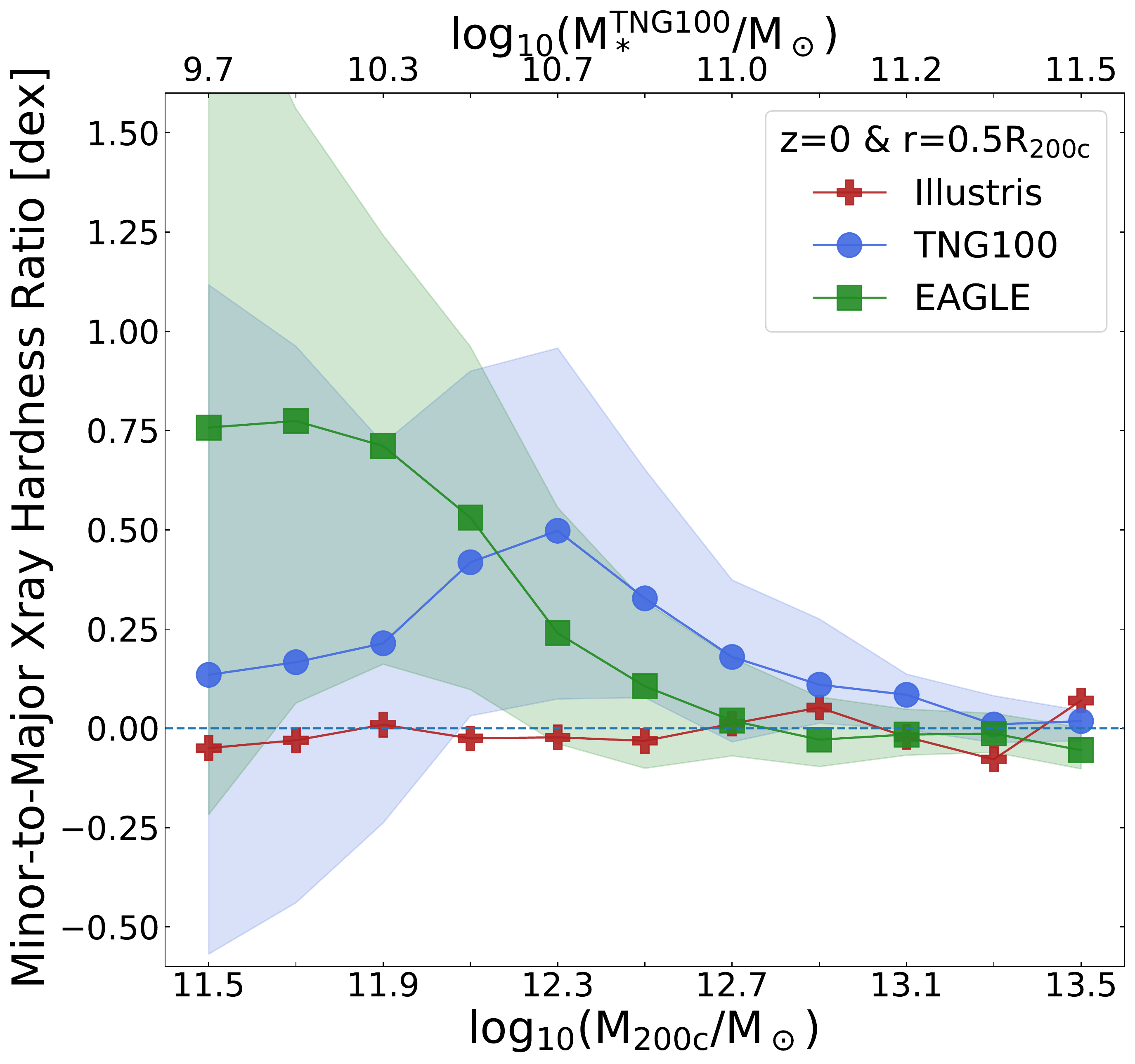}
    \includegraphics[width=0.9\textwidth]{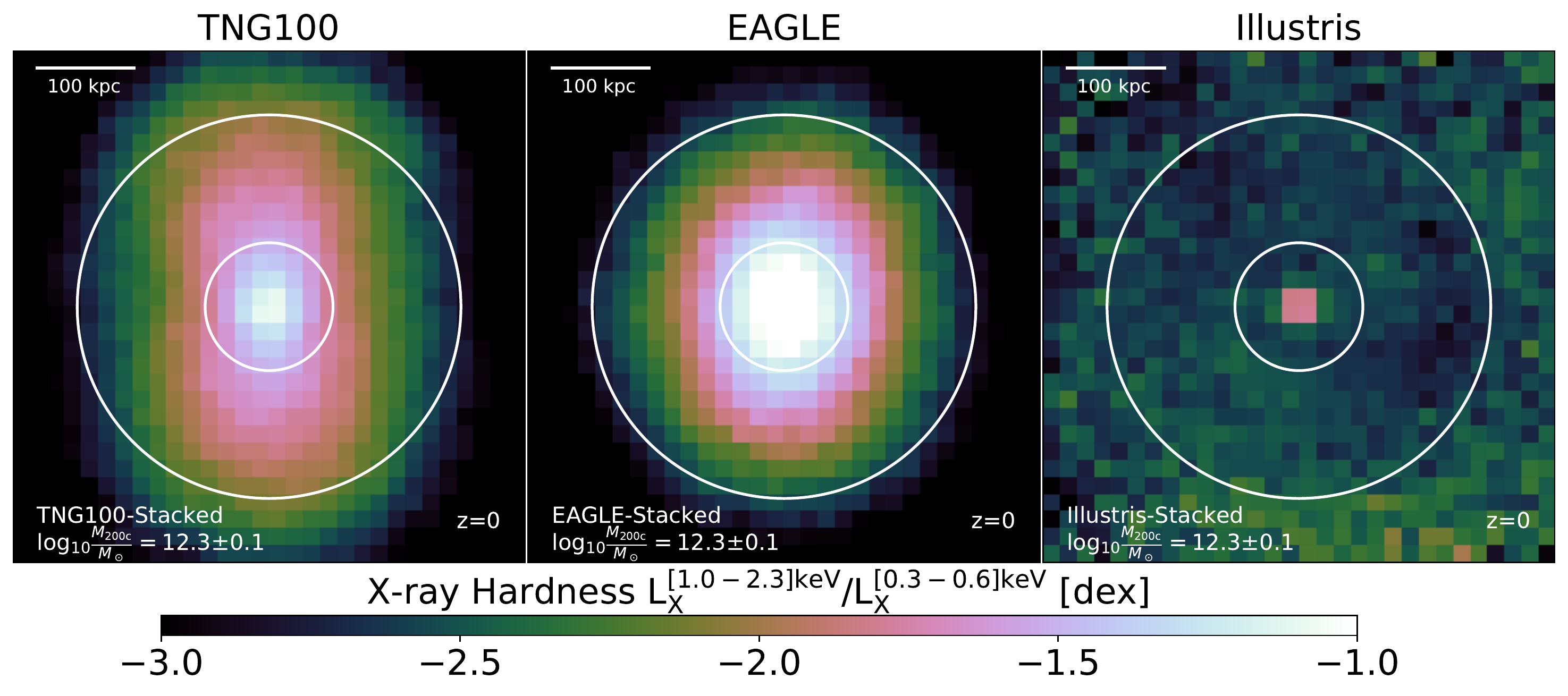}
    \caption{CGM anisotropy in X-ray hardness, for Illustris, TNG, and EAGLE simulated galaxies at $z=0$. {\it Top:} The minor-to-major axes ratio of X-ray hardness as a function of galaxy/halo mass. The hardness is defined as the ratio of the X-ray emission in the hard band ($[1.0-2.3]$ keV) to the soft band ($[0.3-0.6]$ keV). Annotations are as in Figure~\ref{fig:comparison}. For clarity, shaded areas, i.e. the scatter, are given only for TNG and EAGLE. In addition, the top x-axes show median values of stellar mass for TNG galaxies (note the scale is not linear). {\it Bottom:} Stacked edge-on maps of the X-ray hardness in galaxies with $M_{\rm 200c}\sim10^{12.3}M_\odot$, for TNG100, EAGLE and Illustris separately. The two concentric circles specify galactocentric distances of $0.25R_{\rm 200c}$ and $0.75R_{\rm 200c}$. The anisotropy of the X-ray hardness in the CGM is a promising observable to discriminate among different SMBH feedback models.}
    \label{fig:Xray_hardness}
\end{figure*}
The results above clearly suggest that another X-ray observable may be a key probe of the CGM anisotropy: the X-ray hardness. In Figure~\ref{fig:Xray_hardness} we therefore show the angular modulation of the X-ray hardness, defined as the ratio of X-ray emission in the hard $[1.0-2.3]$-keV band to the soft band, $[0.3-0.6]$ keV. 

The {\it top} panel quantifies the minor-to-major ratio of the X-ray hardness as a function of $M_{\rm 200c}$ at $r\sim0.5R_{\rm 200c}$. The main result is that the anisotropy of the X-ray hardness faithfully reflects the trends and angular modulations of the CGM temperature, except that the amplitude of the former is significantly larger ($\sim0.5$ dex vs. $\sim0.1$ dex). The Illustris model predicts, as expected, generally no anisotropic hardness. On the other hand, the signals in TNG and EAGLE are somewhat distinct, with the hardness anisotropy in the former peaking at the usual transitional mass scale (with approximate amplitude of $\sim0.5$ dex) and that in EAGLE being most prominent toward the low-mass end ($\sim0.6-0.8$ dex). At high mass, the TNG signal is stronger than in EAGLE. The X-ray hardness anisotropy predicted by the different models at additional radii is shown in Figure~\ref{fig:a3}.

Importantly, we note that the anisotropic signals in the X-ray hardness presented in Figure~\ref{fig:Xray_hardness}, as well as the signals in the X-ray surface brightness of Figure~\ref{fig:Xray}, are preserved, if not larger, when the minor-to-major ratios are measured based on median stacked maps instead of individual galaxy maps (as described in Section~\ref{sec:ratio}). The former method is particularly relevant for the observational opportunities discussed in the next Section: as shown there, in practice it will not be possible to detect the anisotropic signal in X-ray at the individual-galaxy level, but only via stacking of a sufficiently-large number of galaxies.

To visually illustrate the different predictions from the three simulation models, the bottom panels of Figure~\ref{fig:Xray_hardness} show stacked maps of the X-ray hardness for $M_{\rm 200c}=10^{12.3\pm0.1}M_\odot$ haloes. We hence focus on the regime where SMBH feedback is at the origin of the differences among the models (see arguments in Sections~\ref{sec:smbh_connection} and \ref{sec:comparison}). Visually, we clearly see that TNG predicts a larger angular modulation in X-ray hardness than EAGLE, with less central concentration, whereas there is no significant anisotropy in the Illustris map.

\subsection{X-ray observational opportunities}
\label{sec:detectability}

\begin{figure}
    \includegraphics[width=0.47\textwidth]{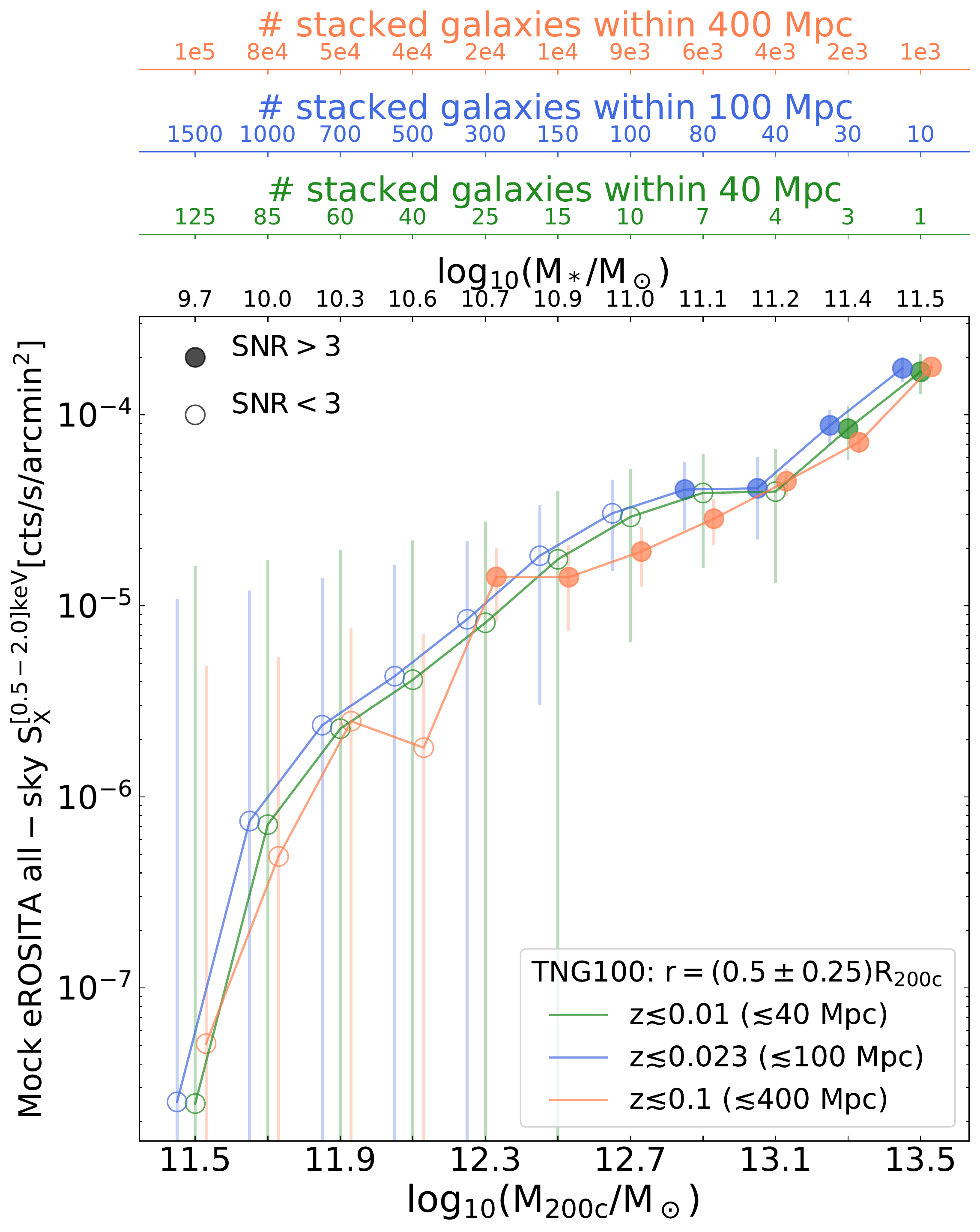}
    \caption{Mock eROSITA survey count rates as a function of halo mass for TNG100 galaxies, based on different assumed surveys/redshifts (three colors). The circles represent the median count rate at each mass bin, and the error bars denote $\sim1\sigma$ Poisson noise based on both source and background photon counts. Filled (empty) circles specify signal-to-noise ratio (SNR) above (below) the value of 3. The values of SNR and Poisson noise are computed for the intentionally detected area (see text) and based on the photon counts obtained from stacking the number of galaxies available within a certain distance (colored top axes), assuming a 2ks observing time per galaxy.}
    \label{fig:Xray_brightness}
\end{figure}

The predictions from the three simulation models studied in this paper are most discrepant either at the low-mass end ($M_{\rm 200c}\sim10^{11.5}M_\odot$) or at the transitional mass range ($M_{\rm 200c}\sim10^{12.0-12.5}M_\odot$), where the TNG anisotropic signals peak. In the following, we show that the latter is within the reach of upcoming X-ray observations. 

We consider a number of stacking experiments to examine the observability of the predicted X-ray signals via the eROSITA all-sky survey. We are interested in signals from the CGM region bracketed by ${\rm [0.25-0.75]R_{200c}}$, which is most relevant to the anisotropic signals presented in this paper. In addition, to study the azimuthal dependence of the signals, we aim to characterize CGM areas spanning at most $\sim1/12$ of the $2\pi$ circumference, in order to constrain how the X-ray emission varies between the major and minor axes for three different azimuthal angles.

In Figure~\ref{fig:Xray_brightness}, we show the eROSITA mock X-ray count rates in the $[0.5-2.0]$ keV band for TNG galaxies, as a function of halo mass and galaxy stellar mass (top axis). The median count rates are obtained based on TNG100 galaxies at each of the three indicated redshifts following the method described in Section~\ref{sec:properties}. To assess the detectability we also derive Poisson noise (error bars) and the signal-to-noise ratio for each mass bin (denoted by filled and empty cicles) based on the source and background photon counts. The latter is calculated assuming an X-ray eROSITA background level of ${\rm 2.1\times10^{-3}counts/s/arcmin^2}$ in the energy band of ${\rm [0.5-2.0]}$ keV (\citealt{merloni.etal.2012}). The photon counts are obtained by stacking a certain number of galaxies in each mass bin, corresponding to the total number of  galaxies available up to the given redshift (specified at the extra axes on the top of the plot\footnote{The number of available galaxies within a certain distance is calculated based on the TNG100 mass function, corrected by the relevant volume factor. We also exclude face-on galaxies ($\sim1/3$ of the total number of galaxies), which are not useful for detecting anisotropic signals.}) and assuming 2ks observing time. In addition, the Poisson noise (error bars) and signal-to-noise ratio are computed for the intentionally-detected area; i.e. $1/12$ of the ${\rm [0.25-0.75]R_{200c}}$ annulus, accounting for the fact that we treat the 4 quadrants identically. 

According to Figure~\ref{fig:Xray_brightness}, if we stack $\sim10000$ galaxies out to redshift $z=0.1$, then TNG predicts a detectable X-ray signal down to the mass range of $M_{\rm200c}\sim10^{12.3}M_\odot\ (M_*\sim10^{10.7}M_\odot)$ with ${\rm SNR\gtrsim3}$ (orange), corresponding to roughly $400$ source photons detected per ${\rm arcmin^2}$. This number of galaxies is within the reach of current galaxies survey, e.g. the Sloan Digital Sky Survey (SDSS, \citealt{alam.etal.2015,saulder.etal.2016}), while larger future spectroscopic galaxy surveys will increase the sample statistics further. 
Based on our findings summarized in Figure~\ref{fig:Xray_brightness}, the upcoming eROSITA all-sky survey therefore has the potential to verify the X-ray anisotropy in the CGM predicted by state-of-the-art cosmological simulations for intermediate- and high-mass galaxies ($M_{\rm 200c}\gtrsim10^{12.0-12.5}M_\odot$), directly discriminating among different models of SMBH feedback. On the other hand, despite the improved spatial resolution, stacking on nearby galaxies out to 40 Mpc (green) or 100 Mpc (blue) with only 2 ks of exposure would result in a detectable signal achievable only for significantly more massive halos, due to the decreased statistics.

Finally, observations for the low-mass end ($M_{200c}\sim10^{11.5}M_\odot$ or $M_*\sim10^{10}M_\odot$) will be significantly more challenging. Galaxy samples larger than those of Figure~\ref{fig:Xray_brightness} are needed to ensure detection via stacking: this in turn will require extending  the observations to farther distances ($z>0.1$). Unfortunately, at such distances, eROSITA will not be able to spatially resolve their anistropic signal. Observations of low-mass systems will be perhaps possible only with next-generation X-ray observatories \citep[e.g.][]{simionescu21}. This regime probes distinct anisotropy signals, as predicted by the three simulations considered in this paper, and would further constrain the impact of stellar feedback models on the circumgalactic medium of galaxies. 

\section{Summary and Conclusions}
\label{sec:conclusion}

In this paper we have used the IllustrisTNG simulations to provide theoretical, quantitative predictions as to how feedback from the central regions of galaxies affects the thermodynamical and chemical properties of surrounding gas -- the circumgalactic medium (CGM). In particular, we study how CGM properties -- density, temperature, metallicity, as well as pressure, entropy, and X-ray emission -- depend on azimuthal angle, defined as the galactocentric angle with respect to the central galaxy's stellar minor axis.

We have studied galaxies across the galaxy stellar mass range of $10^{10-11.5}M_\odot$ and thus focused on the high-mass regime ($\gtrsim 10^{10.5}M_\odot$) where SMBH feedback is expected to be the dominant mechanism that drives galactic outflows and where testing opportunities via X-ray observations are feasible. We have quantified the CGM anisotropy at $z=0$ via stacking and through population averages, as a function of galactocentric distance, across a wide range of galaxy and halo masses, and have compared the TNG predictions to those from two other simulations, Illustris and EAGLE.

Our main findings are as follows:

\begin{itemize}
    \item The TNG model predicts that the CGM of massive galaxies at the transitional mass regime of $M_{*}\simeq10^{10.5}-10^{11.0}M_\odot$ is markedly anisotropic (Figures~\ref{fig:signals_11_0} and \ref{fig:BH_dependence}), with gas along the minor axis of galaxies being more diluted, hotter and more metal rich than the gas along galactic planes. The amplitude of this angular modulation is non-negligible ($0.1-0.3$ dex), is clearly seen by stacking across large number of galaxies, and persists out to the virial radius of haloes ($\sim260$ kpc on average). Our work suggests that the Milky Way's mass range is the sweet spot where the CGM anisotropy is most prominent because here two fundamental physical conditions occur: i) The presence of sufficiently-strong, feedback-driven outflows, as those produced in the TNG model by the kinetic SMBH feedback; ii) the presence of a gaseous disk in the inner regions of galaxies that can (re)direct outflows in bi-polar directions.\\ 
    
    \item For $M_* \simeq10^{10.5}M_\odot$ TNG galaxies, the CGM anisotropy in gas density and temperature is remarkably more pronounced for galaxies that are quenched, host an overmassive SMBH with respect to the average, or host a SMBH that has injected a larger-than-average amount of energy in the kinetic feedback mode, which in turn produces strong SMBH-driven winds (Figure~\ref{fig:BH_dependence}). This supports the picture whereby, in TNG,  SMBH-driven outflows are the physical driver of the angular modulation of CGM density and temperature at the high-mass end. In fact, the strength of the gas-metallicity anisotropy exhibits opposite correlations with those galaxy properties that are a measure of SMBH activity.\\
    
    \item According to the TNG model, the CGM is indeed anistropic in its thermodynamical properties and chemical content over a large galaxy mass range, $M_{*}\simeq10^{10}-10^{11.5}M_\odot$ (Figure~\ref{fig:mass_dependence}), and it is generally more pronounced at smaller galactocentric distances. In TNG the metallicity angular modulation is stronger for lower-mass galaxies, and overall decreases with increasing galaxy mass. In contrast, the density and temperature anisotropies are maximal at the aforementioned transitional mass scale; i.e. at or slightly above the Milky Way-mass scale, corresponding to halo masses of $M_{\rm 200c}\sim10^{12.0 -12.5}M_\odot$ and SMBH masses of $M_{\rm BH}\sim10^{8.0-8.5}M_\odot$. We speculate that the CGM is more isotropic at both higher and lower masses because of the less effective stellar and SMBH-driven thermal feedback at the low-mass end and of the relative isotropy of the inner gas distribution at the high-mass end.\\
    
    \item In general the CGM is more diluted along the minor axis, but its temperature is enhanced, making the halo gas pressure essentially isotropic. On the other hand, the CGM entropy maximally captures, via its strong angular modulation, the imprints of feedback (Figure~\ref{fig:P_K}).\\
    
    \item In comparison to TNG, the original Illustris simulation returns negligible, if not, vanishing levels of angular dependence of the CGM properties, throughout the studied mass range (Figure~\ref{fig:comparison}). \\
    
    \item On the other hand, the EAGLE model, despite the very different implementation of both stellar and SMBH feedback, shows many qualitatively-similar results: gas density is generally lower, and temperature and metallicity higher, orthogonal to galaxy disks. However, quantitatively, EAGLE and TNG predict different levels of CGM anisotropy and very different mass trends (Figure~\ref{fig:comparison}). EAGLE produces more prominent anisotropies in the CGM density and temperature at the low-mass end, but also weaker, if not vanishing, CGM angular modulations for massive galaxies, $M_{\rm 200c}\gtrsim10^{12}M_\odot$.
\end{itemize}
    
Our paper demonstrates that the $z=0$ angular modulation of the CGM properties is highly sensitive to the way energy from stellar and SMBH feedback is injected into the surrounding gas. It thus provides a promising test-bed to discriminate among galaxy feedback models and to (in)validate the outcome of current and future state-of-the-art cosmological galaxy simulations. This is particularly the case for massive galaxies and hence for SMBH feedback. For $M_{\rm 200c}\gtrsim10^{12.3}M_\odot$ or $M_{*}\gtrsim10^{10.7}M_\odot$, the eROSITA telescope, via its all-sky survey, should be able to detect X-ray emission from the CGM via stacking, across the relevant halo scales (Figure~\ref{fig:Xray_brightness}). In particular, in terms of X-ray observational signatures, we find that: 
     
\begin{itemize}
    \item The anisotropy of CGM density, temperature and metallicity imprints complex and informative signatures in the angular modulation of the X-ray surface brightness of the simulated galaxies (Figure~\ref{fig:Xray}). Whereas Illustris predicts isotropy, the CGM of TNG (EAGLE) $M_{*}\simeq10^{10.5-11}M_\odot$ galaxies is $\sim1.8$ (1.6) times \emph{brighter} along the minor versus major axis for the $[0.6-1.0]$ keV band, with angular modulation increasing towards harder bands. \\

    \item The X-ray hardness of the CGM, defined as the ratio of X-ray emission between hard (e.g. $[1.0-2.3]$ keV) and soft (e.g. $[0.3-0.6]$ keV) bands, is a promising observable to discriminate among different SMBH feedback models (Figure~\ref{fig:Xray_hardness}), exhibiting a marked angular dependence of $\sim 0.5$ dex at the transitional mass scale.
\end{itemize}

Within the interpretative framework of IllustrisTNG, the CGM anisotropy that we have quantified in this paper — if in place also in the Universe — would be the manifestation across the galaxy population of the bipolar bubbles that eROSITA has detected in X-ray emission above and below the disk of our Galaxy \citep{predehl.etal.2020}, which themselves trace the Fermi bubbles \citep{Su.2010}. In fact, the TNG model predicts coherent, dome-like features of over-pressurized gas that impinge into the CGM above and below the disks of most Milky Way- and Andromeda- like galaxies \citep{pillepich.etal.2021} of the TNG50 simulation \citep{pillepich.etal.2019, nelson.etal.2019b}, whereby the gas within the bubbles is under dense, $10^{6-7}$ K hot and enriched to metallicities of $0.5-2~Z_{\odot}$.

The detection of X-ray anisotropy in the CGM of Milky Way-mass galaxies in the Universe would constitute an indirect confirmation of the predicted ubiquity of eROSITA-like bubble features in external individual galaxies \citep{pillepich.etal.2021}, and place stringent constraints on the current implementation, or at least manifestations,  of SMBH feedback in simulations.

In fact, the novel phenomenon uncovered in SDSS data by \citealt{martin-navarro.etal.2021}, dubbed ``anisotropic satellite galaxy quenching'', and in place also in the TNG but not in the original Illustris simulations, already provides compelling support to the ejective character of SMBH feedback. The lack of anisotropy in the CGM density of Illustris galaxies (Figure~\ref{fig:comparison}, top panel, and Figure~\ref{fig:a1}) further supports the interpretation that the anisotropic satellite-quenching signal is driven by the clearing of halo gas preferentially in the direction of the minor axis of central galaxies in groups because of the activity of their SMBHs.

\section*{Data Availability}

The IllustrisTNG simulations are publicly available and accessible at \url{www.tng-project.org/data} \citep{nelson19a}. Similarly, data from the Illustris and EAGLE projects are available at \url{www.illustris-project.org/data} \citep{nelson15} and e.g. \url{http://eagle.strw.leidenuniv.nl/} \citep{Mcalpine.etal.2016}, respectively. Data directly related to this publication and its figures is available on request from the corresponding author.

\section*{Acknowledgements}

AP and NT thank Elad Zinger for useful conversations. DN acknowledges funding from the Deutsche Forschungsgemeinschaft (DFG) through an Emmy Noether Research Group (grant number NE 2441/1-1) and NT and AP acknowledge funding by the Deutsche Forschungsgemeinschaft (DFG, German Research Foundation) -- Project-ID 138713538 -- SFB 881 (``The Milky Way System'', subproject A01). NW is supported by the GACR grant 13491X. The primary TNG simulations were carried out with compute time granted by the Gauss Centre for Supercomputing (GCS) under Large-Scale Projects GCS-ILLU and GCS-DWAR on the GCS share of the supercomputer Hazel Hen at the High Performance Computing Center Stuttgart (HLRS). Additional simulations and analyses had been carried out on the Isaac machine of the Max Planck Institute for Astronomy (MPIA) and on the other systems at the Max Planck Computing and Data Facility (MPCDF). 

\bibliographystyle{mnras}
\bibliography{refs}

\appendix
\section{Measurements at different galactocentric distances}
\label{sec:app_a}
\begin{figure}
  \includegraphics[width=0.42\textwidth]{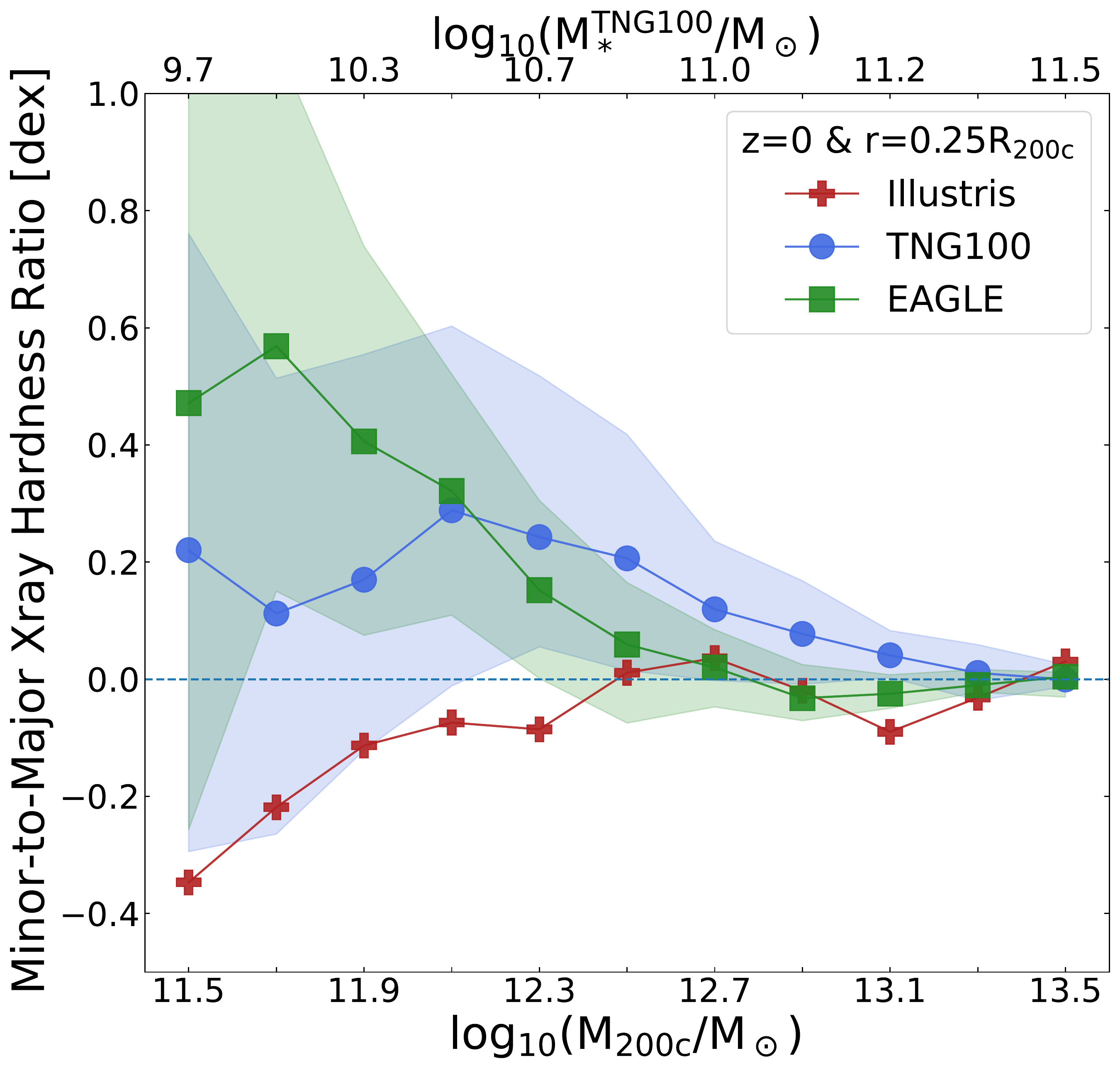}
  \includegraphics[width=0.42\textwidth]{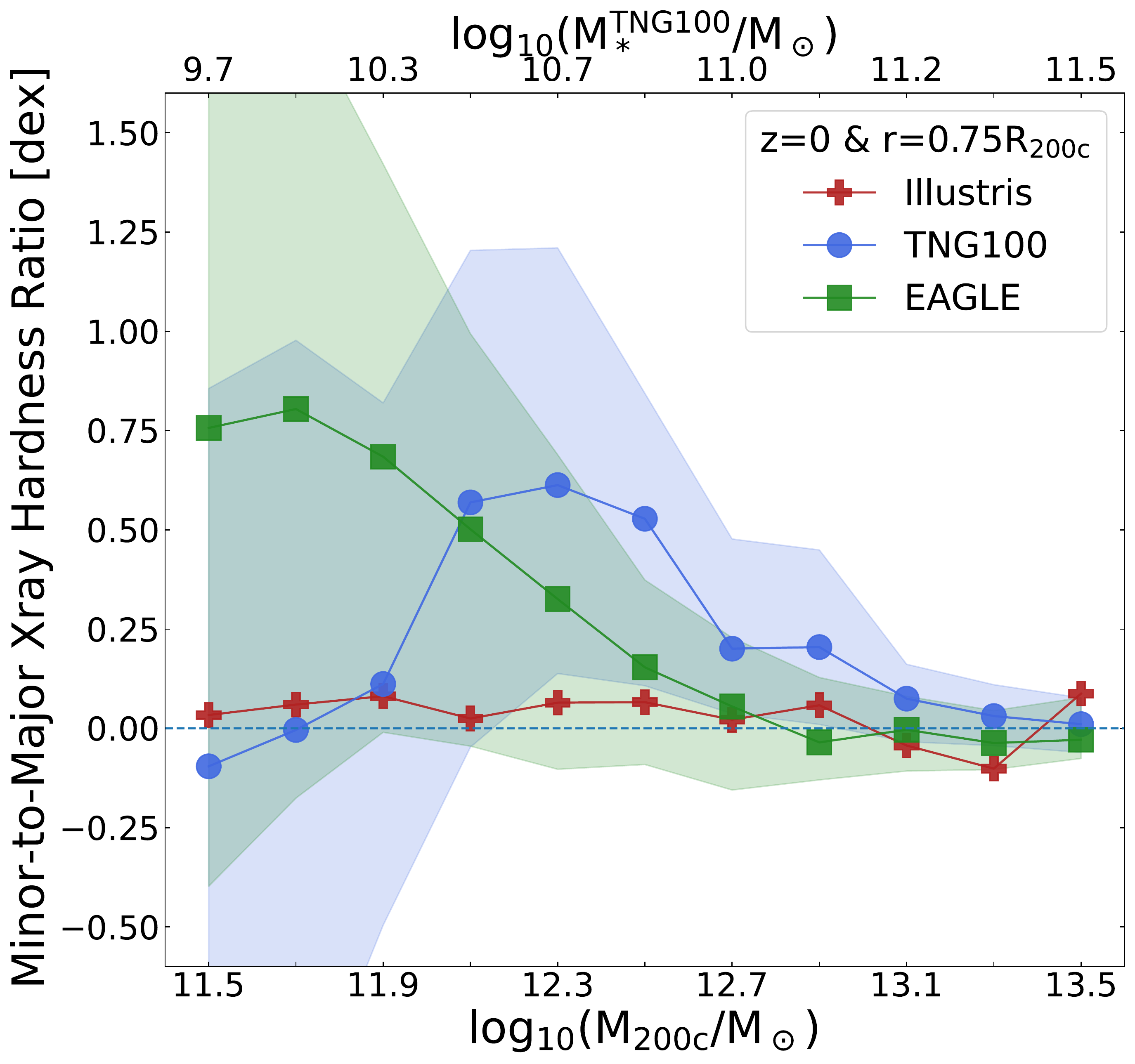}
   
  \caption{Similar to Figure~\ref{fig:Xray_hardness} but for the measurements at other distances: $r=0.25R_{\rm 200c}$ ({\it top}) and $r=0.75R_{\rm 200c}$ ({\it bottom}).}
  \label{fig:a3}
\end{figure}

\begin{figure*}
    \includegraphics[width=0.33\textwidth]{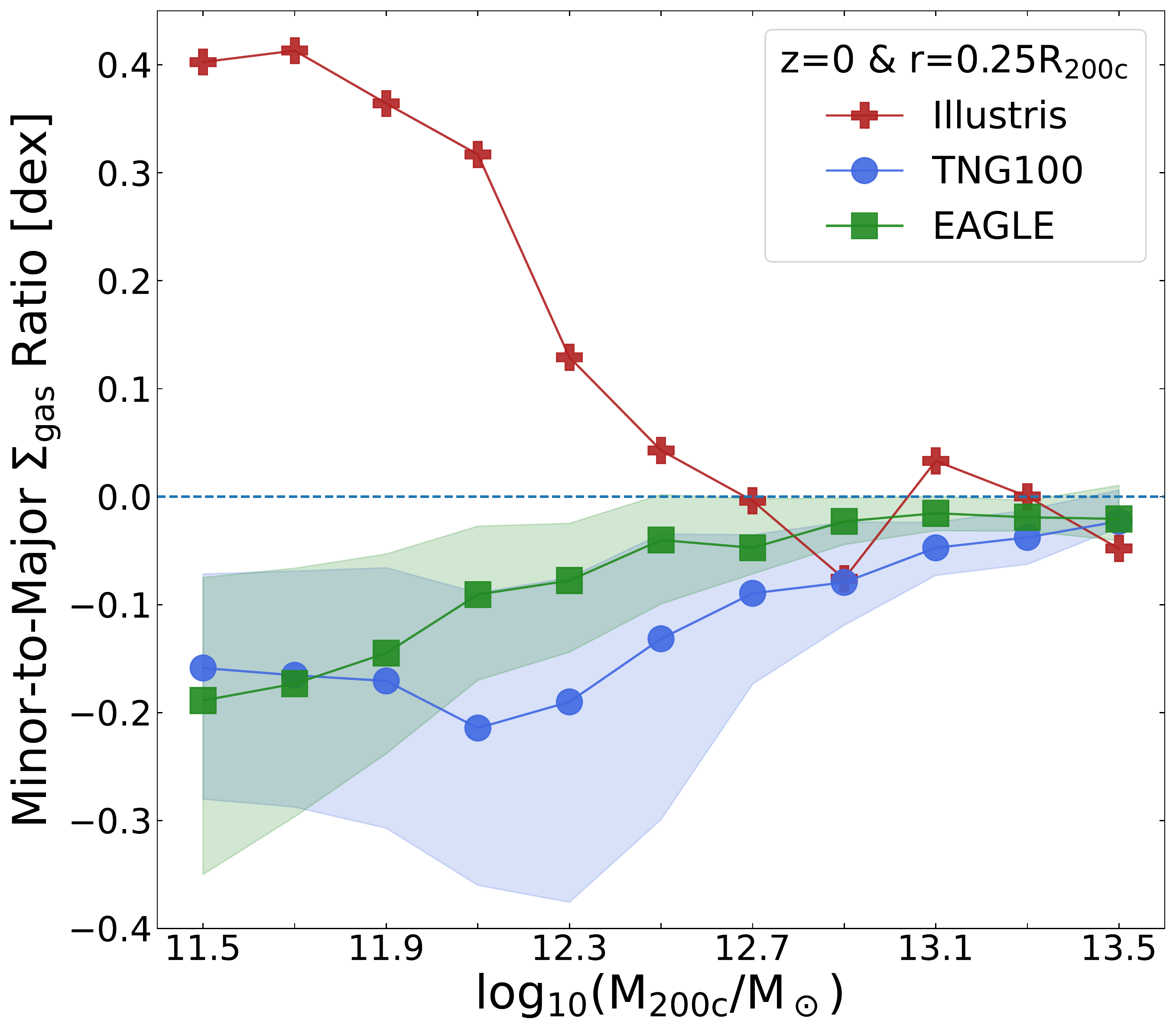}
    \includegraphics[width=0.33\textwidth]{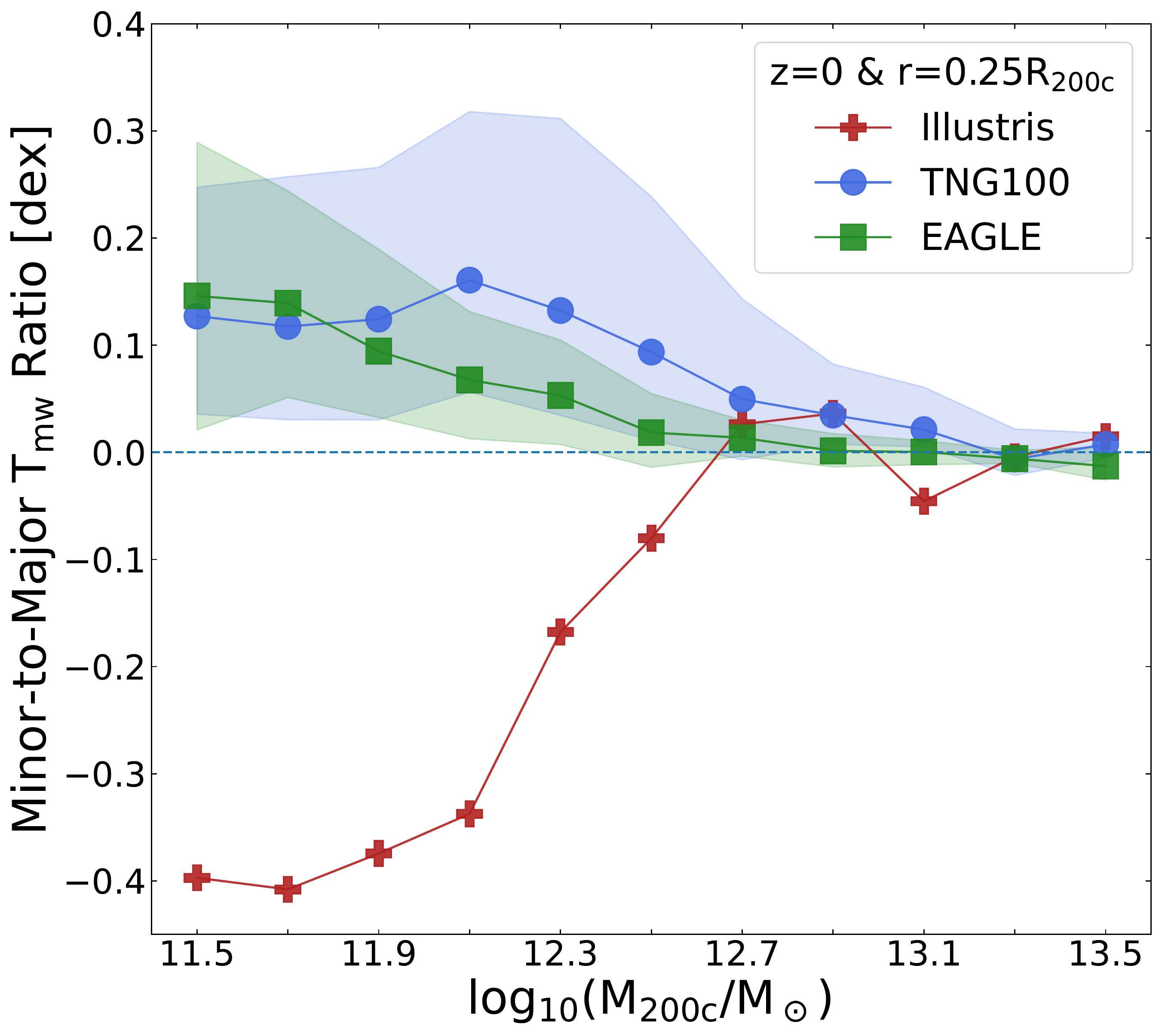}
    \includegraphics[width=0.33\textwidth]{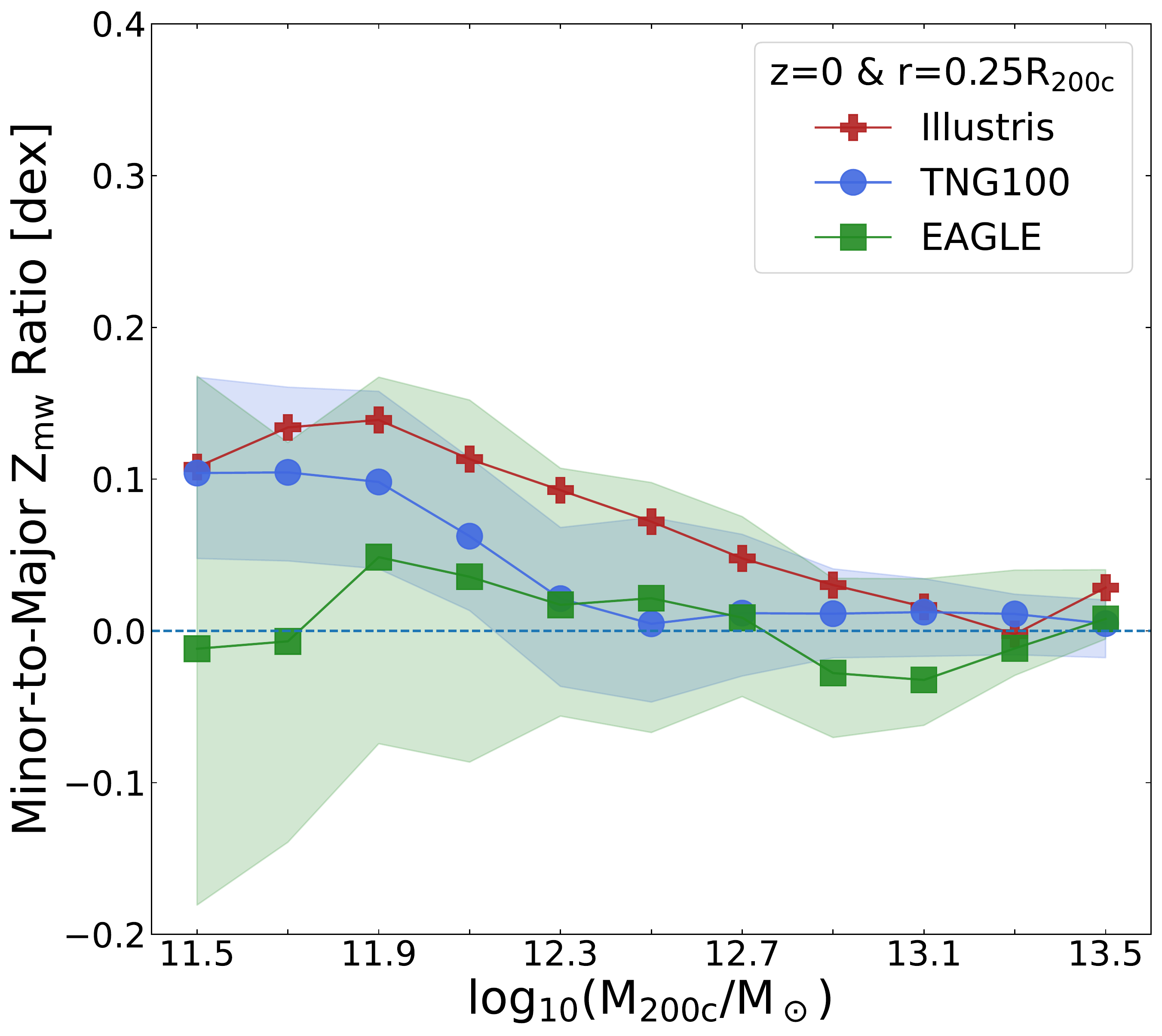}
    \includegraphics[width=0.33\textwidth]{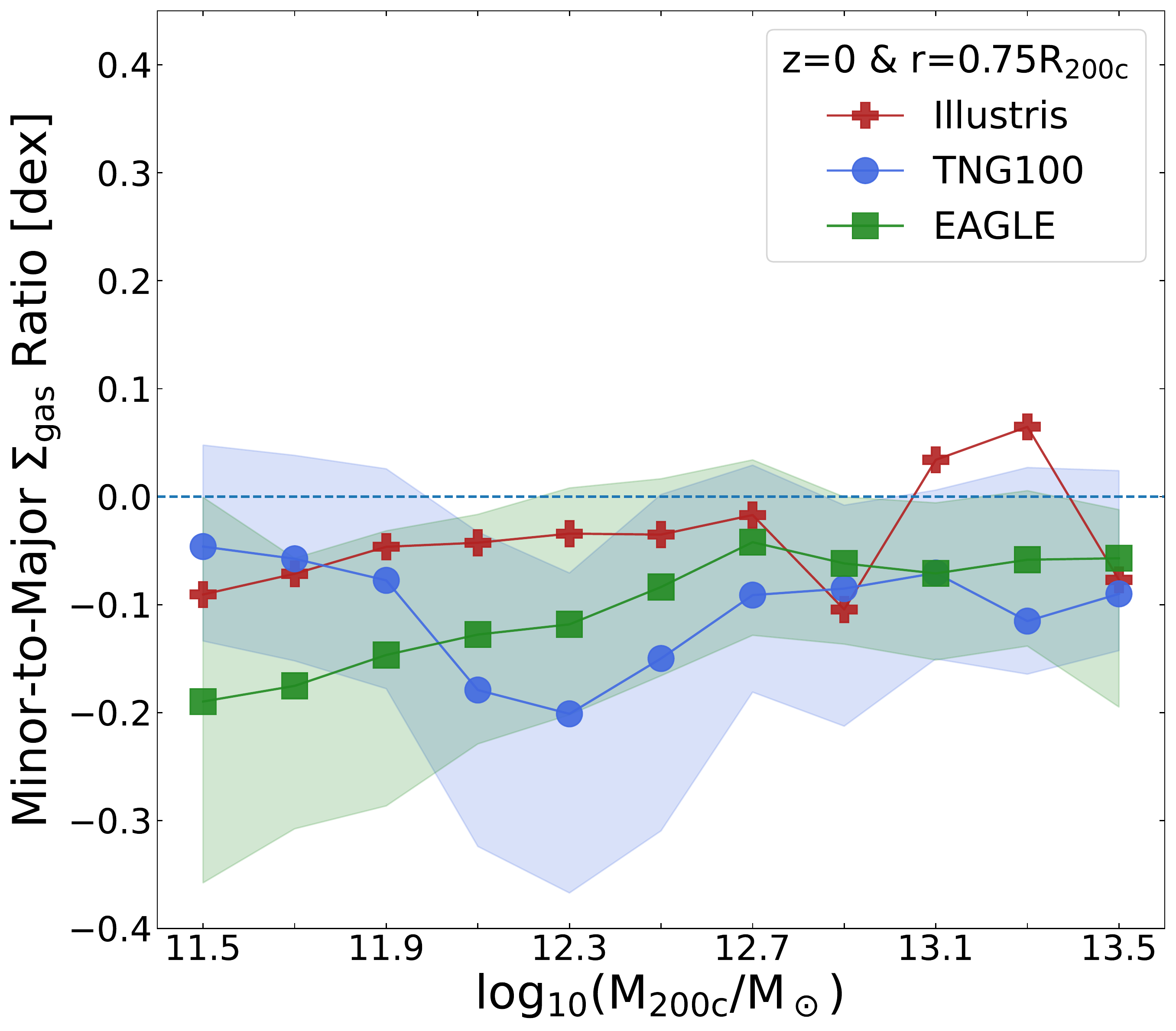}
    \includegraphics[width=0.33\textwidth]{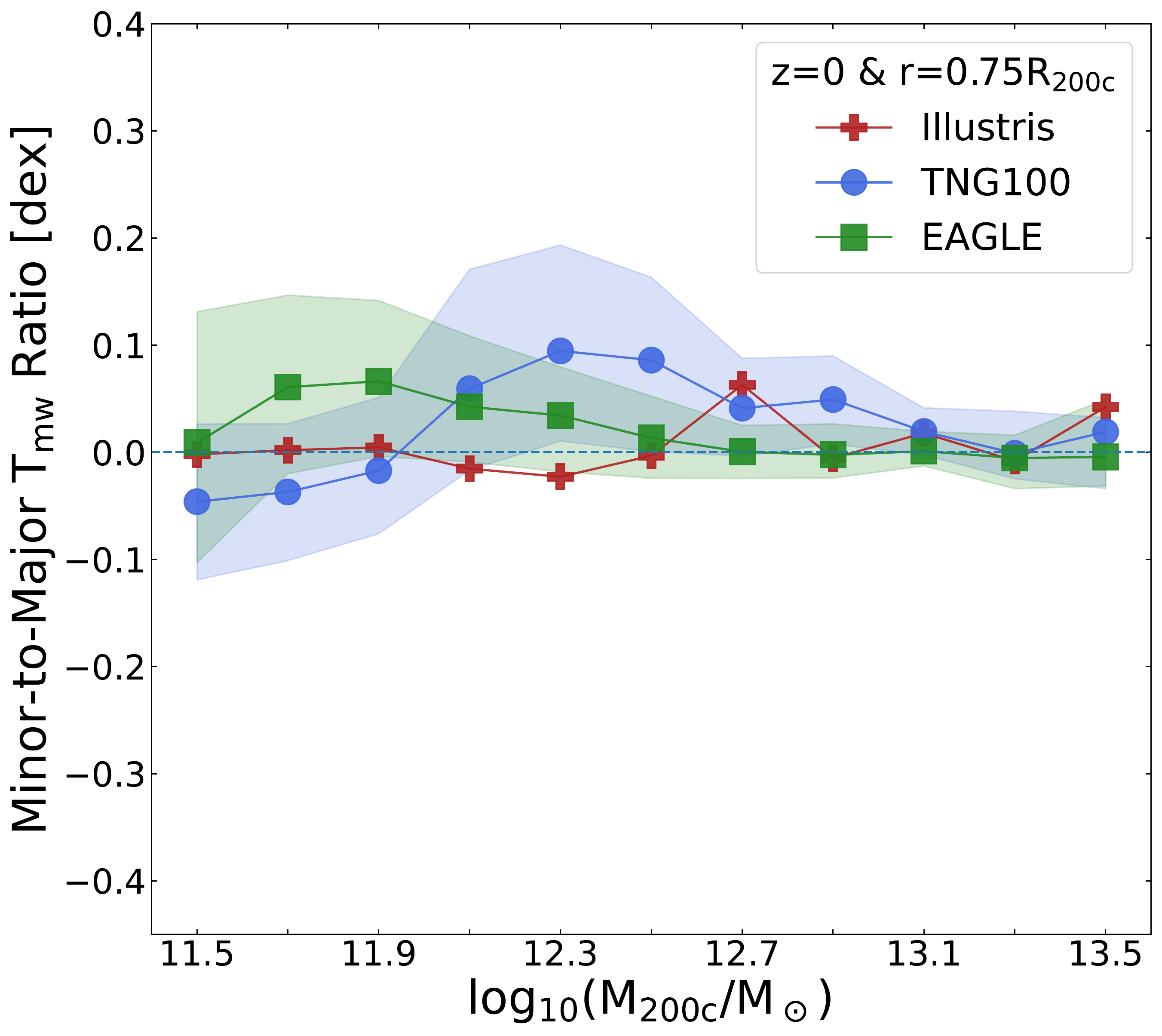}
    \includegraphics[width=0.33\textwidth]{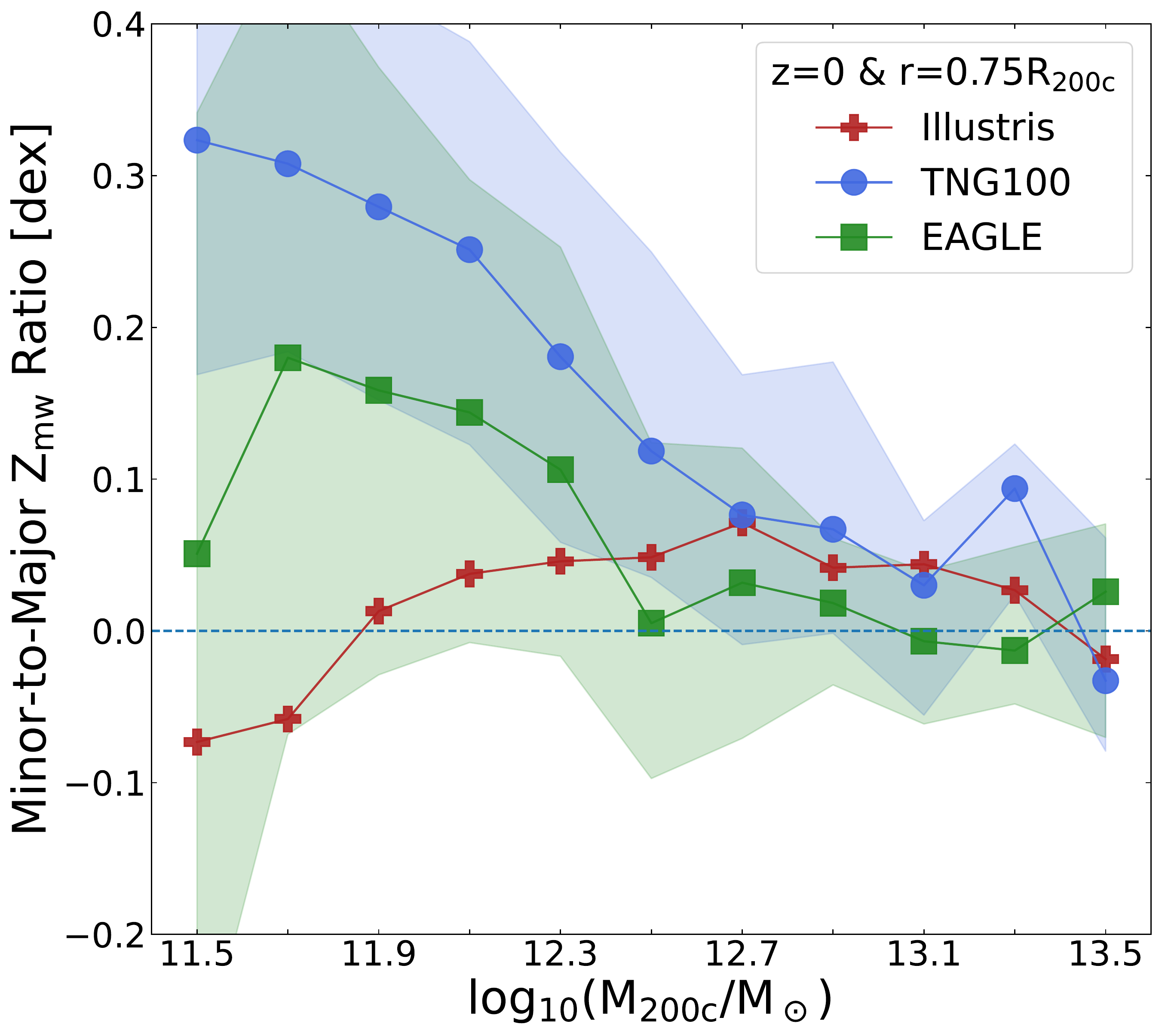}
\caption{Similar to Figure~\ref{fig:comparison} but for the measurements at different galactocentric distances: $r=0.25R_{\rm 200c}$ ({\it top row}) and $r=0.75R_{\rm 200c}$ ({\it bottom row}).}

\label{fig:a1}
\end{figure*}
\begin{figure*}
  \includegraphics[width=0.99\textwidth]{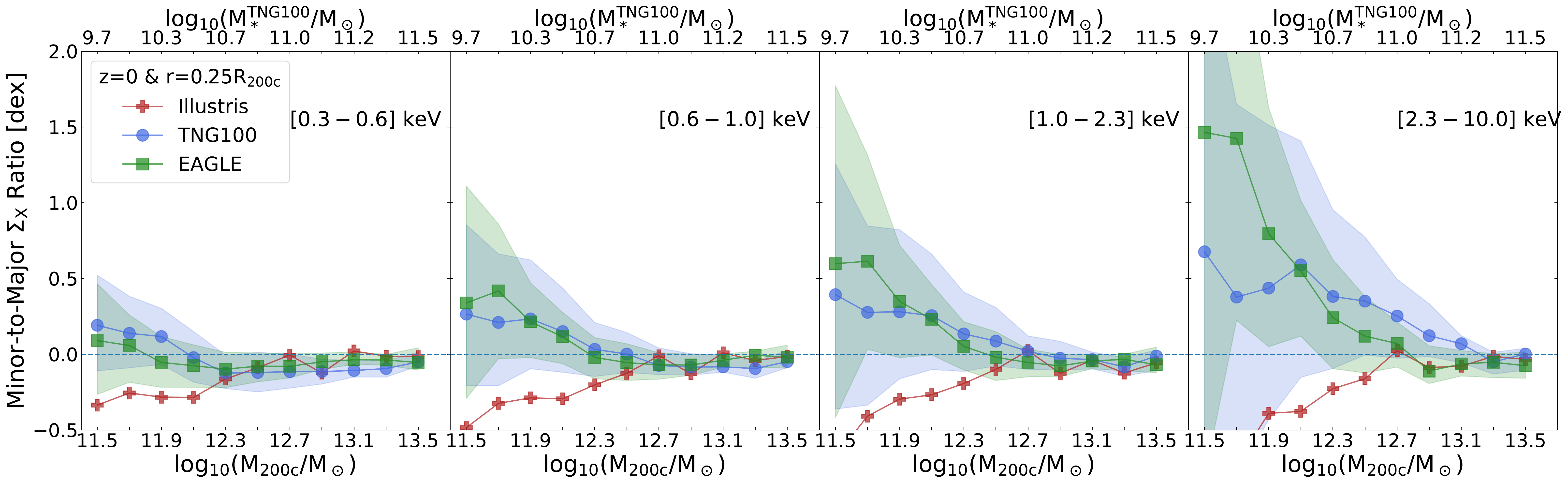}
  \includegraphics[width=0.99\textwidth]{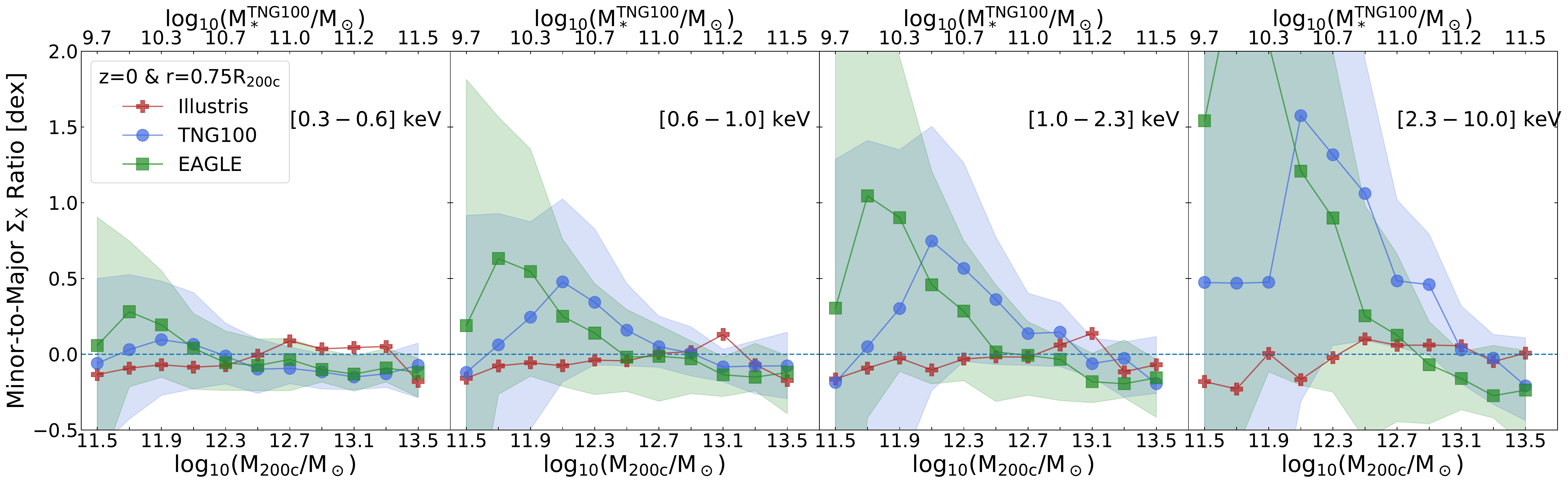}
   
  \caption{Similar to Figure~\ref{fig:Xray} but for the measurements at other distances: $r=0.25R_{\rm 200c}$ ({\it top}) and $r=0.75R_{\rm 200c}$ ({\it bottom}).}
  \label{fig:a2}
  \end{figure*}

In this Section we provide additional comparisons between TNG, EAGLE, and Illustris for measurements at other galactocentric distances of $0.25R_{\rm 200c}$ and $0.75R_{\rm 200c}$: X-ray hardness (Figure~\ref{fig:a3}), thermodynamics and metallicity content (Figure~\ref{fig:a1}), and X-ray surface brightness (Figure~\ref{fig:a2}).

Qualitatively, the comparison among the three simulations at large distance ${\rm r\sim0.75R_{200c}}$ is consistent with the results presented in the paper at ${\rm r\sim0.5R_{200c}}$. On the other hand, at smaller radii ${r\sim0.25R_{200c}}$, the Illustris simulation predicts anisotropic signals with significantly larger amplitudes at the low-mass end ($M_{\rm200c}\lesssim10^{12.3}M_\odot$) compared to the results at the other two radii, where Illustris produces almost isotropic distribution of thermodynamic properties as well as metal content. For instance, concerning the gas temperature ({\it top middle} panel of Figure~\ref{fig:a1}), the Illustris value of minor-to-major ratio at $M_{\rm200c}\sim10^{11.5}M_\odot$ is about $-0.4$ dex, compared to $\sim0.15$ dex value of TNG and EAGLE, namely Illustris predicts that at small radii the gas is colder along the minor axis in contrast with TNG and EAGLE predictions. The opposite predictions of Illustris, in comparison to the other two simulations, could be explained by its stellar feedback model, in which cold gas is deposited preferably along the minor axis (see also discussion in Section~\ref{sec:comparison}).

\section{X-ray emission-weighted quantities}
\label{sec:app_b}
\begin{figure*}
    \includegraphics[width=0.49\textwidth]{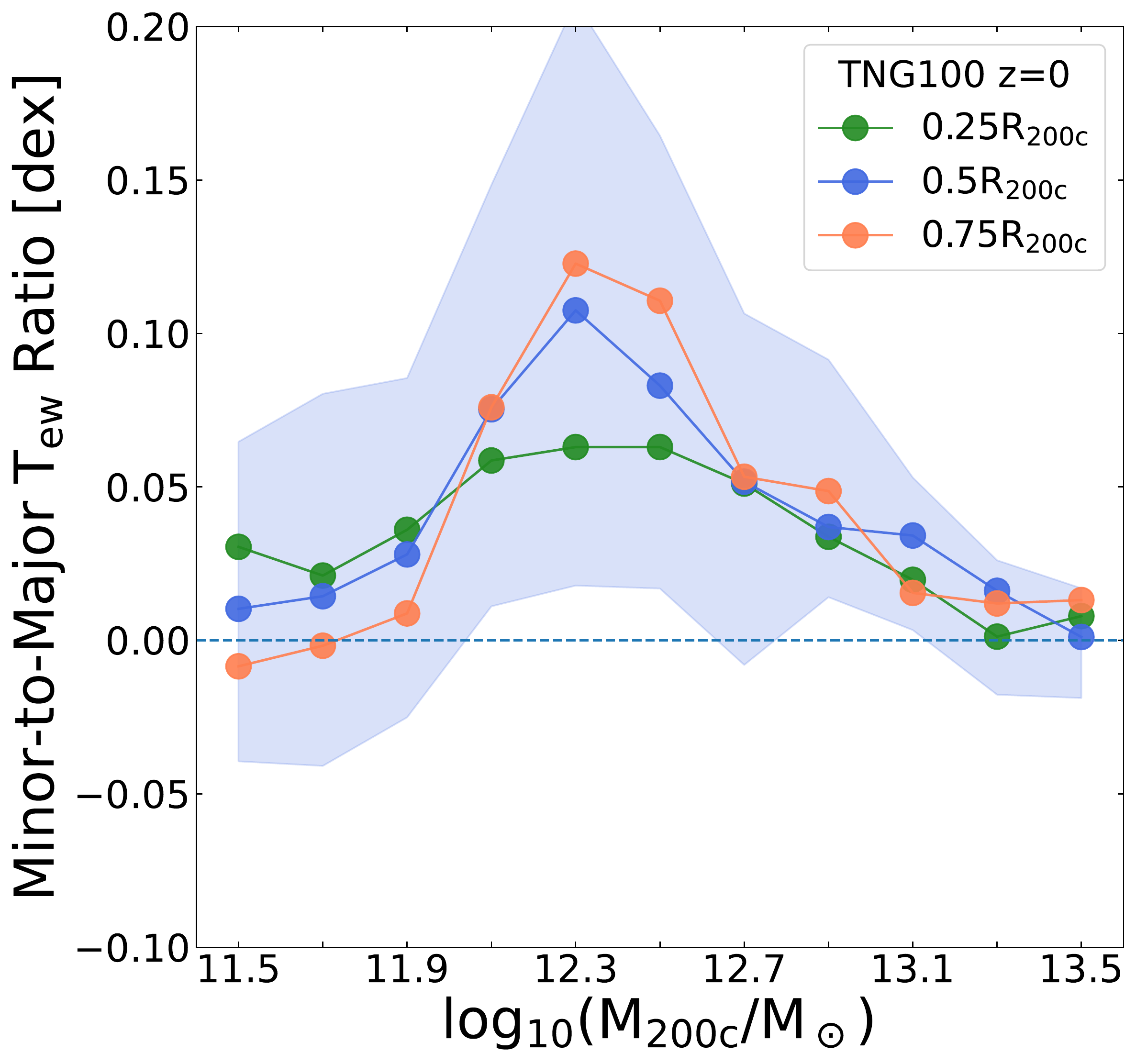}
    \includegraphics[width=0.49\textwidth]{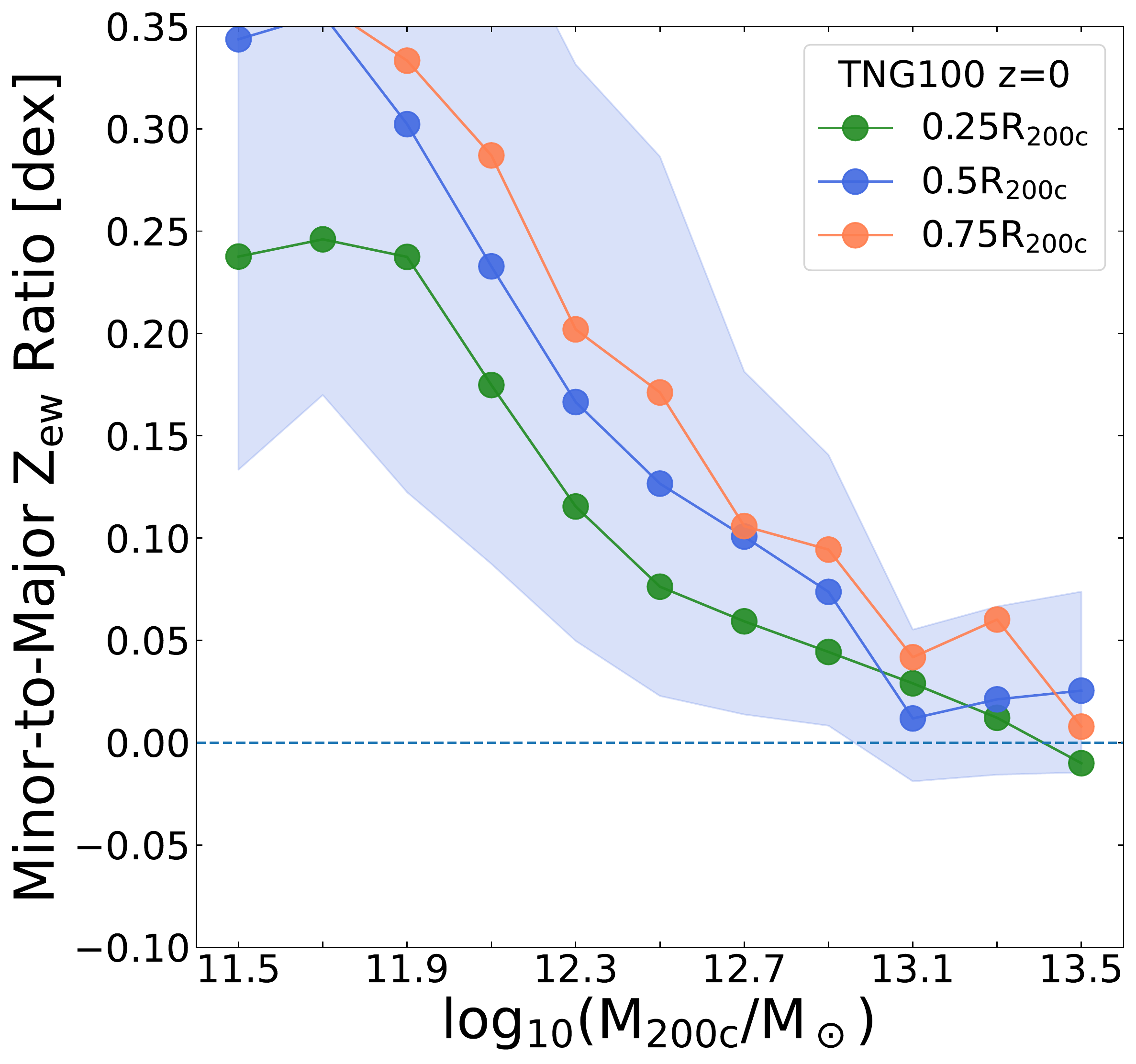}
    
\caption{Mass dependence of the anisotropic signals in gas temperature ({\it left}) and metallicity ({\it right}) according to the TNG100 simulation. This figure is similar to Figure~\ref{fig:mass_dependence} (middle column), but gives results for CGM quantities that are averaged with weights given by the X-ray emissivity of the gas, in the $[0.5-2.0]$ keV band.}
\label{fig:b1}
\end{figure*}
We examine the anisotropic signals in gas temperature and metallicity when their averaged values are taken as X-ray emission-weighted mean. The mass dependence of those signals, which is similar to that shown in Figure~\ref{fig:mass_dependence} (middle column), are shown in Figure~\ref{fig:b1}. The X-ray emission is computed in the energy band $[0.5-2.0]$ keV. 

In comparison to the mass-weighted quantities, the anisotropic signals in X-ray emission-weighted gas temperature and metallicity exhibit similar patterns. However, it is worth noting that the amplitude of emission-weighted temperature (metallicity) is overall larger by about 0.03 (0.1) dex except for temperature signal measured at small radii ($r\sim0.25R_{200c}$).
\label{lastpage}
\end{document}